\documentclass[lineno]{JFM-FLM_Au}

\usepackage{subfigure}
\usepackage{overpic}
\usepackage{array}
\usepackage{multirow}
\usepackage{xcolor}
\usepackage{makecell}
\definecolor{myRED}{HTML}{ff7f82}
\definecolor{myGREEN}{HTML}{4db578}
\definecolor{myBLUE}{HTML}{80aeff}
\definecolor{myPURPLE}{HTML}{ba7dff}

\lefttitle{H. Ji, Y. Luo, H. Zhou and Y. Zhao}
\righttitle{Journal of Fluid Mechanics}

\title{Progressive Mixture-of-Experts with autoencoder routing for continual RANS turbulence modelling}

\author{Haoyu Ji\aff{1,2},
  Yinhang Luo\aff{1},
  Hanyu Zhou\aff{1,2} \and Yaomin Zhao\aff{1,2}
  }

\affiliation{
	\aff{1}HEDPS, Center for Applied Physics and Technology, School of Mechanics and Engineering Science, Peking University, Beijing 100871, China
	\aff{2}State Key Laboratory for Turbulence and Complex Systems, School of Mechanics and Engineering Science, Peking University, Beijing 100871, China
}
 
\corresau{Yaomin Zhao, \email{yaomin.zhao@pku.edu.cn}}

\begin{document}
\maketitle

\begin{abstract}
Developing Reynolds-averaged Navier–Stokes (RANS) turbulence models that remain accurate across diverse flow regimes is a long-standing challenge. In this work, we propose a novel framework, termed the progressive mixture-of-experts (PMoE), designed to enable continual learning for RANS turbulence modelling. The framework employs a modular autoencoder-based router to associate each flow scenario with a specialised turbulence model, referred to as an expert. When a new flow regime cannot be adequately represented by the existing router and expert set, a new expert together with its routing component can be introduced at low cost, without modifying or degrading previously trained ones, thereby naturally avoiding catastrophic forgetting. The framework is applied to a range of flows with distinct physical characteristics, including airfoil wake, channel, periodic hill, and square duct flows. The resulting PMoE model effectively integrates multiple experts and achieves improved predictive accuracy across both seen and unseen test cases that differ in operating conditions or configurations. Owing to sparse activation, model expansion does not incur additional computational cost during inference. The proposed framework therefore provides a scalable pathway towards lifelong-learning turbulence models for industrial computational fluid dynamics.

\end{abstract}



\section{Introduction}\label{sec:Introduction}

Reynolds-averaged Navier-Stokes (RANS), supported by turbulence closure models, remains a predominant computational fluid dynamics (CFD) approach in industry.
RANS maintains a balance between computational efficiency and accuracy for predicting mean flow quantities, making it critical for applications across aerospace, automotive, and energy sectors \citep{rumsey2014turbulence_anayl,bush2019recommendations_anayl}. 
The main difficulty, however, lies in closing the Reynolds stress tensor, especially for complex flows such as boundary layer transition, separation and secondary flows.
Traditional closures are based on physical arguments, empirical tuning, or, more recently, data-driven calibration \citep{Launder1974KEpsilon,spalart1992SA,wilcox2008komega,ling2016TBNN}.

The demand for improving turbulence models has led to the use of machine learning methods \citep{Duraisamy2019Data_age,Brunton2020annurev}.
Many machine learning methods have been introduced into turbulence modelling, including Bayesian optimization \citep{xiao2016Bayesian}, ensemble Kalman filter \citep{Zhang2022EnKF}, adaboost decision trees \citep{Ling2015feature}, random forests \citep{Wang2017PIML}, tensor basis neural networks \citep{ling2016TBNN}, symbolic regression (SR) \citep{Weatheritt2016GEP,zhao2020GEP,schmelzer2020Sparse_SR}, among others.
In these studies, the modelling strategies vary considerably, from correcting the parameters of linear eddy viscosity models to forming completely new nonlinear eddy viscosity models, with the optimization targets varying from predicting transition \citep{duraisamy2015transition,zhang2023transition} to characterizing boundary layer separation \citep{zhu2019separated,wu2023separated}.  
These models, which are typically trained to improve predictive accuracy for specific cases, can thus be termed as expert models. 

Despite considerable advances, achieving robust generalization for data-driven turbulence models remains not only an active research area but also a significant challenge.
Based on the accumulated experience from industrial turbulence simulations, evidence suggests that a universal, simple, and local turbulence model may be difficult to achieve \citep{rumsey2022nasa}. 
To enhance generalization for traditional turbulence models, one approach is to perform parameter calibration in various application scenarios. 
The GEKO model, for example, provides empirical parameter combinations for the $k-\omega$ model under a wide range of cases \citep{menter2025GEKO}. 
This cross-case generalization capability, \emph{i.e.} the capability to make accurate predictions in diverse flow scenarios, is particularly important and even more challenging for models trained by machine learning methods \citep{sandberg2022machine}. 

To achieve generalization across diverse flow scenarios, researchers have explored numerous approaches that leverage model architecture design, training data diversity, feature engineering, and physical constraints.
These efforts can be broadly categorized into three distinct paradigms.
The first strategy focuses on parameter calibration using data-driven approaches to enhance adaptability. For instance, \citet{bin2023AIAA_SA} constructed a unified model by fine-tuning the parameters of the Spalart-Allmaras (SA) model, thereby extending its applicability across various flows.
The second approach seeks to improve generalization through multi-case training strategies, where the model is exposed to distinct geometries and physical characteristics during the learning phase. Notable examples include the symbolic-regression-based training developed by \citet{fang2023Multicase}, the multi-case surrogate optimization by \citet{Amarloo2023separation}, and the progressive data-augmented  framework to incrementally enhance cross-case generalization by \cite{rincon2025generalisable}.
The third category involves model aggregation or ensemble methods, which utilize machine-learning-based weighting functions to blend multiple baseline models. 
In a series of recent studies \citep{de2024XMA,cherroud2025XMA,oulghelou2025XMA}, machine learning methods were employed to train weighting functions, successfully leveraging the strengths of different data-driven models to achieve improved accuracy across multiple cases.
Parallel to these RANS-based developments, significant progress has also been made in wall-modelled large eddy simulations (WMLES) \citep{Lozano2023BBF,arranz2024BBF}, in which a model trained on several canonical flows exhibits good agreement with reference data for both canonical tests and realistic aircraft configurations.

It should be noted that most existing data-driven models, either enhanced by multi-case joint training or model aggregation, are considered reliable mostly within the vicinity of their training distributions, while applying to completely unknown flow regimes remains a critical challenge.
In practical CFD applications, users are frequently confronted with novel flow configurations and operating conditions distinct from prior experiences.
Directly training the existing model with newly obtained data often degrades its original performance, which is known as catastrophic forgetting \citep{french1999catastrophic}.
A common approach is to retrain the model jointly on both pre-existing and new data, but this approach might face a series of challenges. 
As new regimes arise continuously in realistic applications, joint retraining can be hindered by complex optimisation tasks, data accessibility issues, and performance degradation on previously well-modelled cases, all of which motivate continual learning.
In order to achieve continual learning for WMLES, \cite{zhang2025KIA} proposed an additive framework for training wall models, showing promising success in a series of cases.
Nevertheless, novel frameworks with the capability of continual learning are still needed for developing generalisable RANS models.

To achieve a framework capable of sustainable generalization, we propose a novel approach termed progressive mixture-of-experts (PMoE) in the present study.
The PMoE method is founded on the mixture-of-experts (MoE) architecture \citep{Jacobs1991MoE}, which has been the subject of much research in the field of machine learning \citep{shazeer2017sparse_gating,Fedus2022MoE} and becomes very popular in the field of large language models due to its scalability \citep{du2022glam,dai2024deepseekmoe}.
Building on the typical architecture consisting of a router and a group of experts, the MoE offers a distinct advantage for turbulence modelling, \emph{i.e.} the decomposition of a complex, high-dimensional physical problem into manageable sub-tasks.
Central to this architecture is the mechanism of sparse activation.
Unlike monolithic models, the router in an MoE network activates only a specific subset of expert models for a given local flow state, leaving the majority silent during the inference process.
This property is crucial for industrial CFD applications, as it does not incur the prohibitive computational cost usually associated with large-scale neural networks, while achieving sufficient representational expressiveness in order to deal with various scenarios featuring different flow physics.

One important novel feature of the present PMoE framework, which is missing for the generic MoE structure, is the continual-learning capability. 
This is in order to achieve sustainable generalization, enabling the model to incorporate newly introduced datasets without full retraining.
Accordingly, the PMoE is designed to be equipped with a modular autoencoder-based router.
The PMoE framework is applied to develop a turbulence model by successively introducing the data of four flow regimes, including airfoil wake, channel, periodic hill, and square duct flows.
The trained PMoE model is then extensively assessed, validating its performance in training cases and unseen cases with various operating conditions and configurations.
The objective is mainly to show the ability of PMoE to incorporate new flow scenarios without retraining or degrading existing experts.

Before introducing the framework in detail, it is important to highlight three key issues that need to be addressed to develop a PMoE turbulence model generalisable for various flow regimes.
First, the PMoE framework needs to classify diverse flow scenarios in an unsupervised manner using only RANS data, and this requirement brings a significant challenge regarding the selection and extraction of appropriate physical features for the pre-processing procedure to serve as effective model inputs. 
Second, the router architecture must support continual-learning capabilities, determining whether a new case belongs to a known flow regime or represents a novel one. 
For the latter, the architecture must allow for the rapid addition of new modules without the need for retraining. 
Finally, the distinct nature of turbulence across different flows implies that a unified model form may be insufficient \citep{Duraisamy2019Data_age}, and different flow regimes often require fundamentally different correction forms. 
Consequently, the design of specific expert architectures tailored to these varying physical characteristics is crucial for the success of a truly generalisable model.
These topics will be addressed in detail in the following sections.

This paper is organized as follows. 
The structure of the new framework is discussed in \textsection~\ref{sec:Methodology}. 
The training process and the setting of cases are introduced in \textsection~\ref{sec:PMoE}. 
The model assessment, validation and computational cost analysis are given in \textsection~\ref{sec:validation}. 
Finally, conclusions are offered in \textsection~\ref{sec:conclusion}.

\section{Methodology}\label{sec:Methodology}
The MoE framework, which has the potential to develop generalisable turbulence models, is introduced in \textsection~\ref{subsec:GMoE}.
Moreover, the router part of a generic MoE framework is typically implemented by multilayer perceptron (MLP), which enables multi-task learning but requires further improvement for continual learning.
In \textsection~\ref{subsec:PMoE}, we introduce the PMoE structure, demonstrating how the newly proposed framework facilitates continual generalization of turbulence modelling.

\subsection{Introduction to the generic mixture-of-experts framework}\label{subsec:GMoE}
\begin{figure}
    \hspace*{-0.2cm}
  \centerline{
  \begin{overpic}[height=0.42\textwidth]{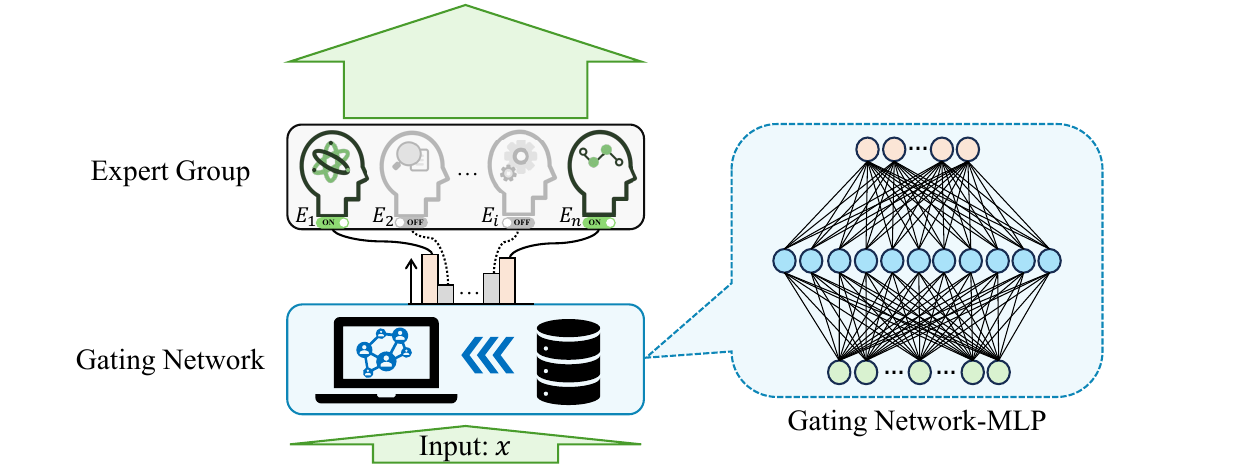} 
        \put(19.3,27.4){$(a)$}
        \put(56,27.4){$(b)$}
        \put(60,30){$G_{\sigma}(\boldsymbol{x})=\mathrm{Softmax}\left(\boldsymbol{x} \mathsfbi{W}_g\right)$}
        \put(25,15){$G(\boldsymbol{x})_{i}$}
        \put(28.8,30.5){$\sum_{i=1}^{n} G(\boldsymbol{x})_{i} E_{i}(\boldsymbol{x})$}
    \end{overpic}
    }
  \captionsetup{justification=justified, singlelinecheck=true}
  \caption{Schematic of the generic MoE framework with a MLP softmax gating network. (a) The whole process of the MoE framework. (b) Zoom-in view of a gating network based on MLP.}
\label{fig:MoE}
\end{figure}

As shown in figure~\ref{fig:MoE}, the generic MoE consists of a set of $n$ experts $\{E_1, \dots,E_n\}$ and a gating network (also named as a router) $G$, and the key advantage of this framework is the decomposition of a complex, high-dimensional problem into manageable sub-tasks dealt by different experts.
For turbulence modelling, each expert $E_i$ can be viewed as a specialized closure model tailored for a specific flow regime, while the router $G$ functions as a flow regime classifier.

For a set of $n$ experts, the output $\boldsymbol{y}$ for a given input flow state $\boldsymbol{x}$ is the weighted sum of expert predictions:
\begin{equation}
    \boldsymbol{y} = \sum_{i=1}^{n} G(\boldsymbol{x})_i E_i(\boldsymbol{x}).
    \label{eq:moe_sum}
\end{equation}
Here, $G(\boldsymbol{x})_i$ is the $i^{\mathrm{th}}$ component of the output vector of $G$ and represents the probability (or confidence) that the corresponding flow $\boldsymbol{x}$ belongs to the regime governed by expert $i$.
A simple choice of the gating function \citep{Jordan1994MoE}, as shown in figure~\ref{fig:MoE}(b) for example, is to multiply the input $\boldsymbol{x}$ by a trainable weighting matrix $\mathsfbi{W}_g$ and then apply the $\mathrm{Softmax}(\cdot)
$ function, as
\begin{equation}
    G_{\sigma}(\boldsymbol{x})=\mathrm{Softmax}\left(\boldsymbol{x} \mathsfbi{W}_{g}\right),
    \label{Router_MLP}
\end{equation}
where the subscript $\sigma$ denotes the Softmax nonlinearity.

The key advantage of the MoE framework lies in its sparse activation mechanism, which dramatically reduces computational costs while maintaining model capacity. 
Typically, sparse MoE employs a top-$k$ routing strategy where only the most relevant $k$ experts are activated for each input. 
For example, we set $k=1$ in our current implementation, thus the router selects the one most relevant expert per input token, ensuring that the rest $n-1$ experts remain completely dormant during model inference.
With this feature of sparse activation, the computational cost of the MoE model, despite having a large pool of experts $\{E_1, \dots,E_n\}$, scales only with the size of a single expert rather than the entire ensemble. 
Consequently, the inference cost remains nearly constant regardless of the total number of experts, enabling the deployment of extremely large-scale models with billions or even trillions of parameters with relatively low computational requirements \citep{shazeer2017sparse_gating,Fedus2022MoE}. 

While the generic MoE offers a path toward modular turbulence modelling, applying it to industrial CFD requires a capability that standard architectures lack, \emph{i.e.} continual learning.
A truly generalisable turbulence model must be able to assimilate new flow physics (e.g., transitioning from simple shear flows to complex corner separations) without ''forgetting'' previously learned regimes.
However, standard MLP-based routers in figure~\ref{fig:MoE} are highly susceptible to catastrophic forgetting \citep{french1999catastrophic}, as updating the weights to accommodate a new flow regime often degrades the classification accuracy for existing ones, highlighting the urgent need to extend the framework with continual learning capability.

In the broader machine learning literature, continual-learning is attracting more and more interests, and the corresponding strategies can be generally classified into three paradigms \citep{wang2024comprehensiveContinal}: (i) regularisation-based methods, which constrain weight updates to preserve important parameters \citep[e.g.][]{kirkpatrick2017EWC}; (ii) replay-based methods, which retain or generate representative samples from earlier tasks for joint retraining \citep[e.g.][]{rolnick2019ER}; and (iii) architecture-based methods, which dedicate separate model components to different tasks to prevent interference \citep[e.g.][]{Rusu2016PNN}. 
Among architecture-based approaches, \citet{aljundi2017expert} proposed Expert Gate, a lifelong learning framework in which each task is associated with a dedicated autoencoder gate and expert network. 
At test time, the autoencoder with the lowest reconstruction error routes the input to its corresponding expert, while the reconstruction errors also quantify inter-task relatedness to guide knowledge transfer when training new experts. 
This design demonstrated effective sequential task learning for image classification. 
The present PMoE framework builds on this broad principle of modular autoencoder-based routing, while introducing several novel strategies essential for turbulence modelling, as discussed in the following sections.

\subsection{Progressive mixture-of-experts framework for turbulence modelling}\label{subsec:PMoE}

\begin{figure}
    \hspace*{-0.2cm}
  \centerline{\includegraphics[height=0.40 \textwidth]{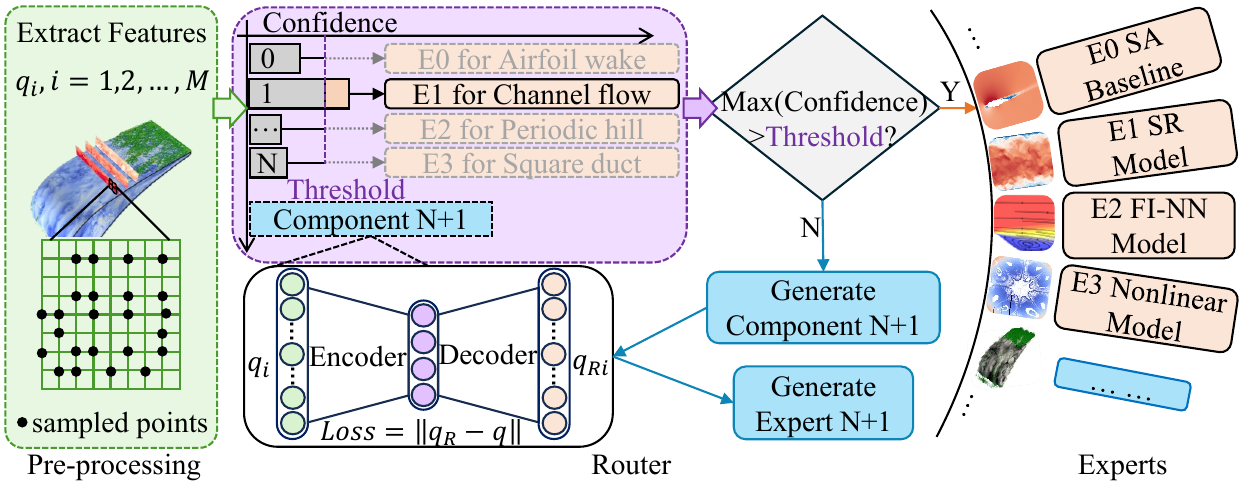}}
  \captionsetup{justification=justified, singlelinecheck=true}
  \caption{Schematic of the proposed PMoE framework.The framework consists of three parts: (i) a pre-processing stage that extracts physics-informed features from the baseline RANS field; (ii) a modular autoencoder-based router that classifies the incoming flow regime by comparing reconstruction errors across components, and triggers model expansion when no existing component achieves sufficient confidence; and (iii) a sparsely activated group of heterogeneous experts, each tailored to a specific flow regime.}
\label{fig:PMoE}
\end{figure}

As shown by the schematic in figure~\ref{fig:PMoE}, the PMoE framework consists of three key components, including the procedure of pre-processing to extract flow features as model inputs, the router structure responsible for classifying various types of flows, and a group of experts each trained for a specified flow scenario.
In particular, the PMoE replaces the MLP router with a modular bank of autoencoders, each dedicated to recognizing a specific flow signature.
This ensures that the introduction of a new expert and its corresponding router module does not interfere with the existing modules.
Accordingly, the PMoE framework is expected to develop progressively with new datasets introduced successively, and the details are discussed in the following.

\subsubsection{Feature extraction and physics-informed sampling}\label{subsubsec_mask}

During the pre-processing stage, the input to the PMoE framework relies on local flow features extracted from a baseline RANS calculation.
In the present study, we use the SA model \citep{spalart1992SA} as the baseline due to its wide applications.
Following previous data-driven RANS efforts \citep{Ling2015feature,Wang2017PIML}, we construct an input vector $\boldsymbol{x} = [q_1, q_2, \dots, q_M]$ consisting of $M=7$ flow features, as detailed in table~\ref{tab:features}.
Note that the listed variables are normalized in a way of $A/({\left |A \right |+\left |B \right |})$ to ensure consistent scaling across different flow regimes.

\begin{table}
  \begin{center}
    \begin{tabular}{
      >{\centering\arraybackslash}p{1.9cm}
      >{\centering\arraybackslash}p{4.8cm}
      >{\centering\arraybackslash}p{2.3cm}
      >{\centering\arraybackslash}p{3.5cm}
    }
      \textbf{Feature $(q_i)$} & \multicolumn{1}{c}{\textbf{Description}} & \textbf{Raw feature $(\hat{q}_i)$} & \textbf{Normalization factor $(q_i^*)$} \\
      [3pt]
      \hline
      $q_1$ & Vorticity magnitude & $\displaystyle \|\mathsfbi{W}\|$ & $\displaystyle \|\mathsfbi{S}\|$ \\
      [10pt]
      $q_2$ & Ratio of excess rotation rate to strain rate (Q criterion) &$\displaystyle \frac{1}2{}({\|\mathsfbi{W}\|^2 - \|\mathsfbi{S}\|^2})$ & $\displaystyle \| \mathsfbi{S}\|^2$\\
      [14pt]
      $q_3$ & Ratio of pressure normal stresses to shear stresses & $\displaystyle \sqrt{\frac{\partial \boldsymbol{P}}{\partial x_{i}} \frac{\partial \boldsymbol{P}}{\partial x_{i}}}$ & $\displaystyle \frac{1}{2} \rho \frac{\partial U_{k}^{2}}{\partial x_{k}}$\\
      [16pt]
      $q_4$ & Distance to the nearest wall &$\displaystyle d$ & $\displaystyle L_\mathrm{ref}$\\
      [14pt]
      $q_5$ & SA viscosity &$\displaystyle \tilde{\nu}$ & $\displaystyle 100\nu$\\
      [14pt]
      $q_6$ & Pressure gradient along streamline & $\displaystyle U_{l} \frac{\partial \boldsymbol{P}}{\partial x_{l}}$ & $\displaystyle \sqrt{\frac{\partial \boldsymbol{P}}{\partial x_{j}} \frac{\partial \boldsymbol{P}}{\partial x_{j}} U_{i} U_{i}}$\\[19pt]
      $q_7$ & Nonorthogonality between velocity and its gradient \citep{gorle2012ransq8} & $\displaystyle \left|U_{i} U_{j} \frac{\partial U_{i}}{\partial x_{j}}\right|$ & $\displaystyle \sqrt{U_{l} U_{l} U_{i} \frac{\partial U_{i}}{\partial x_{j}} U_{k} \frac{\partial U_{k}}{\partial x_{j}}}$ \\
    \end{tabular}
    \captionsetup{justification=justified, singlelinecheck=true}
    \caption{Flow features used as input in the framework. $\mathsfbi{W}$ is the rotation rate tensor, $\mathsfbi{S}$ is the strain rate tensor, $\rho$ is the fluid density, $\nu$ is the fluid viscosity, $\tilde{\nu}$ is the SA modified turbulent eddy viscosity, $d$ is the distance to the nearest wall, $L_\mathrm{ref}$ is the reference height, and $U_i$ is the mean velocity.
    The normalized feature $q_i$ is obtained by normalizing the corresponding raw features value $\hat{q}_i$ with normalization factor $q_i^*$ according to $q_i={\hat{q}_i}/({\left |\hat{q}_i \right |+\left |q_i^* \right |})$.}
    \label{tab:features}
  \end{center}
\end{table}

We note that the features listed in table~\ref{tab:features} are not universally effective across all flow regimes. 
Prior studies \citep{yin2020feature,de2024XMA,cherroud2025XMA} suggest that a subset of these variables may often be sufficient for specific cases, and methods such as permutation feature importance (PFI) \citep{wu2025framework_PFI} can be employed to identify the relevant features for a specific case. 
Nevertheless, to accommodate future applications that may involve a broad range of flow regimes with distinct critical features, we adopt the full set as the model input.

A critical challenge in training a generalisable router is ensuring that the training data represents the dynamically active regions of the flow, rather than being dominated by the free stream or quiescent zones. Therefore, we employ a physics-informed sampling strategy during the pre-processing stage. Rather than uniformly sampling the entire computational domain, the sampling region is restricted based on the underlying flow physics. Grid points are selected only where either the strain-rate magnitude $\|\mathsfbi{S}\|$ or the vorticity magnitude $\|\mathsfbi{W}\|$ exceeds a specific threshold relative to the domain statistics. Specifically, we define the sampling mask $\mathcal{M}$ such that a point is included if
\begin{equation}
\|\mathsfbi{W}_j\| > \epsilon \cdot \mathrm{median}( \|\mathsfbi{W}\|) \quad \mathrm{or} \quad \displaystyle \|\mathsfbi{S}_j\| > \epsilon \cdot \mathrm{median}( \|\mathsfbi{S}\|),
\label{eq:mask}
\end{equation}
where $\|\mathsfbi{W}_j\|$ and $\|\mathsfbi{S}_j\|$ denotes the magnitudes of vorticity and strain-rate at the $j^{\mathrm{th}}$ sampled point, respectively, and $\epsilon$ is a thresholding parameter set to $0.05$ in the present study. A sensitivity test has been performed and the results are robust within $\epsilon \in \{0.01, 0.05, 0.1\}$.
This criterion ensures that the selected samples are concentrated in boundary layers, shear layers, and separation zones, etc., effectively filtering out grid points of limited physical significance such as those in freestream regions. 
From this masked region, $N$ points are randomly sampled to form the input matrix $\mathsfbi{X} \in \mathbb{R}^{N \times M}$.
The masking criterion in equation~\ref{eq:mask} is applied only during the sampling stage for the PMoE router. 
In regions with limited physical significance such as freestream regions, the baseline RANS predictions are generally adequate to provide accurate predictions, thus no extra corrections are needed.

\subsubsection{Modular autoencoder-based router for continual learning}\label{subsubsec_modular}

\begin{figure}
    \hspace*{-0.1cm}
  \centerline{
  \begin{overpic}[height=0.40\textwidth]{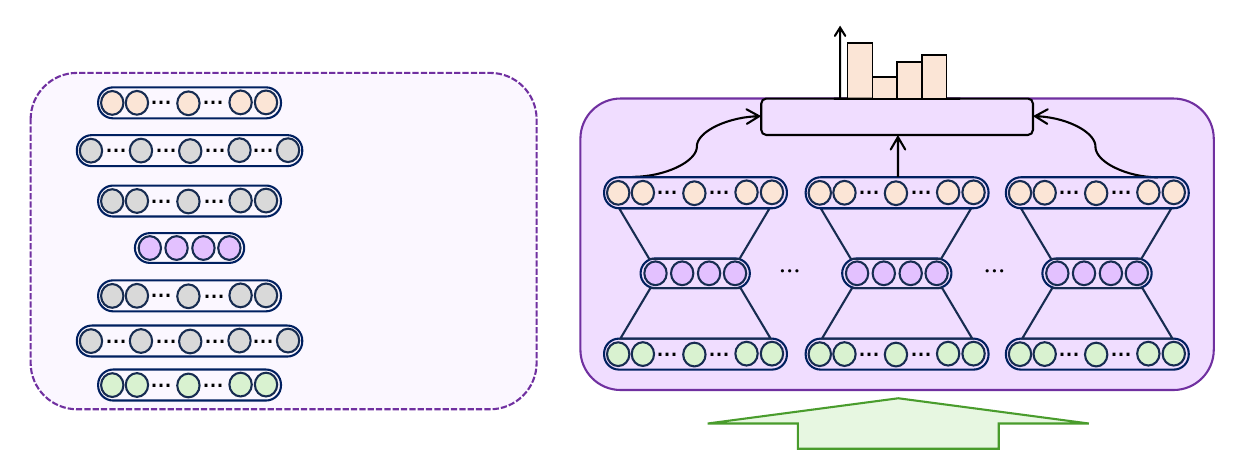} 
        \put(0.3,31){$(a)$}
        \put(44.6,31){$(b)$}
        \put(6,6){$7$}
        \put(6,29){$7$}
        \put(27,6){Input Layer}
        \put(26,29){Output Layer}
        \put(3,9.6){$16$}
        \put(6,13.3){$8$}
        \put(27.2,11){Encoder $f_\theta$}
        \put(3,25){$16$}
        \put(6,21){$8$}
        \put(27.2,23){Decoder $g_\phi$}
        \put(9,17.2){$4$}
        \put(24,17.2){Bottleneck Layer}
        \put(9,2){Structure of the Component}
        \put(65,2.7){Input Features}
        \put(51.7,11.7){Encoder}
        \put(51.5,18.7){Decoder}
        \put(67.9,11.7){Encoder}
        \put(67.8,18.7){Decoder}
        \put(84,11.7){Encoder}
        \put(83.8,18.7){Decoder}
        \put(62,11.7){$x_i$}
        \put(61.5,18.7){$\hat{x}_i$}
        \put(63,27.9){$\mathrm{Loss}=\| \boldsymbol{\hat{x}} - \boldsymbol{x} \|$}
        
    \end{overpic}
    }
  \captionsetup{justification=justified, singlelinecheck=true}
  \caption{Structure of the router based on autoencoder. (a) Neural network architecture design for the autoencoder component. (b) The modular architecture of the router.  }
\label{fig:Router}
\end{figure}

The core of the PMoE framework is its ability to determine whether the incoming case belongs to a known scenario or represents a novel one that requires model expansion.
As illustrated in figure~\ref{fig:Router}, the PMoE router features a modular architecture, in which each router component adopts a standard autoencoder architecture following the design of \cite{hinton2006autoencoder}, aiming to compress and reconstruct the feature space of a specific flow regime.
The idea of learning internal representations through minimizing input reconstruction traces back to the backpropagation framework of \cite{rumelhart1986autoencoder}.
The network in this study consists of an encoder $f_{\theta}$ mapping the input $\boldsymbol{x}$ to a latent representation $\boldsymbol{z}$, and a decoder $g_{\phi}$ reconstructing it as $\hat{\boldsymbol{x}}$.
The network parameters of the encoder and the decoder shown in figure~\ref{fig:Router}(a) are presented by $\theta$ and $\phi$ respectively.
The training objective for the autoencoder is to minimize the reconstruction loss over the $N$ sampled points, as
\begin{equation}
    \boldsymbol{z}=f_{\theta}(\boldsymbol{x}), \quad \boldsymbol{\hat{x}}=g_{\phi}(\boldsymbol{z}), \quad \theta, \phi=\arg \min _{\theta, \phi} \mathcal{L}(\boldsymbol{x}, \boldsymbol{\hat{x}}).
    \label{equation_autoencoder}
\end{equation}
Here, the reconstruction error $\mathcal{L}(\boldsymbol{x},\boldsymbol{\hat{x}})$ is assessed as 
\begin{equation}
\mathcal{L}(\boldsymbol{x}, \boldsymbol{\hat{x}})=\frac{1}{N}\sum_{j}^{N}L_j=\frac{1}{N}\sum_{j}^{N}\sqrt{\frac{1}{M}\sum_{i}^{M} (\hat{x}_{i,j}-x_{i,j}) ^2  },
\label{equation:Loss}
\end{equation}
where for the $j^\mathrm{th}$ point, $x_{i,j}$ is the $i^{\mathrm{th}}$ component of its input feature vector, and $L_j = \sqrt{ \frac{1}{M} \sum_{i=1}^{M} \left( \hat{x}_{i,j} - x_{i,j} \right)^2 }$ refers to its corresponding reconstruction error over the $M$ features.

For every router component, a compact architecture is adopted, consisting of several hidden layers with moderate width to balance representational capacity and generalization as shown in figure~\ref{fig:Router}(a). 
Leaky Rectified Linear Unit (LeakyReLU) \citep{maas2013LeakyReLU} activations are employed between layers to introduce nonlinearity, and the Adam optimizer \citep{Kingma2014AdamAM} is used with an initial learning rate of $10^{-2}$.
A step-based learning rate scheduler is employed, where the learning rate is reduced by a factor of $0.98$ every $100$ epochs to improve convergence stability.
The number of training epochs is capped at $5000$ based on empirical convergence studies, while an early-stopping strategy is employed to improve efficiency and mitigate overfitting.
For all component trainings, the loss consistently reaches a plateau well before the maximum epoch count, and extending the training further does not yield noticeable performance improvement.
After the training of the $k^\mathrm{th}$ router component $C_k$ is converged, we set the $99.9\mathrm{th}$ percentile of the reconstruction error of the training data as the corresponding threshold $T_k$, in order to eliminate the misleading influence of points with extreme distributions.
A sensitivity test has been performed and the results are robust with $T_k>99\%$.

During the inference process, the router must determine if a new input flow belongs to a known regime or represents a novel physical scenario through the reconstruction error. 
For a new input case with $N$ sampled points, a point $j$ is considered recognized by component $C_k$ if its local reconstruction error satisfies $L_{j} < T_k$. 
The global confidence level $p_k$ of expert $k$ regarding the current flow is defined as the relative frequency of recognized points 
\begin{equation}
p_k = \frac{N_k}{N}, \quad \text{where } N_k = \sum_{j}^{N} \mathbb{I}(L_j < T_k),
\end{equation}
where $\mathbb{I}(\cdot)$ is the indicator function.

The framework employs a \textit{winner-takes-all} or top-$1$ gating strategy. 
The router selects the expert $E_{\mathcal{K}}$ corresponding to the component with the maximum confidence as
\begin{equation}
\mathcal{K} = \arg\max_{k} p_k.
\end{equation}
However, activation is conditional and governed by a global acceptance threshold $T_{\mathrm{accept}}$. 
When the maximum confidence $p_\mathcal{K}$ exceeds a threshold of $T_{\mathrm{accept}}=90\%$, the expert $E_\mathcal{K}$ will be activated.
Conversely, if $p_\mathcal{K}$ falls below $T_{\mathrm{accept}}$, the flow is deemed unknown, which triggers the continual learning process.
In this case, a new autoencoder component $C_{\mathrm{new}}$ and a new expert $E_{\mathrm{new}}$ are initialized and trained on the new dataset. 
It should be emphasised that the expansion process is strictly additive. When a new regime is introduced, the corresponding router component and expert are trained while all previously learned components and experts remain frozen.
This modular design ensures isolation between old and new components during training, thus inherently overcoming catastrophic forgetting.
Note while a threshold of $T_{\mathrm{accept}}=90\%$ is found to be effective for the distinct regimes in the present study, future industrial applications may require this hyperparameter to be tuned based on the desired sensitivity to novel flow physics.

The rationale for employing an autoencoder for flow classification rests on the premise that, the hierarchical importance of physical features varies significantly across different flow scenarios, which can be seen from the feature important analysis in appendix~\ref{appA}.
During the routing stage, an input feature matrix $\mathsfbi{X}\in\mathbb{R}^{N\times M}$ is compressed into a low-dimensional bottleneck representation $\mathcal{H}_\theta = f_\theta(\mathsfbi{X})$ to extract the feature combinations most critical for reconstructing the local flow physics.
Following the information bottleneck (IB) principle proposed by \cite{tishby2000informationbottleneck}, the training objective is formulated as
\begin{equation}
\min_{\theta}\left[I(\mathsfbi{X};\mathcal{H}_\theta)-\gamma I(\mathcal{H}_\theta;\mathsfbi{Y})\right],
\label{IB}
\end{equation}
where $\theta$ denotes the encoder parameters, $I(\cdot;\cdot)$ represents mutual information, and $\gamma$ governs the trade-off between compression and relevance. For the reconstruction task, the output matrix $\mathsfbi{Y}\in\mathbb{R}^{N\times M}$ is taken to be consistent with the input $\mathsfbi{X}$. The IB formulation implies the existence of a minimal informative feature subset $S \subseteq \{1, 2, \dots, M\}$ such that the sub-matrix $\mathsfbi{X}_S\in\mathbb{R}^{N\times |S|}$ satisfies
\begin{equation}
I(\mathsfbi{X}_S;\mathsfbi{Y})=I(\mathsfbi{X};\mathsfbi{Y}),\quad
I(\mathsfbi{X}_{S'};\mathsfbi{Y})<I(\mathsfbi{X};\mathsfbi{Y}),\quad \forall S'\subset S .
\label{submatrix}
\end{equation}
This implies that the autoencoder effectively isolates feature combinations that are strictly relevant to flow reconstruction. 
Consequently, for distinct flow scenarios, the network relies on different minimal subsets. 
This mechanism enables flow identification by exploiting the dependence of the latent representation $\mathcal{H}_\theta$ on flow-specific informative features.
Note that the dimension of the latent bottleneck layer is a critical parameter which determines how much the input data can be compressed.
We set the latent layer to have four nodes in the present study as shown in figure~\ref{fig:Router}(a), which is based on a series of sensitivity tests.

\subsubsection{Sparsely activated expert group with heterogeneous formulation}
\label{subsec_heterogeneous_experts}

A distinct advantage of the PMoE framework is its explicit exploitation of heterogeneous expert formulations.
Unlike data-driven models that typically enforce a single-form  closure correction, e.g., a neural network predicting Reynolds stress anisotropy \citep{ling2016TBNN}, the PMoE allows each expert $E_k$ to adopt the mathematical structure best suited for its specific flow regime. 
This flexibility ensures that simple flows can be modelled by interpretable algebraic corrections, while complex non-equilibrium flows can leverage high-capacity neural networks.
We remark that although conventional MoE architectures can in principle accommodate different expert structures, standard implementations almost universally employ homogeneous experts due to the requirement of joint end-to-end training. 
In the PMoE framework, each expert is trained independently, removing this constraint and enabling fundamentally different closure strategies.

In the present implementation, we utilize the Spalart--Allmaras (SA) model \citep{spalart1992SA} as the baseline model. 
The standard transport equation for the modified eddy viscosity $\tilde{\nu}$ is given by
\begin{equation}
  \frac{\mathrm{D} \tilde{\nu}}{\mathrm{D} t} = \underbrace{C_{b1} \tilde{S} \tilde{\nu}}_{\mathcal{P}} + \underbrace{\frac{1}{\sigma}\left[\bnabla\bcdot((\nu+\tilde{\nu}) \bnabla \tilde{\nu})+C_{b2}(\bnabla \tilde{\nu})^{2}\right]}_{\mathcal{T}} - \underbrace{C_{w1} f_{w}\left(\frac{\tilde{\nu}}{d}\right)^{2}}_{\mathcal{D}},
  \label{SA_equation}
\end{equation}
where $\mathcal{P}$, $\mathcal{T}$, and $\mathcal{D}$ represent the production, transport, and destruction terms, respectively. 
The standard definitions for the model constants (such as $C_{b1}$, $C_{b2}$, $\sigma$ and $C_{w1}$) and closure functions like $f_w$ are retained from \citet{spalart1992SA}, and we only present the terms that will be modified here for brevity. 
In particular, the turbulent eddy viscosity is computed from $\nu_{t}=\tilde{\nu} f_{\nu 1}$ and $f_{\nu 1}$ is defined as 
\begin{equation}
  f_{\nu 1}=\frac{\chi^{3}}{\chi^{3}+c_{\nu 1}^{3}}, \quad \chi \equiv {\tilde{\nu}}/{\nu},
  \label{equ:f_nu1}
\end{equation}
where $c_{\nu 1}=7.1$ and $\chi$ represents the ratio between $\tilde{\nu}$ and the molecular viscosity $\nu$.
Moreover, $\tilde{S}$ is the modified vorticity magnitude.
Within the PMoE framework, we demonstrate the integration of three distinct expert formulations.

\textbf{Type I: parameter correction for wall-attached flows.} 
For equilibrium boundary layers where the baseline model structure is sound but parameter calibration is suboptimal, the expert provides a scalar correction to existing coefficients. For the wall-attached expert $E_1$, we employ a symbolic regression approach \citep{Weatheritt2016GEP} to recalibrate the damping function $f_{v1}$ as suggested by \citet{bin2023AIAA_SA}. 
The expert predicts a corrected functional form which replaces the standard definition in (\ref{equ:f_nu1}) to better capture the logarithmic layer velocity profile.

\textbf{Type II: correction of production term for flow separation.} 
For flows dominated by separation and strong pressure gradients, the equilibrium assumption in the linear eddy viscosity hypothesis often fails. 
In such regions, conventional models may mischaracterize the eddy viscosity distribution near separation and within recirculation zones, which in turn affects the prediction of separation and reattachment. 
This behavior is associated with the difficulty of linear eddy-viscosity models in representing the non-equilibrium balance between production and dissipation under adverse pressure gradients and rapidly varying strain rates.
Here, the expert $E_2$ is formulated as a neural network derived from the field inversion and machine learning (FIML) method. 
FIML is chosen because it has been widely applied and demonstrated success in separation corrections   \citep{Eric2016FI}, produces spatially varying correction fields that are well suited for flows where modelling errors are localized, and is compatible with standard RANS solvers.
It imposes a spatially varying correction term $\beta$ into the production term of the transport equation:
\begin{equation}
  \mathcal{P}^* = \beta(\boldsymbol{x}) \cdot C_{b1} \tilde{S} \tilde{\nu}.
  \label{SA_equation_beta}
\end{equation}
This multiplicative correction allows the model to locally suppress or enhance turbulence production in recirculation zones without altering the model behavior in the free stream.

\textbf{Type III: constitutive relation modification for secondary flows.} 
For corner flows where linear eddy viscosity models fail to predict secondary motions due to the isotropy assumption, which constrains Reynolds stress anisotropy and suppresses cross-stream momentum redistribution, the expert $E_3$ modifies the stress-strain relationship itself. 
We adopt a data-driven calibration of the quadratic constitutive relation (QCR) \citep{spalart2000QCR}:
\begin{equation}
  \tau_{ij,\mathrm{QCR}}=\tau_{ij}^{linear} - C_\mathrm{cr1}(\boldsymbol{x})\left(O_{ik} \tau_{jk}+O_{jk} \tau_{ik}\right),
  \label{QCR_equation}
\end{equation}
where $O_{ik}$ is the normalized rotation tensor. Unlike the standard QCR where $C_\mathrm{cr1}$ is constant, the expert $E_3$ predicts a spatially varying $C_\mathrm{cr1}$ field to accurately capture corner vortices.

This modularity allows the PMoE framework to progressively integrate experts ranging from concise symbolic expressions to complex neural mappings without structural conflict, effectively decoupling the model architecture from the constraints of a single baseline formulation.
This flexibility allows developers to select the most suitable expert design strategy for a given flow regime and to update or expand the expert group in step with advances in the literature, thereby further enhancing the practicality, extensibility, and efficiency of the framework.

While the formulation of individual experts reflects physical modelling considerations, the assignment of flow regimes during training and deployment is entirely feature-driven. 
No manual flow-type specification is required when processing a new case. 
The sampling strategy is likewise based on local kinematic indicators, ensuring minimal dependence on prior physical labeling.

\section{Continuously learning PMoE model with various cases introduced progressively}\label{sec:PMoE}

\begin{figure}
    \hspace*{-1.0cm}
  \begin{overpic}[height=0.44\textwidth]{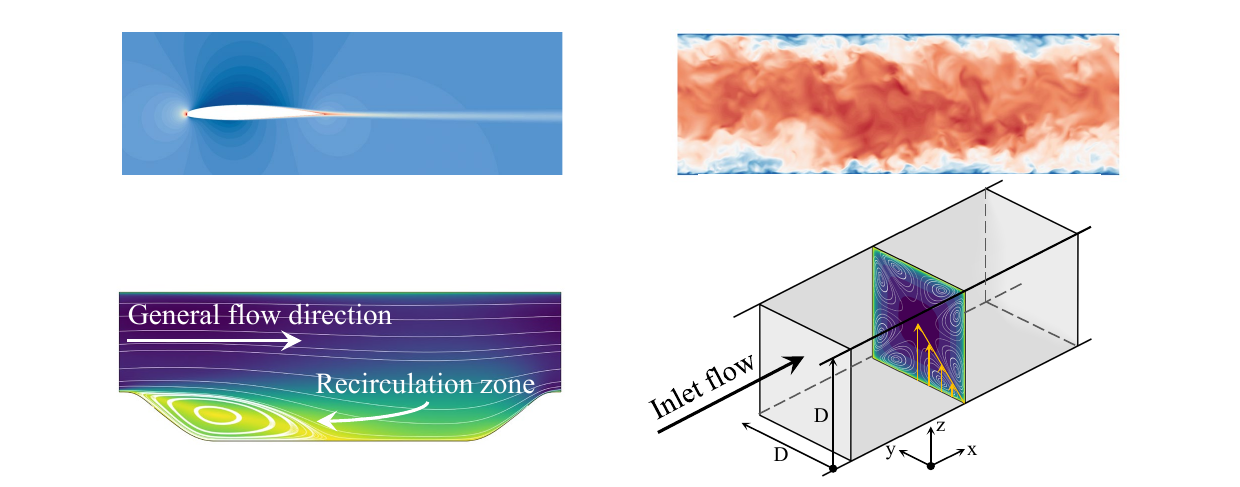} 
        \put(8.8,37.5){$(a)$}
        \put(53.3,37.5){$(b)$}
        \put(8.8,16.7){$(c)$}
        \put(53.3,16.7){$(d)$}
    \end{overpic}
  \captionsetup{justification=justified, singlelinecheck=true}
  \caption{Schematic diagrams of typical examples of various flow regimes. (a) 2DANW. (b) Fully developed channel flow. (c) Periodic hill flow. (d) Square duct flow.}
\label{fig:flow_regimes}
\end{figure}

To demonstrate the sustainable generalization capability of the PMoE framework, we employ a curriculum learning strategy. 
As shown in figure~\ref{fig:flow_regimes}, each dataset is introduced sequentially on a case-by-case basis in order of increasing geometric complexity, starting from the airfoil wake case (free-shear turbulence away from solid walls) and progressing to the channel flow cases (equilibrium wall-bounded flow with parallel flat-wall), periodic hill cases (curved-wall featuring separation), and the square duct case (featuring corner-induced secondary flow).
It should be noted that the identification of whether a new dataset belongs to a known or novel regime is performed automatically by the router via the reconstruction-error-based confidence metric. 
However, the order in which datasets are introduced and the design of the corresponding expert formulation at each stage are prescribed by the user, analogous to a practitioner sequentially encountering new flow scenarios in an industrial workflow.
This obviously relies on modeller's experience and physical insight.
Importantly, because each router component and expert is trained independently with all previous modules frozen, the final model is invariant to the order of regime introduction.

The evolution of the model is denoted by stages $S_0$ through $S_3$, with the corresponding trained PMoE denoted as PMoE-S0, PMoE-S1, PMoE-S2 and PMoE-S3, respectively. 
At each stage, the router evaluates the novelty of the incoming data. 
We extract features from sampled points in the baseline RANS flow field, and construct input vector $\boldsymbol{x}$ and matrix $\mathsfbi{X}\in\mathbb{R}^{N\times M}$, where $M=7, N=10000$ in our training process.
The input data is then evaluated through the existing router modules.
If the maximum confidence $p_{\mathcal{K}}$ falls below the acceptance threshold $T_{\mathrm{accept}} = 90\%$, the system triggers the expansion mechanism: a new autoencoder component is trained to recognize the regime, and a specialized expert is trained to close the RANS equations. The specific datasets and their role in the curriculum are detailed in table \ref{tab:flow_cases}.
It should be noted each router (and expert model) is designed to handle a range of flow conditions over the same geometry. Thus for each flow regime, only a portion of the working conditions are used for the training, while the rest are left for model validation in \textsection~\ref{sec:validation}.
To demonstrate the feasibility of the PMoE framework while maintaining minimal computational cost, the current training dataset is intentionally kept as small as possible. Further reducing the training set will inevitably degrade the model's generalizability.
For each flow regime, the router and the corresponding expert are trained using the same subset of cases. 
The numerical setup of RANS calculation for each case is summarized in appendix~\ref{appB}.

While completing the expansion of the router from PMoE-S0 to PMoE-S3, we also need to synchronously train the corresponding expert model for each newly identified flow regime. 
It should be particularly pointed out that the regimes are distinguished by the router automatically rather than pre-specified by humans during the training process.

\begin{table}
  \begin{center}
    \begin{tabular}{
      >{\centering\arraybackslash}p{2.5cm}
      >{\centering\arraybackslash}p{3.3cm}
      >{\centering\arraybackslash}p{1.8cm}
      >{\centering\arraybackslash}p{1.5cm}
      >{\centering\arraybackslash}p{3.1cm} 
    }
    \textbf{Flow Regime} & \textbf{Case} & \textbf{Abbreviation} & \textbf{Stages} & \textbf{Reference} \\
    \hline
    \multirow{1}{=}{\centering Stage 0: Wake}
    & 2D Airfoil Near-Wake & 2DANW & training & \cite{nakayama1985ANW} \\
    \hline
    \multirow{6}{=}{\centering Stage 1: Channel}
    & Fully developed channel at $\Rey_\tau = 2000$ & C2000 & training  & \multirow{6}{=}{\centering \cite{lee2015C5200,yamamoto2018channel8000}}\\
    & Fully developed channel at $\Rey_\tau = 5200$ & C5200 & training &  \\
    & Fully developed channel at $\Rey_\tau = 8000$ & C8000 & validation & \\
    \hline
    \multirow{4}{=}{\centering Stage 2: Periodic hill}
      & Periodic hill with $\alpha = 0.8$ & PH0p8 & validation & \multirow{4}{=}{\centering \cite{xiao2020PH}}\\
    & Periodic hill with $\alpha = 1.0$ & PH1p0 & training &  \\
    & Periodic hill with $\alpha = 1.2$ & PH1p2 & validation & \\
    & Periodic hill with $\alpha = 1.5$ & PH1p5 & validation & \\
    \hline
    \multirow{3}{=}{\centering Stage 3: Square duct}
      & Square duct at $Re = 2500$ & SD2500 & training & \multirow{3}{=}{\centering \cite{pinelli2010SquareDuct3500,vinuesa2018SquareDuct} }\\
    & Square duct at $Re = 3500$ & SD3500 & validation & \\
    & Square duct at $Re = 5693$ & SD5693 & training & \\
  \end{tabular}
  \caption{Summary of flow cases used for training and validation.}
  \label{tab:flow_cases}
  \end{center}
\end{table}

\subsection{Stage 0: Baseline initialization}

\begin{figure}
    \hspace*{-0.4cm}
  \begin{overpic}[height=0.9\textwidth]{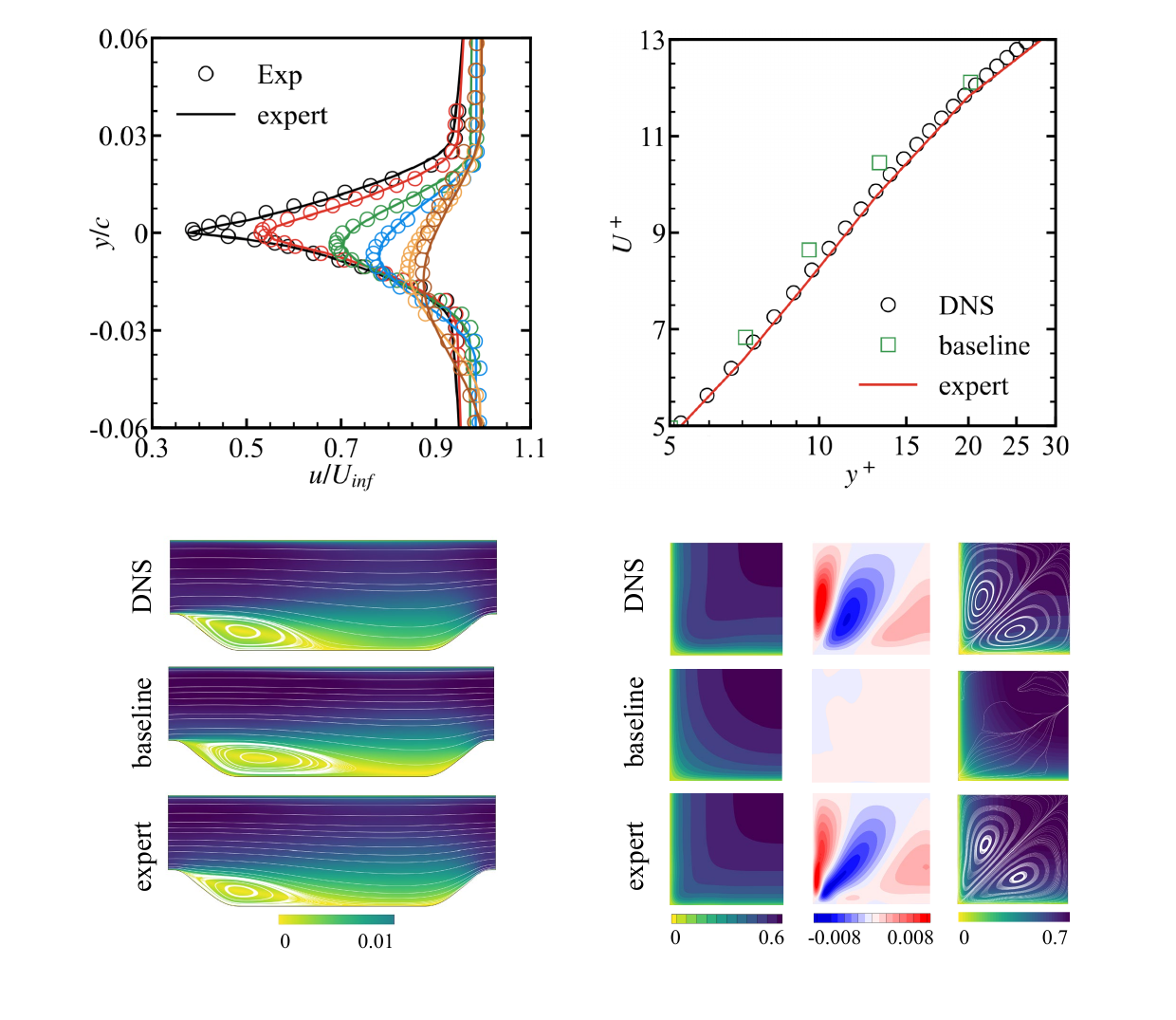} 
        \put(11.9,84.3){$(a)$}
        \put(56.6,84.3){$(b)$}
        \put(12.5,41.4){$(c)$}
        \put(55.2,41.4){$(d)$}
        \put(24.8,42.0){$\left \| u \right \| $}
        \put(60.5,42.5){$u_x \hspace{1.3cm} u_y \hspace{1.2cm} \left \| u \right \|$}
    \end{overpic}
  \captionsetup{justification=justified, singlelinecheck=true}
  \caption{Velocity profiles and contours predicted by the baseline SA model, the expert model and high-fidelity data. 
  (a) Expert for S0 trained by 2DANW. (b) Expert for S1 trained by C5200. (c) Expert for S2 trained by PH1p0. (d) Expert for S3 trained by SD2500 and SD5693.}
\label{fig:experts}
\end{figure}

The initial model, PMoE-S0, is established using the 2D Airfoil Near-Wake (2DANW) case.
This case is a verification of the airfoil wake flow, with a uniform inflow, no-slip boundary conditions on the airfoil surface, and a far-field free-stream condition.
Detailed case configuration are referred to the NASA web page \footnote{Data available online at  \url{https://tmbwg.github.io/turbmodels/airfoilwakeverif500c.html}}, and the grid used for the present RANS calculations is given in appendix~\ref{appB}. 
This regime represents simple free-shear turbulence where the standard SA model is known to perform adequately \citep{bin2023AIAA_SA}.
Consequently, we designate the SA model as baseline expert $E_0$, and its performance in the 2DANW case is presented in figure~\ref{fig:experts}(a). 
Furthermore, the first router component $C_0$ is trained on the wake features, establishing the baseline latent space distribution. 
As shown in table \ref{tab:training_process}, the router identifies this regime with $99.9\%$ confidence, serving as the anchor for future expansions.

\begin{table}
  \begin{center}
    \begin{tabular}{
      >{\centering\arraybackslash}p{4.8cm}
      >{\centering\arraybackslash}p{1.7cm}
      >{\centering\arraybackslash}p{1.9cm}
      >{\centering\arraybackslash}p{1.4cm}
      >{\centering\arraybackslash}p{2.3cm} 
    }
                                    &\textbf{$S0$}    & \textbf{$S0\to S1$} & \textbf{$S1\to S2$} & \textbf{$S2\to S3$}\\
      \hline
      \textbf{New Regime}           & 2DANW           & C2000, C5200   & PH1p0          & SD2500, SD5693 \\
      \textbf{MAX confidence with existing router}      &$-$            & $0.0\%$       & $58.4\%$        & $56.7\%$\\
      \textbf{Introduced Expert}           &SA Baseline    & SR Expression  & FI-NN          & FI-NN     \\
      \textbf{Confidence of new router component}  &$99.9\%$     & $99.5\%$      & $93.5\%$      & $97.0\%$ \\
    \end{tabular}
    \captionsetup{justification=justified, singlelinecheck=true}
    \caption{
    Progressive construction of the PMoE model. 
    }
    \label{tab:training_process}
  \end{center}
\end{table}

\subsection{Stage 1: Adaptation to wall-bounded turbulence}

In the second phase, the model is exposed to fully developed turbulent channel flows at $Re_\tau = 2000$ and $5200$, which are driven by a volumetric body force, with cyclic boundary conditions at the inlet and outlet and no-slip walls. 
The grid sizes used for the present RANS calculations are provided in appendix~\ref{appB}.
When this data is fed into PMoE-S0, the router yields a confidence of $0.00\%$ as presented in table \ref{tab:training_process}, correctly identifying that the physics of the wall turbulence differs fundamentally from the free-shear wake.

This triggers the creation of a new expert $E_1$. 
As discussed in \S\ref{subsec_heterogeneous_experts}, we employ offline symbolic regression to derive a correction for the damping function $f_{\nu 1}$. 
The resulting expert formulation is:
\begin{equation}
  f_{\nu 1}^* = 0.985 \tanh\left(\frac{\chi^{1.3}}{31.1}\right).
  \label{Expert_Channel}
\end{equation}
Applying this expert to RANS calculations, the corrected model is able to provide more accurate wall-normal profiles compared to the baseline SA, agreeing well with the DNS result \citep{lee2015C5200} as shown in figure~\ref{fig:experts}(b). 
Simultaneously, a new router component $C_1$ is trained. 
The updated model, PMoE-S1, subsequently recognizes the channel data with 
$99.5\%$ confidence, demonstrating its ability to capture wall-bounded flow characteristics (as represented here by channel flow), while retaining Expert 0 for wakes.

\subsection{Stage 2: Capturing flow separation}
\label{sec:separation}

Most industrial flows exhibit features such as strong pressure gradients, streamline curvature and separation, which violate the equilibrium assumption underlying conventional RANS models.
The periodic hill flow is usually adopted as a canonical benchmark for separated turbulence, as it combines adverse pressure gradients, separation and reattachment in a simple, well-defined geometry, with the accurate prediction of the separation bubble remaining a key modelling challenge.
The Reynolds number of the periodic hill cases is set as $5600$, consistent with DNS simulation \citep{xiao2020PH}, and the hill slope defined by $\alpha = l/h$ varies across cases, where $l$ and $h$ are the hill width and height, respectively.

The curriculum thus proceeds to the periodic hill case with $\alpha=1.0$ as shown in table~\ref{tab:flow_cases}.
The PMoE-S1 router produces a maximum confidence of $58.4\%$ as presented in table \ref{tab:training_process}. 
Note that this relatively high-level of confidence is likely due to shared wall-bounded features. Nevertheless, it falls below the $90\%$ threshold, flagging the regime as novel.

To address the non-equilibrium physics in the recirculation bubble, we train a new expert $E_2$ using the FIML method. This expert injects a source term $\beta(\boldsymbol{x})$ into the production term as presented in equation \ref{SA_equation_beta}.
The details are elaborated in Appendix \ref{appC}.
As shown in figure~\ref{fig:experts}(c), the trained separation flow expert can more accurately capture the phenomena of separation and reattachment, while the estimation of the separation bubble size by the basic SA model is significantly larger than that of the expert model and DNS results. Furthermore, upon integrating $E_2$ and its corresponding router component, the identification confidence for separated flows (as represented here by periodic hill flow) rises to $93.5\%$, yielding model PMoE-S2.

\subsection{Stage 3: Modelling corner-induced anisotropy}

Finally, the model encounters the square duct flow at $Re=2500, 5693$. 
This case is driven by a volumetric body force, with cyclic inlet and outlet boundaries and no-slip walls on all duct surfaces.
This case is physically distinct due to the secondary motions in the corner region, which are impossible for linear eddy viscosity models.
The training data are obtained from published DNS simulations \citep{pinelli2010SquareDuct3500,vinuesa2018SquareDuct}. 
The Reynolds number is based on the half edge length $h=D/2$ of the square duct and the bulk velocity $U_b$ as shown in figure~\ref{fig:flow_regimes}(d). 

The PMoE-S2 router yields a confidence of $56.7\%$ as presented in table \ref{tab:training_process}, again triggering expansion.
A new expert 3 $E_3$ is then trained to predict the non-linear stress coefficient $C_\mathrm{cr1}(\boldsymbol{x})$ in the QCR formulation in equation \ref{QCR_equation}. 
The details are elaborated in Appendix \ref{appC}.
As shown in the figure~\ref{fig:experts}(d), the trained expert model successfully captures the generation of secondary flow, while the baseline SA model is unable to calculate the secondary flow.
The final model, PMoE-S3, achieves a recognition confidence of $97.0\%$ for this regime (as represented here by square duct flow).

\section{Model validation}\label{sec:validation}

The validation of the PMoE framework focuses on three critical performance metrics: the interpretability of the unsupervised routing, the prevention of catastrophic forgetting on previously learned regimes, and the generalization capability to cases with different operating conditions or configurations. 
All results presented herein utilize the final PMoE-S3 model to demonstrate its cumulative capabilities.

\subsection{Validation for the PMoE router}

\begin{figure}
    \hspace*{0.1cm}
  \begin{overpic}[width=0.97\textwidth]{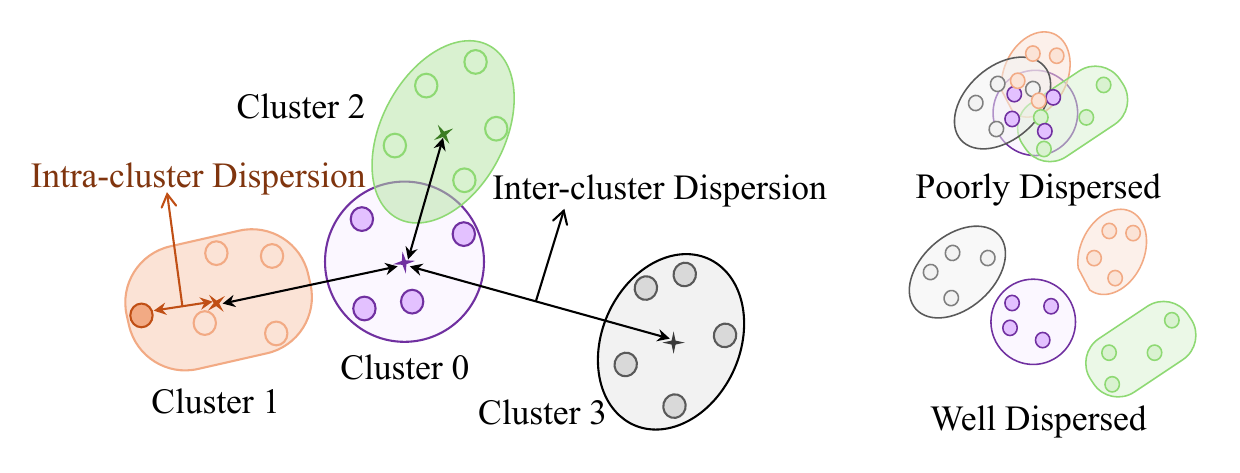} 
    \end{overpic}
  \captionsetup{justification=justified, singlelinecheck=true}
  \caption{
  Schematic illustration of intra- and inter-cluster dispersion based on the Mahalanobis distance.
  }
\label{fig:M_Distance}
\end{figure}

A prerequisite for reliable expert selection is the router's ability to distinguish flow regimes based on local RANS features. We analyze the four-dimensional latent space of the autoencoder components using the Mahalanobis distance \citep{Mahalanobis1936MahalanobisDist}, $D_M$, which measures the distance between a point $\boldsymbol{z}$ and a distribution with mean $\boldsymbol{\mu}$ and covariance $\boldsymbol{\Sigma}$:
\begin{equation}
  D_M^2(\boldsymbol{z}) = (\boldsymbol{z} - \boldsymbol{\mu})^\mathrm{T} \boldsymbol{\Sigma}^{-1} (\boldsymbol{z} - \boldsymbol{\mu}).
\end{equation}
Unlike the Euclidean distance, $D_M$ accounts for the correlation structure of the learned representations. Based on this metric, we compute both the intra-cluster dispersion, which quantifies the compactness of samples within a regime, and the inter-cluster dispersion measuring the separation between regime centroids, as illustrated in figure~\ref{fig:M_Distance}. 
The ratio between the averaged inter- and intra-cluster dispersions is adopted as a diagnostic indicator of the latent-space organization. A high ratio implies that different flow regimes are topologically distinct within the component's latent space, whereas a low ratio suggests possible overlap.

This analysis is carried out for all four components. Each component is exposed to data from all four flow regimes to assess its discriminative behavior, and table \ref{tab:latent_analysis} summarizes these statistics. 
The intra-cluster dispersion remains consistent as $\approx 1.8$ across all components, indicating that the autoencoders compress different flow regimes with comparable efficiency. 
Furthermore, the ratios between the averaged inter- and intra-cluster dispersions are also presented, and all four components exhibit ratios exceeding $3.4$, indicating that the corresponding clusters have little overlap as illustrated by the well-separated configuration in figure~\ref{fig:M_Distance}.
Although the router decision is ultimately based on the reconstruction error along the full encoder–decoder pathway, the latent-space analysis still provides an interpretable metric for its behavior, as the observed clustering patterns reflect the feature representations that govern reconstruction quality and hence routing outcomes.
This mechanism, combined with the competitive winner-takes-all strategy, ensures robust classification. 
As evidenced in table \ref{tab:router_results}, the router achieves $>90\%$ classification accuracy on the validation sets, confirming that the present router successfully distinguishes the tested regimes.

\begin{table}
  \begin{center}
    \begin{tabular}{
      >{\centering\arraybackslash}p{3.4cm}
      >{\centering\arraybackslash}p{2.0cm}
      >{\centering\arraybackslash}p{2.2cm}
      >{\centering\arraybackslash}p{2.2cm}
      >{\centering\arraybackslash}p{2.4cm} 
    }
      \textbf{Dispersion}                &\makecell{$C_0$ \\ \textbf{Wake}} & \makecell{$C_1$ \\ \textbf{Channel}} & \makecell{$C_2$ \\ \textbf{Periodic hill}} & \makecell{$C_3$ \\ \textbf{Square duct}} \\
      \hline
      \textbf{Average intra-cluster} & $1.8207$                      & $1.8383$                  & $1.8438$               & $1.6947$      \\
      [3pt]
      \textbf{Average inter-cluster}    & $6.6845$                           & $6.3698$                  & $7.6183$               & $7.8923$     \\
      [3pt]
      \textbf{inter-cluster/intra-cluster}        & $3.6713$                           & $3.4650$                       & $4.1319$               & $4.6569$        \\
      
    \end{tabular}
    \captionsetup{justification=justified, singlelinecheck=true}
    \caption{Dispersion analysis of the latent space clusters using Mahalanobis distance.}
    \label{tab:latent_analysis}
  \end{center}
\end{table}

\begin{table}
  \begin{center}
    \begin{tabular}{
      >{\centering\arraybackslash}p{1.6cm}
      >{\centering\arraybackslash}p{1.6cm}
      >{\centering\arraybackslash}p{1.7cm}
      >{\centering\arraybackslash}p{2.6cm}
      >{\centering\arraybackslash}p{2.2cm} 
      >{\centering\arraybackslash}p{2.4cm} 
    }
      \textbf{Case} & \textbf{Unknown} &\makecell{$C_0$ \\ \textbf{Wake}} & \makecell{$C_1$ \\ \textbf{Channel}} & \makecell{$C_2$ \\ \textbf{Periodic hill}} & \makecell{$C_3$ \\ \textbf{Square duct}} \\
      \hline
      \textbf{ANW}         & $0.0\%$        & $\boldsymbol{99.7\%}$                      & $0.3\%$                   & $0.0\%$                & $0.0\%$  \\
      \textbf{C2000}       & $0.2\%$         & $0.0\%$                       & $\boldsymbol{98.1\%}$                  & $0.5\%$                & $1.2\%$  \\
      \textbf{C5200}       & $0.0\%$         & $0.0\%$                       & $\boldsymbol{99.4\%}$                  & $0.0\%$                & $0.6\%$  \\
      \textbf{C8000}       & $3.9\%$         & $0.0\%$                       & $\boldsymbol{94.5\%}$                  & $1.3\%$                & $0.3\%$  \\
      \textbf{PH0p8}       & $0.3\%$         & $0.0\%$                       & $5.7\%$                   & $\boldsymbol{91.6\%}$               & $2.5\%$  \\
      \textbf{PH1p0}       & $0.1\%$         & $0.0\%$                       & $5.4\%$                   & $\boldsymbol{91.5\%}$               & $3.0\%$  \\
      \textbf{PH1p2}       & $0.1\%$         & $0.0\%$                       & $4.8\%$                   & $\boldsymbol{92.0\%}$               & $3.2\%$  \\
      \textbf{PH1p5}       & $0.2\%$         & $0.0\%$                       & $4.7\%$                   & $\boldsymbol{91.9\%}$               & $3.2\%$  \\
      \textbf{SD2500}      & $0.4\%$         & $0.0\%$                       & $0.3\%$                   & $0.0\%$                & $\boldsymbol{99.3\%}$  \\
      \textbf{SD3500}      & $0.5\%$         & $0.0\%$                       & $0.0\%$                   & $0.2\%$                & $\boldsymbol{99.3\%}$  \\
      \textbf{SD5693}      & $0.0\%$         & $0.0\%$                       & $2.2\%$                   & $0.8\%$                & $\boldsymbol{97.0\%}$  \\
      
    \end{tabular}
    \captionsetup{justification=justified, singlelinecheck=true}
    \caption{Confidence distribution of the PMoE-S3 router over different components for all flow cases.
    }
    \label{tab:router_results}
  \end{center}
\end{table}

\subsection{Prevention of catastrophic forgetting via in-distribution tests}\label{subsection:validation_trained}

\begin{figure} 
  \hspace{-0.2cm}
    \vspace{0.1cm}
    \centering
    \subfigure{
        \begin{overpic}[height=0.313\textwidth]{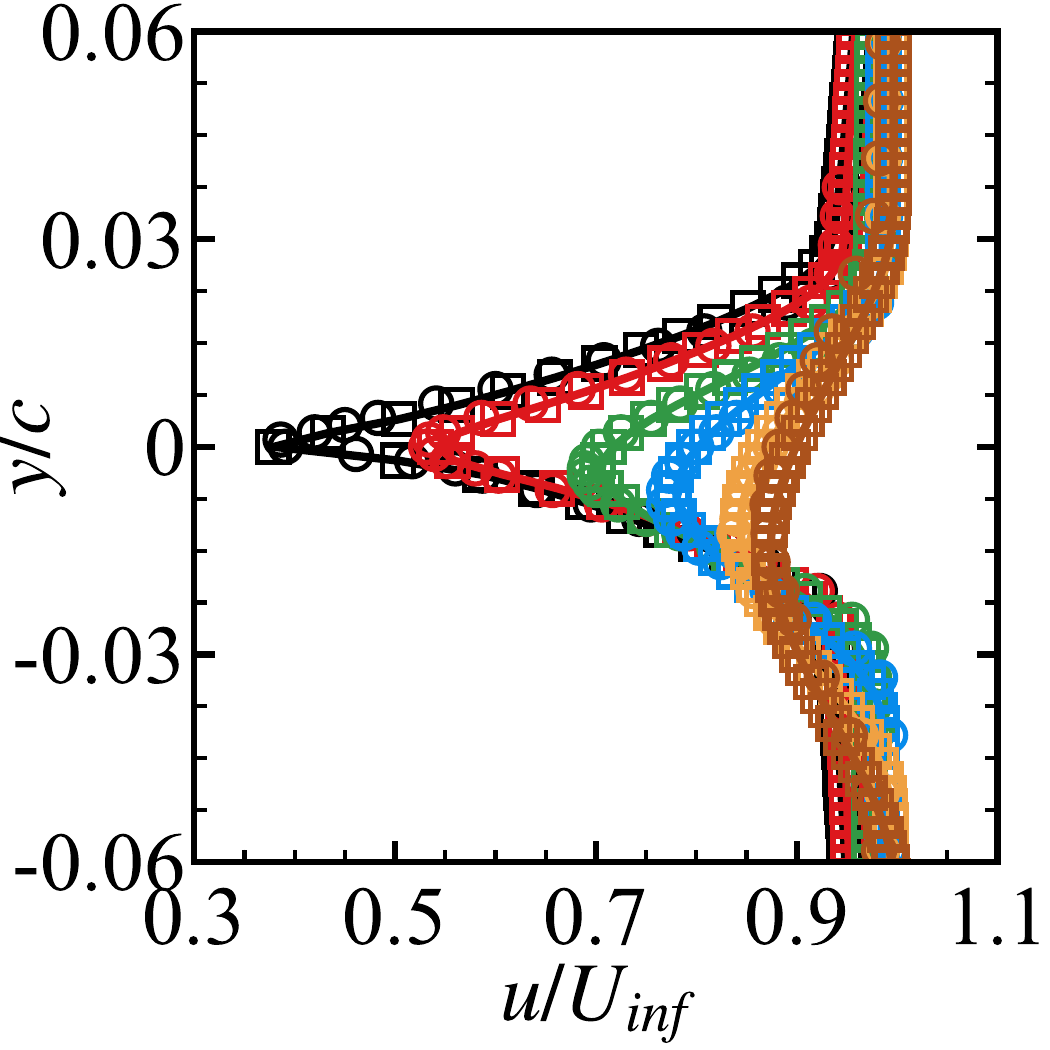} \put(16.8,100){$(a)$}
    	\end{overpic}
        }  
    \subfigure{
        \begin{overpic}[height=0.313\textwidth]{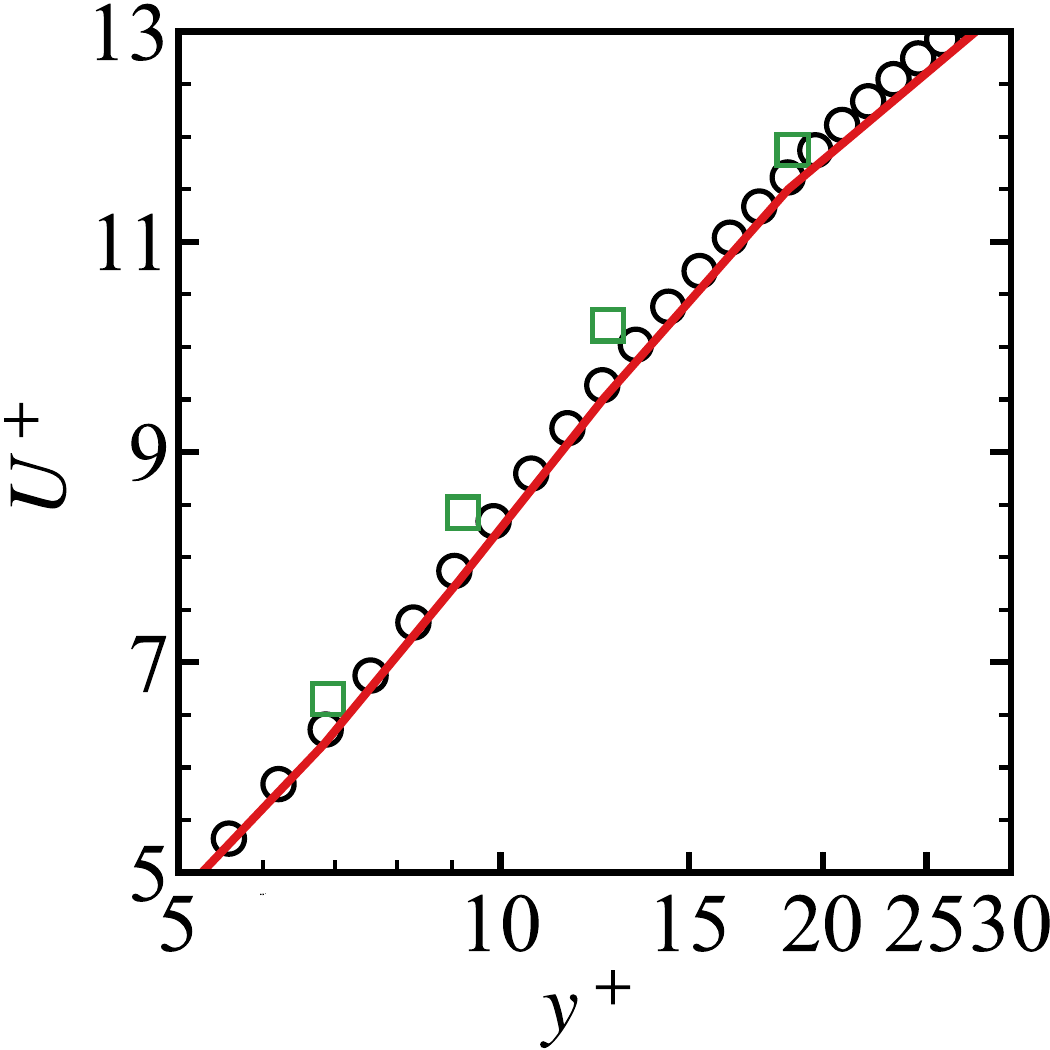} \put(15.7,100){$(b)$}
        \put(19.5,89){\small{$41.7\%$}}
    	\end{overpic}
        }  
    \subfigure{
        \begin{overpic}[height=0.313\textwidth]{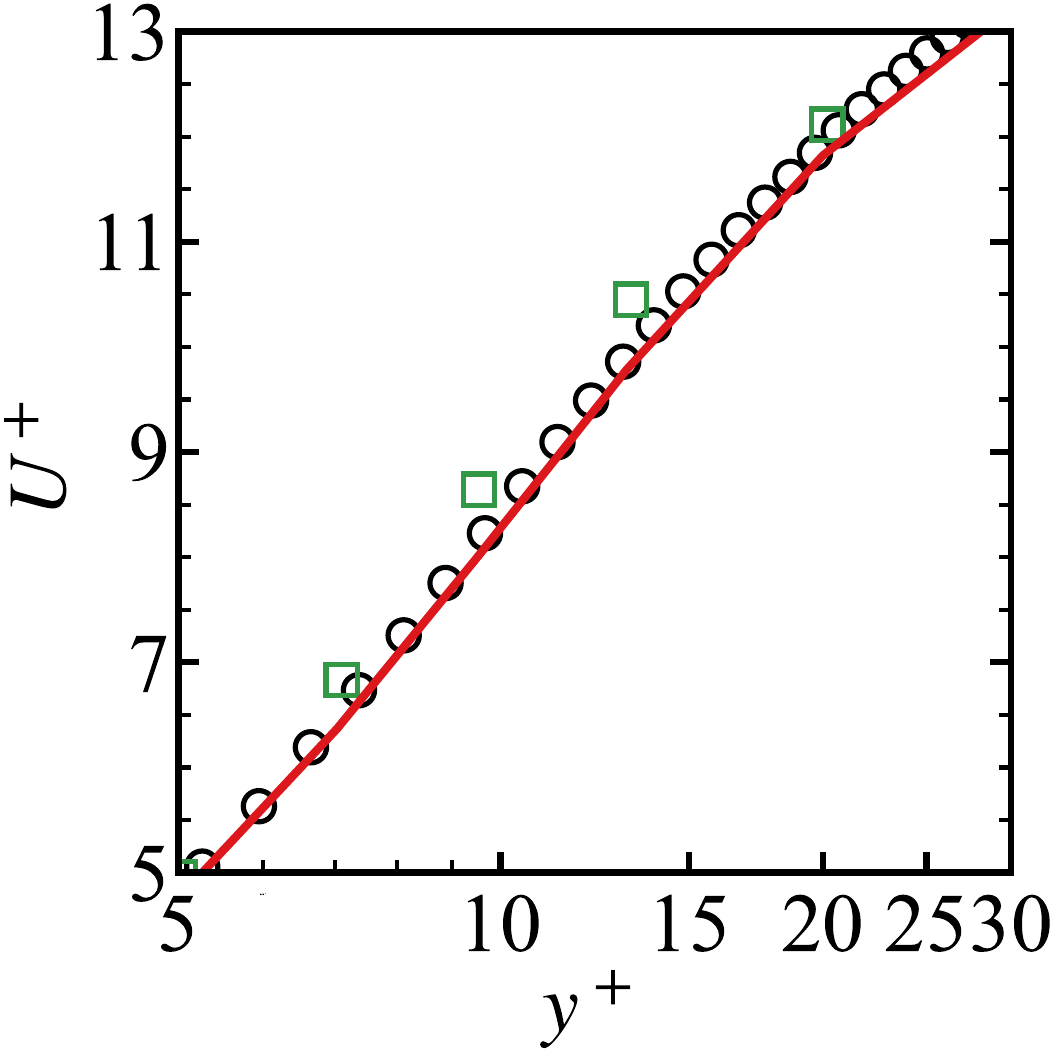} \put(15.5,100){$(c)$}
        \put(19.5,89){\small{$37.7\%$}}
    	\end{overpic}
        }  

    \subfigure{
        \begin{overpic}[height=0.313\textwidth]{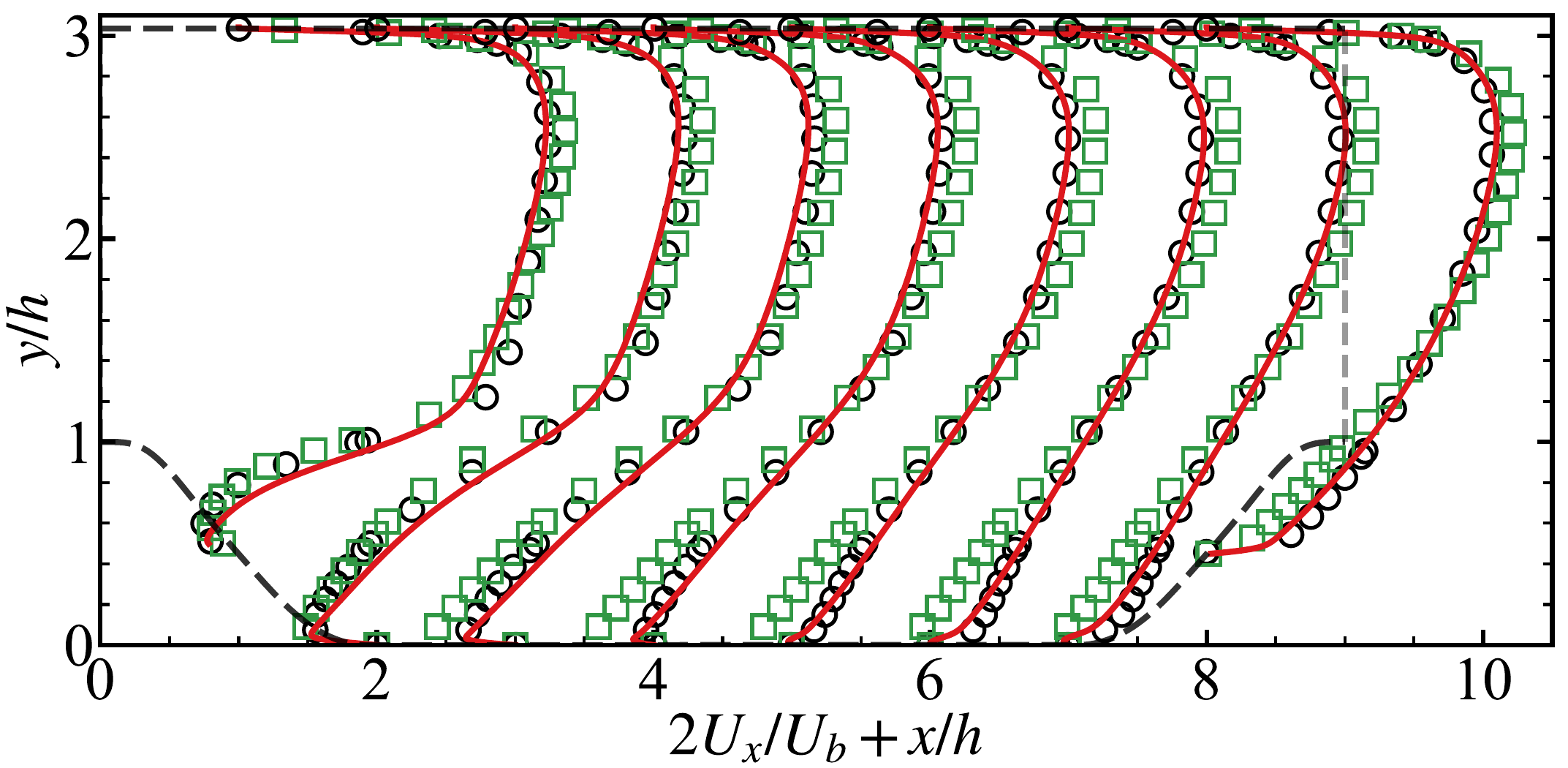} \put(5.8,50){$(d)$}
        \put(7.8,43){\small{$47.2\%$}}
    	\end{overpic}
        }  
    \subfigure{
        \begin{overpic}[height=0.313\textwidth]{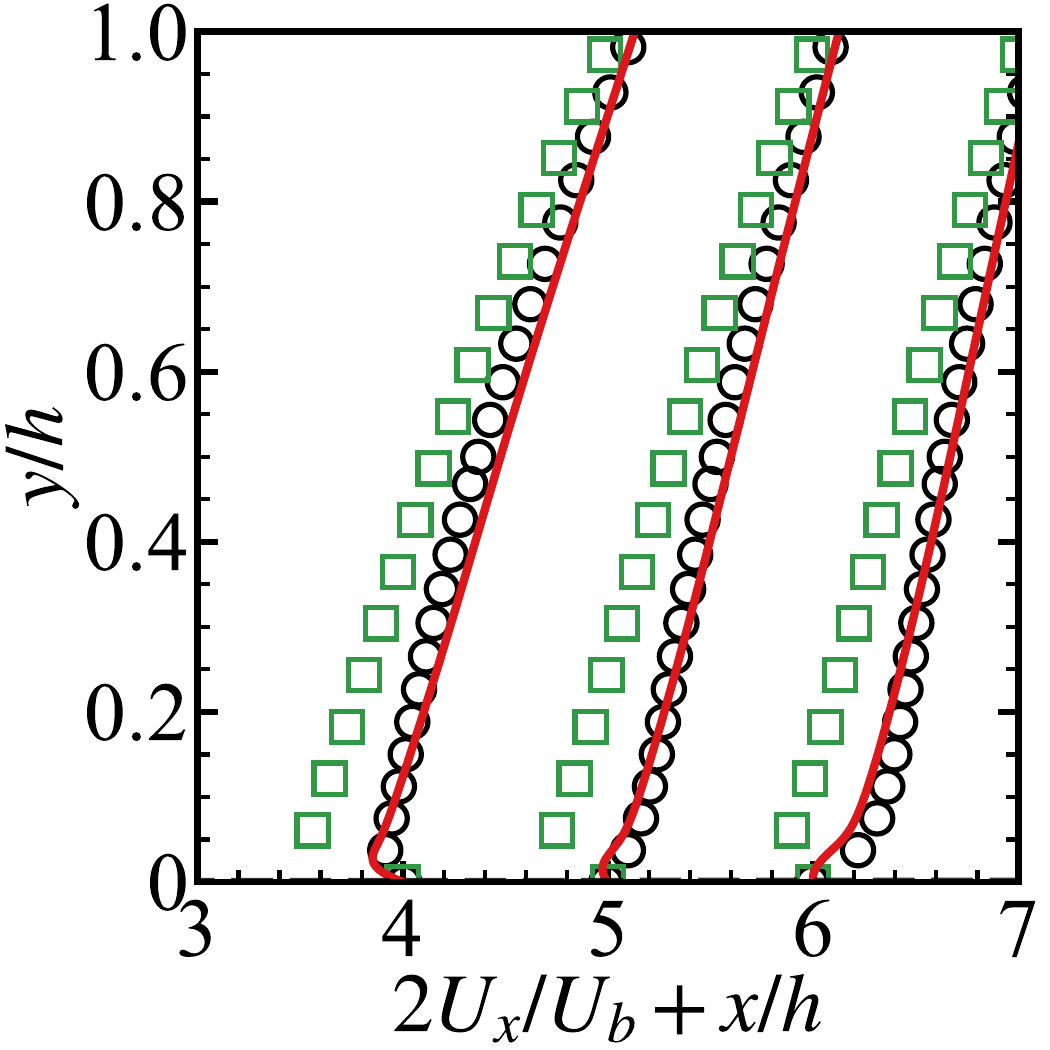} \put(18,100){$(e)$}
        \put(21,89){\small{$76.1\%$}}
    	\end{overpic}
        }

    \subfigure{
        \begin{overpic}[height=0.313\textwidth]{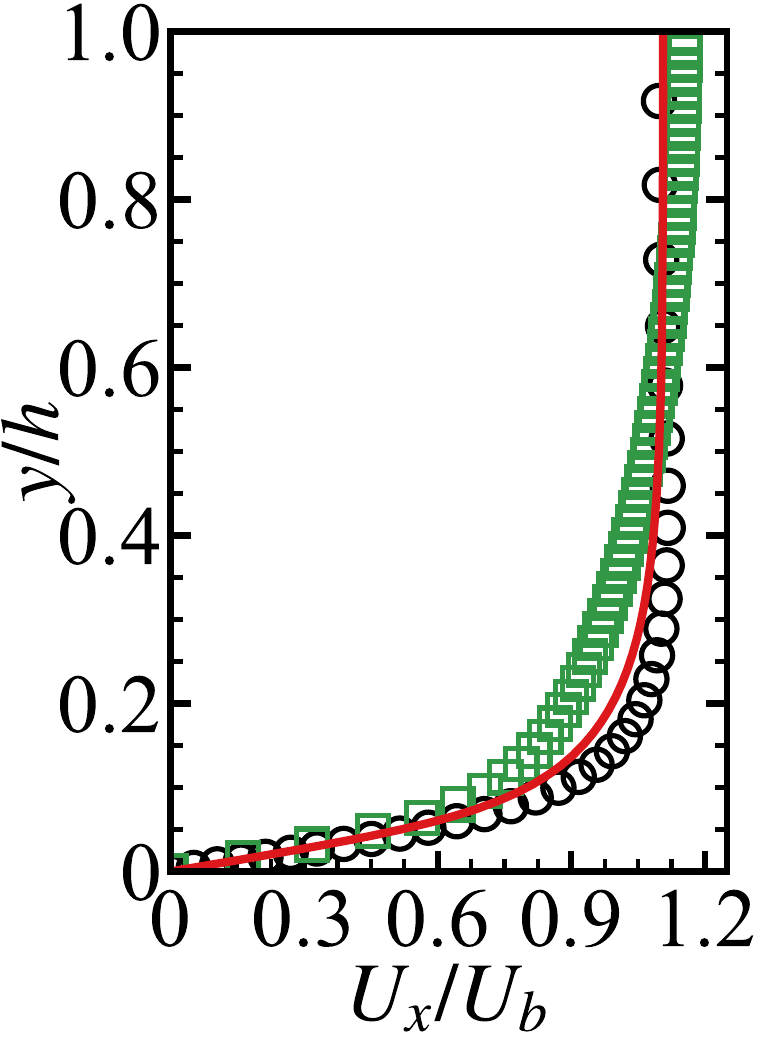} \put(15.5,100){$(f)$}
        \put(18,89){\small{$57.0\%$}}
    	\end{overpic}
        }  
    \subfigure{
        \begin{overpic}[height=0.313\textwidth]{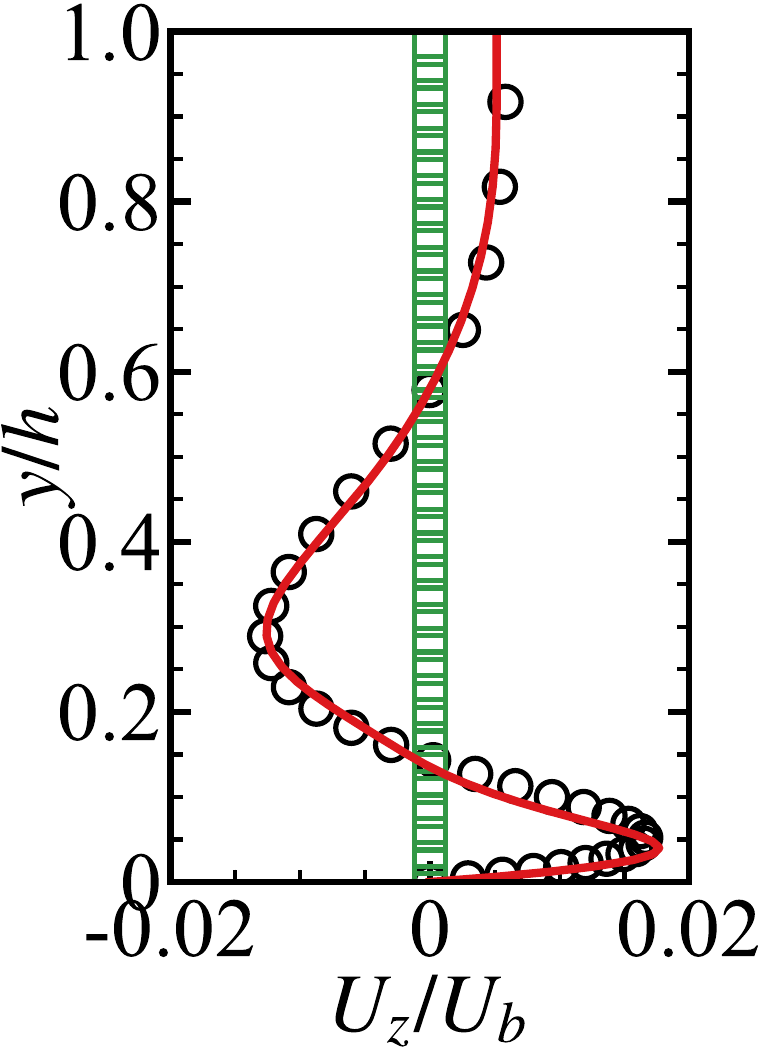} \put(15.5,100){$(g)$}
        \put(18,89){\small{$83.3\%$}}
    	\end{overpic}
        }  
    \subfigure{
        \begin{overpic}[height=0.313\textwidth]{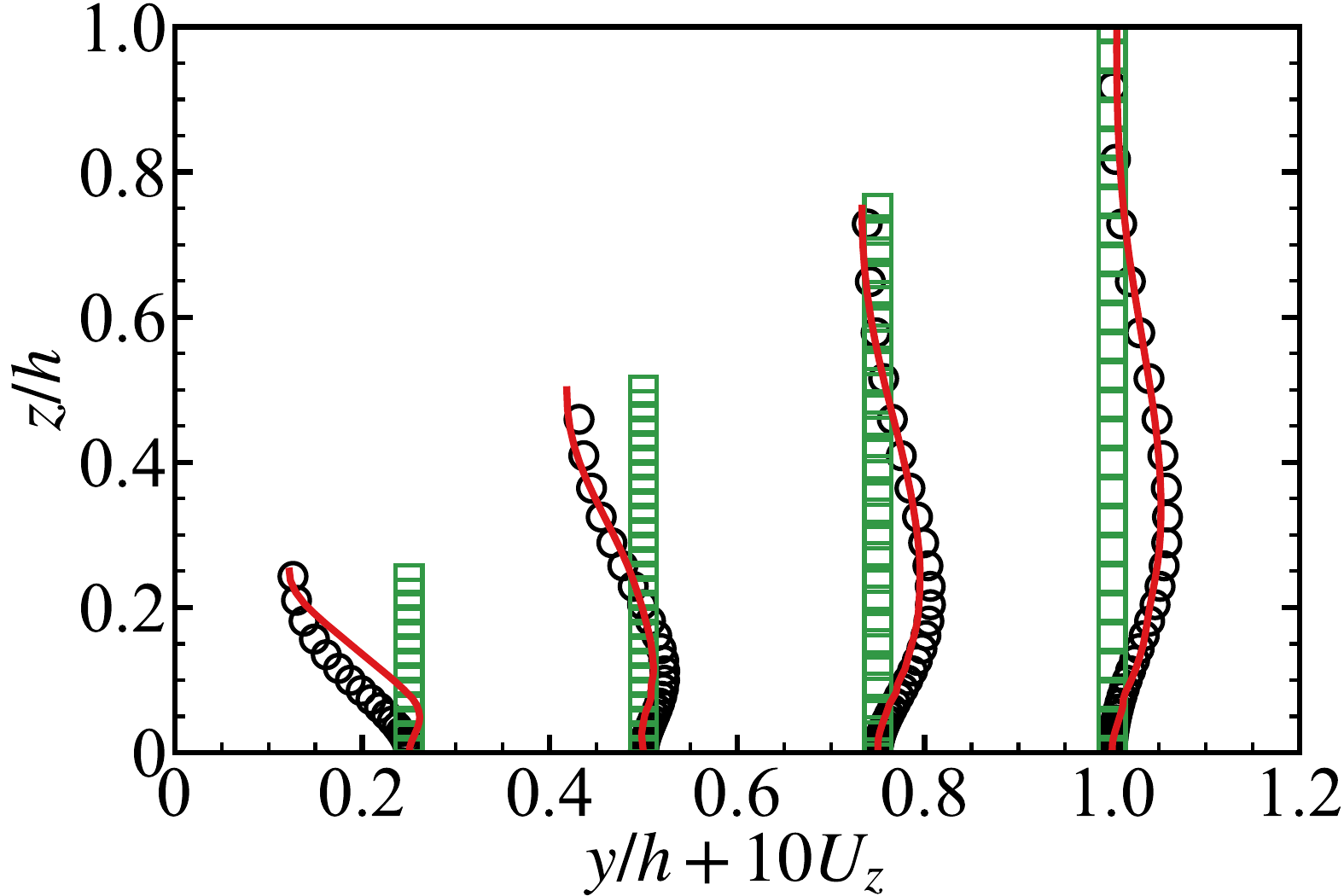} \put(13,66.4){$(h)$}
        \put(15,59){\small{$72.0\%$}}
    	\end{overpic}
        }  

    \subfigure{
        \begin{overpic}[height=0.313\textwidth]{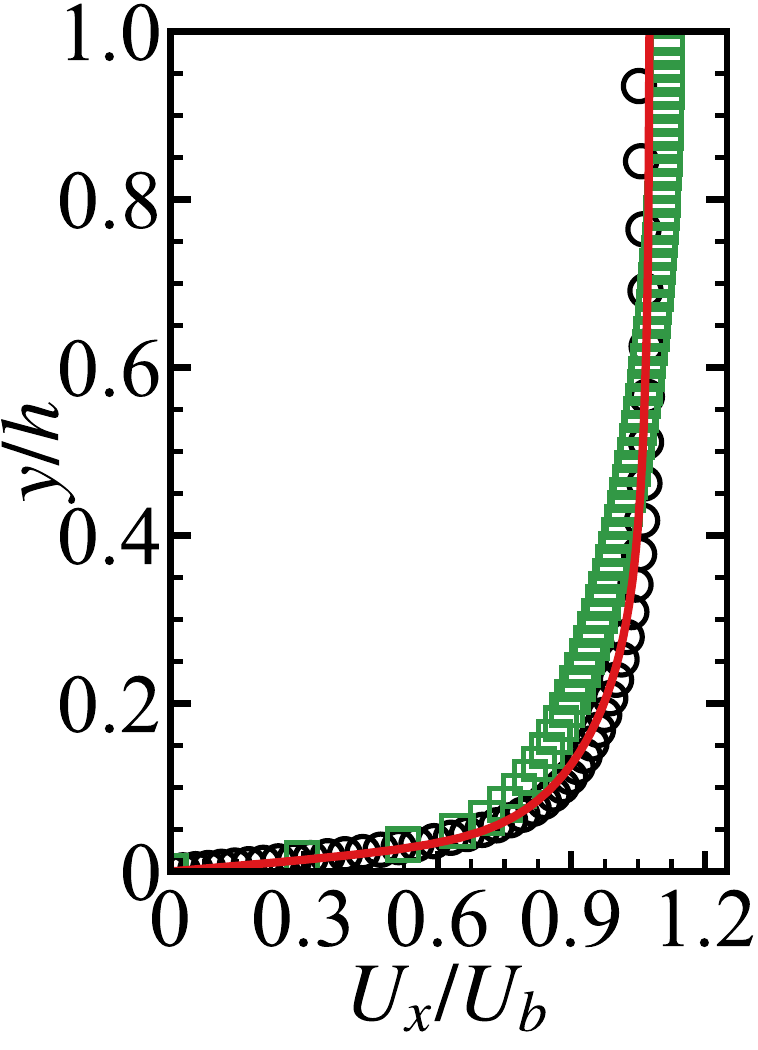} \put(15.5,100){$(i)$}
        \put(18,89){\small{$69.9\%$}}
    	\end{overpic}
        }  
    \subfigure{
        \begin{overpic}[height=0.313\textwidth]{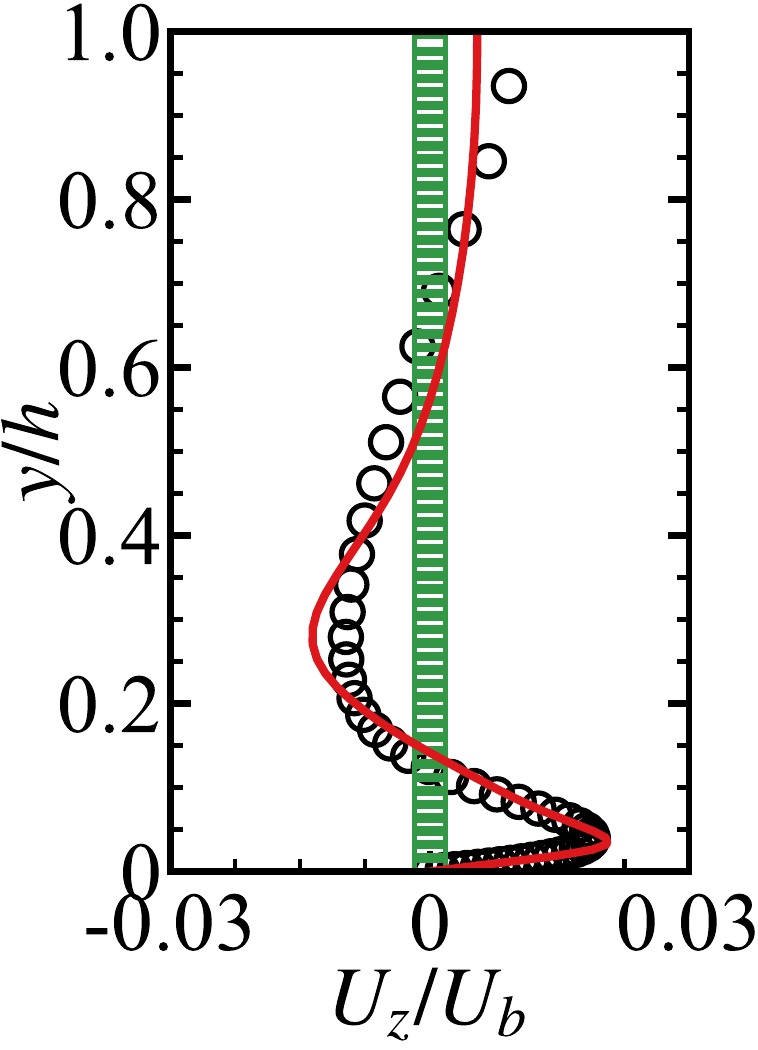} \put(15.5,100){$(j)$}
        \put(18,89){\small{$82.0\%$}}
    	\end{overpic}
        }  
    \subfigure{
        \begin{overpic}[height=0.313\textwidth]{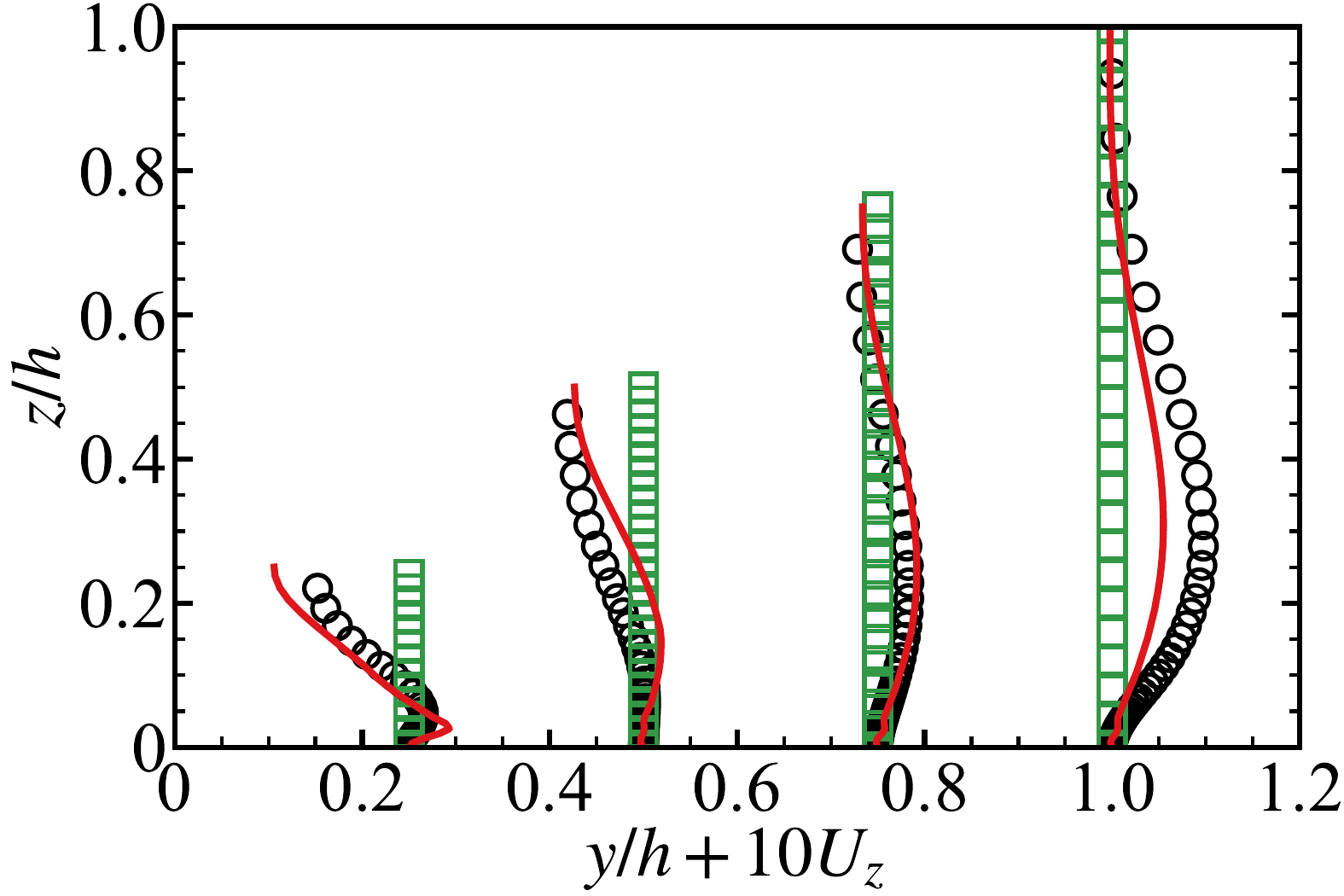} \put(13,66.4){$(k)$}
        \put(15,59){\small{$50.8\%$}}
    	\end{overpic}
        }  

    \captionsetup{justification=justified, singlelinecheck=true}
    \caption{Mean velocity profiles for trained cases. 
    (a) 2DANW (colors indicate streamwise locations). 
    (b) C2000. (c) C5200. (d, e) PH1p0. (f–h) SD2500. (i–k) SD5693. 
    Symbols denote: 
    $\circ$ DNS/Experimental data; 
    \textcolor{green}{$\Box$} Baseline SA; 
    \textcolor{red}{\rule[0.5ex]{1.5em}{0.8pt}} PMoE-S3. 
    The number shown in the top-left corner of each panel is the relative error defined by equation~\ref{RMSE_equ}.}
    \label{fig:validation_trained}
\end{figure}

A primary failure mode in continual learning is catastrophic forgetting, where optimizing for new tasks degrades performance on previous ones \citep{french1999catastrophic}. To assess whether the current PMoE framework is able to avoid catastrophic forgetting, we apply the final PMoE-S3 model back to the initial training cases.
The corresponding results are presented in figure~\ref{fig:validation_trained}. 
The number shown in the top-left corner of each panel denotes the relative RMSE reduction for the PMoE model $\mathrm{RMSE}_{\mathrm{PMoE}}$ with respect to the baseline $\mathrm{RMSE}_{\mathrm{base}}$, defined as 
\begin{equation}
  \mathcal{R} = (\mathrm{RMSE}_{\mathrm{base}} - \mathrm{RMSE}_{\mathrm{PMoE}})/\mathrm{RMSE}_{\mathrm{base}},
  \label{RMSE_equ}
\end{equation}
where the root-mean-square error is given by $\mathrm{RMSE} = \sqrt{\frac{1}{N}\sum_{i=1}^{N}\left(u_i - u_i^{\mathrm{ref}}\right)^2}$.
Here, $u_i$ denotes the model prediction (baseline SA or PMoE, respectively) at the $i$-th reference data location, the error is evaluated for the velocity component shown in each panel, and ${u_i^\text{ref}}_{i=1}^{N}$ are the corresponding reference values, with model predictions interpolated onto these locations when necessary. No value is shown in in figure~\ref{fig:validation_trained}(a) as expert $E_0$ coincides with the baseline.

Figure \ref{fig:validation_trained}(a) presents the wake velocity profiles for the 2DANW case (Stage 0), with each color representing a different observation position: black for $x/c=1.01$, red for $x/c=1.05$, green for $x/c=1.20$, blue for $x/c=1.40$, yellow for $x/c=1.80$ and brown for $x/c=2.19$. Despite the model undergoing three subsequent expansions involving wall-bounded and secondary flows, the prediction for the free-shear wake remains virtually identical to the baseline SA model and agrees well with experimental data. 
This suggests that, in this near-wake region, the flow is dominated by the trailing-edge shear layer and vortex shedding, which are largely insensitive to wall effects.
This confirms that the modular router successfully shields the baseline expert $E_0$ from interference.

Similarly, figures~\ref{fig:validation_trained}(b) and ~\ref{fig:validation_trained}(c) show the mean velocity profiles extracted from the channel flow cases at at $Re_\tau=2000$ and $Re_\tau=5200$ (Stage 1), while figure~\ref{fig:validation_trained}(d) and ~\ref{fig:validation_trained}(e) present the periodic hill at $\alpha=1.0$ (Stage 2).
The final PMoE-S3 model retains the specialized accuracy of experts $E_1$ and $E_2$ in according tests. 
Particularly in the periodic hill case, the reattachment locations are quantified in table~\ref{tab:Separation_location_PH}.
For the training PH1p0 case, the baseline model predicts delayed reattachments near $x/h=7.31$, whereas the PMoE model provides more accurate predictions for the separation bubble size with the reattachment point of $x/h=5.87$.
This improvement reflects the model’s ability to capture the enhanced turbulence production and shear-layer dynamics responsible for earlier flow reattachment.

\begin{table}
  \begin{center}
    \begin{tabular}{
      >{\centering\arraybackslash}p{1.6cm}
      >{\centering\arraybackslash}p{3.2cm} 
      >{\centering\arraybackslash}p{3.0cm} 
      >{\centering\arraybackslash}p{3.0cm}
    }
      \textbf{Case}    & \textbf{DNS} & \textbf{baseline} & \textbf{PMoE-S3} \\
      \hline
      \textbf{PH0p8}   & $5.21$          & $7.34$         & $7.19$    \\
      \textbf{PH1p0}   & $5.04$          & $7.31$         & $5.87$    \\
      \textbf{PH1p2}   & $4.50$          & $7.88$         & $5.36$    \\
      \textbf{PH1p5}   & $4.21$          & $8.32$         & $4.71$    \\
      
    \end{tabular}
    \captionsetup{justification=justified, singlelinecheck=true}
    \caption{Reattachment locations of periodic hill cases.
    }
    \label{tab:Separation_location_PH}
  \end{center}
\end{table}

The capability of the final model PMoE-S3 for trained square duct ($\Rey = 2500 $) is shown in figure~\ref{fig:validation_trained}(f--h). 
Figure~\ref{fig:validation_trained}(f) and \ref{fig:validation_trained}(g) show the velocity profiles at $z/h=0.3, h=D/2$, while the figure~\ref{fig:validation_trained}(h) represents $y/h=0.25, 0.5, 0.75$ and $1.0$ from left to right just as the yellow arrows shown in figure~\ref{fig:flow_regimes}(d). 
The square duct case contains typical corner-induced secondary flow, which cannot be characterized by the SA model based on linear eddy viscosity.
The results of PMoE-S3, however, have a significant improvement compared to the baseline model.
This suggests that the trained PMoE-S3 model can effectively represent the secondary flow in the non-flow direction due to the introduction of nonlinear correction based on QCR.
Moreover, figures~\ref{fig:validation_trained}(i--k) present the results of the SD5693 case. 
The performance of PMoE-S3 remains excellent, with significant improvements in both flow velocity and secondary flow representation compared to the baseline.

\subsection{Generalization to cases with different operating conditions}\label{subsection:validation_unseen}

\begin{figure} 
  \hspace{-0.2cm}
  \vspace{0.1cm}
    \centering
    \subfigure{
        \begin{overpic}[height=0.313\textwidth]{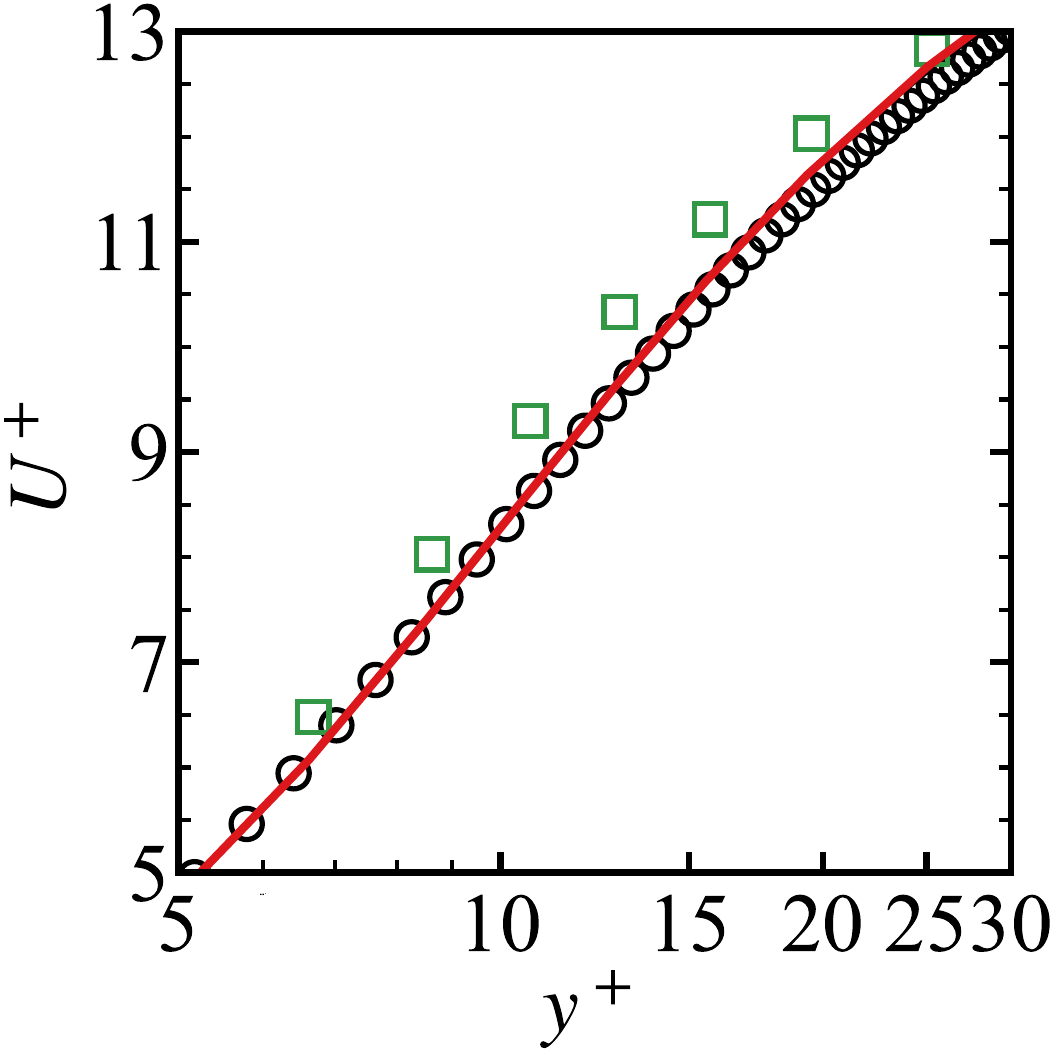} \put(15.7,99){$(a)$}
        \put(18.9,89.2){\small{$77.9\%$}}
    	\end{overpic}
        }  
    \hspace{0.27cm}
    \subfigure{
        \begin{overpic}[height=0.313\textwidth]{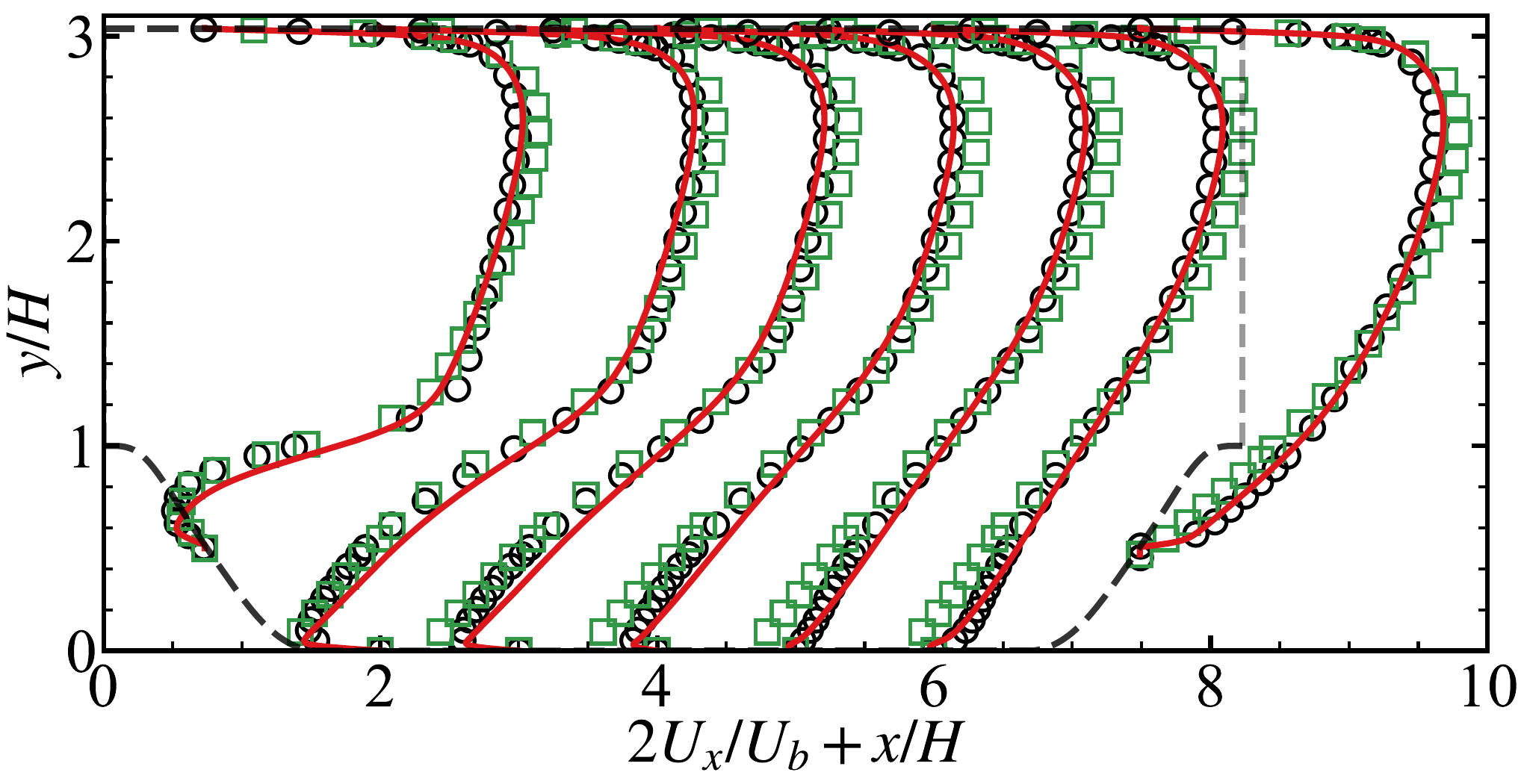} \put(6.4,51.7){$(b)$}
        \put(7.8,45){\small{$28.4\%$}}
    	\end{overpic}
        } 
        
    \subfigure{
        \begin{overpic}[height=0.313\textwidth]{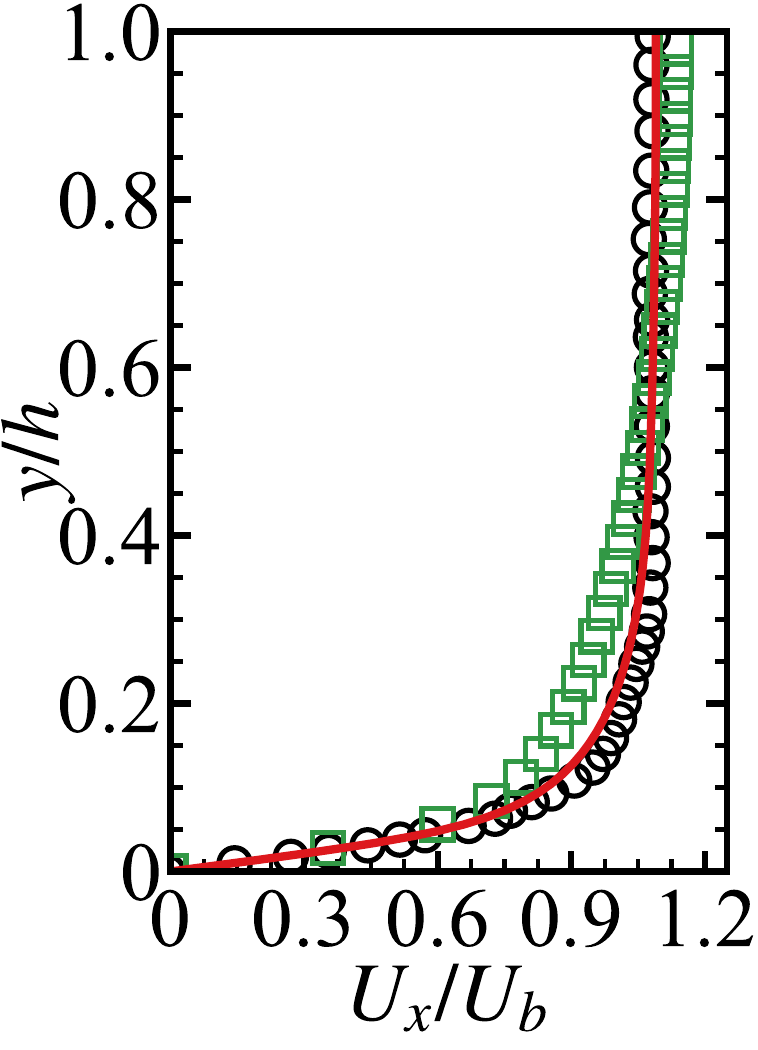} \put(15.5,100){$(c)$}
        \put(18.3,89){\small{$71.7\%$}}
    	\end{overpic}
        }  
    \hspace{0.62cm}
    \subfigure{
        \begin{overpic}[height=0.313\textwidth]{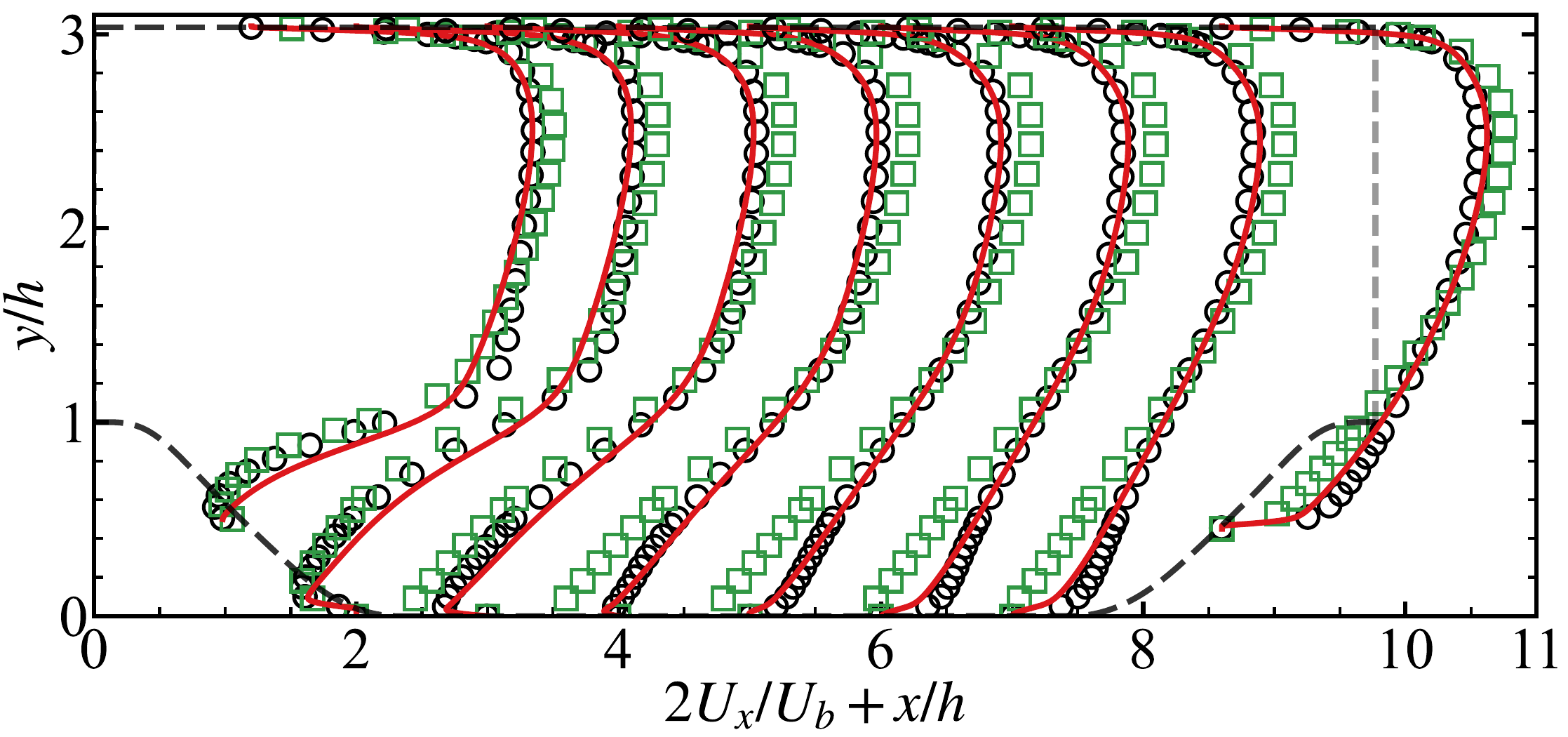} \put(5.3,47.2){$(d)$}
        \put(7.4,41.5){\small{$53.1\%$}}
    	\end{overpic}
        }

    \subfigure{
        \begin{overpic}[height=0.313\textwidth]{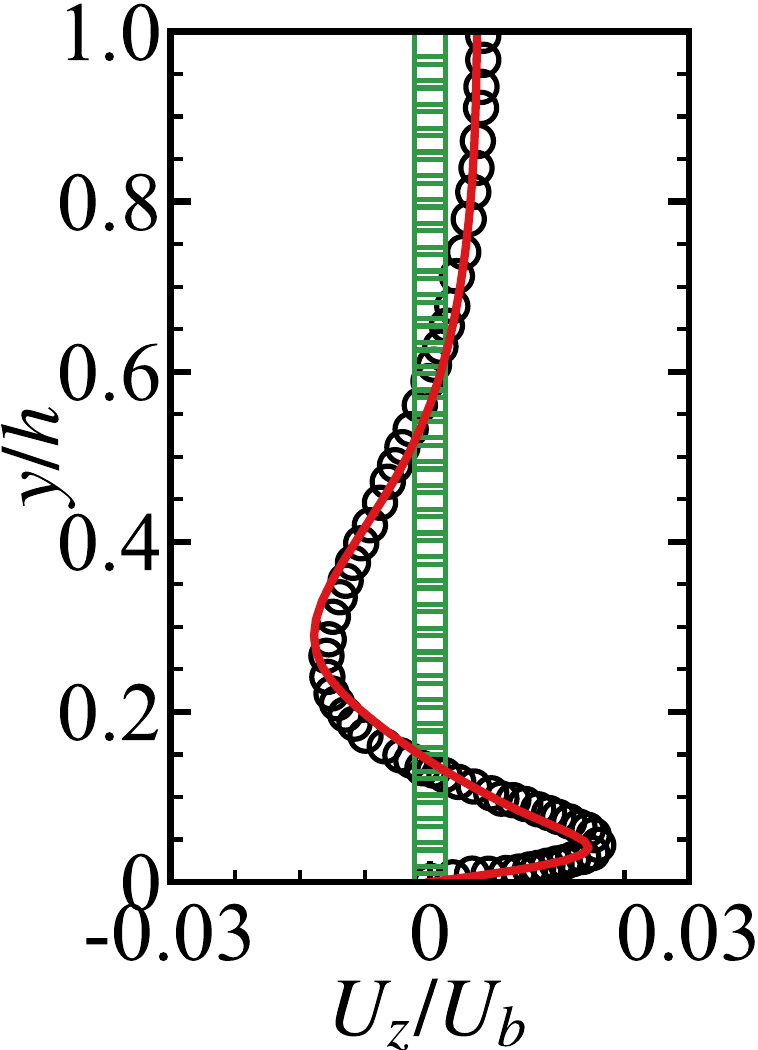} \put(15.5,100){$(e)$}
        \put(18.3,89){\small{$84.3\%$}}
    	\end{overpic}
        }  
    \subfigure{
        \begin{overpic}[height=0.313\textwidth]{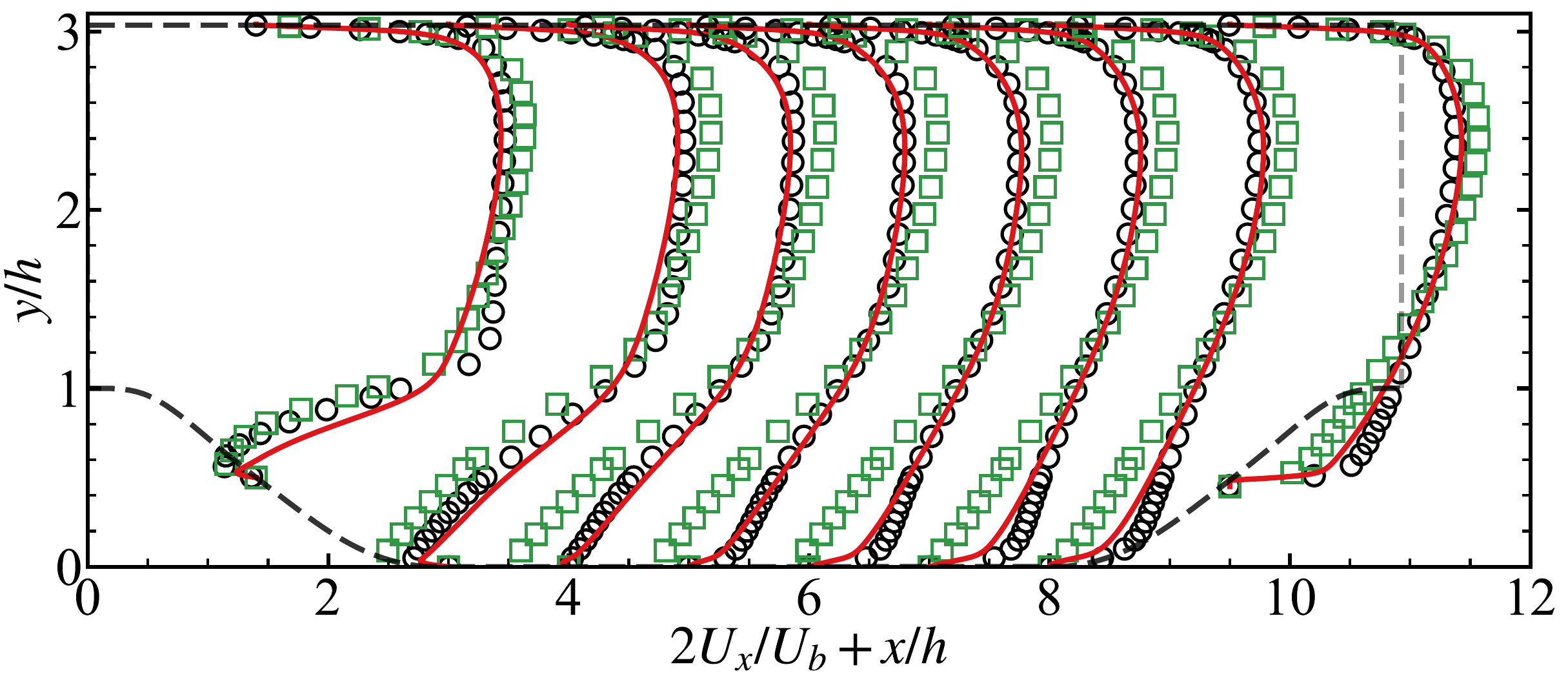} \put(4.8,43.7){$(f)$}
        \put(7.1,38){\small{$57.0\%$}}
    	\end{overpic}
        }
        
    \vspace{0.01cm}

    \subfigure{
        \begin{overpic}[height=0.313\textwidth]{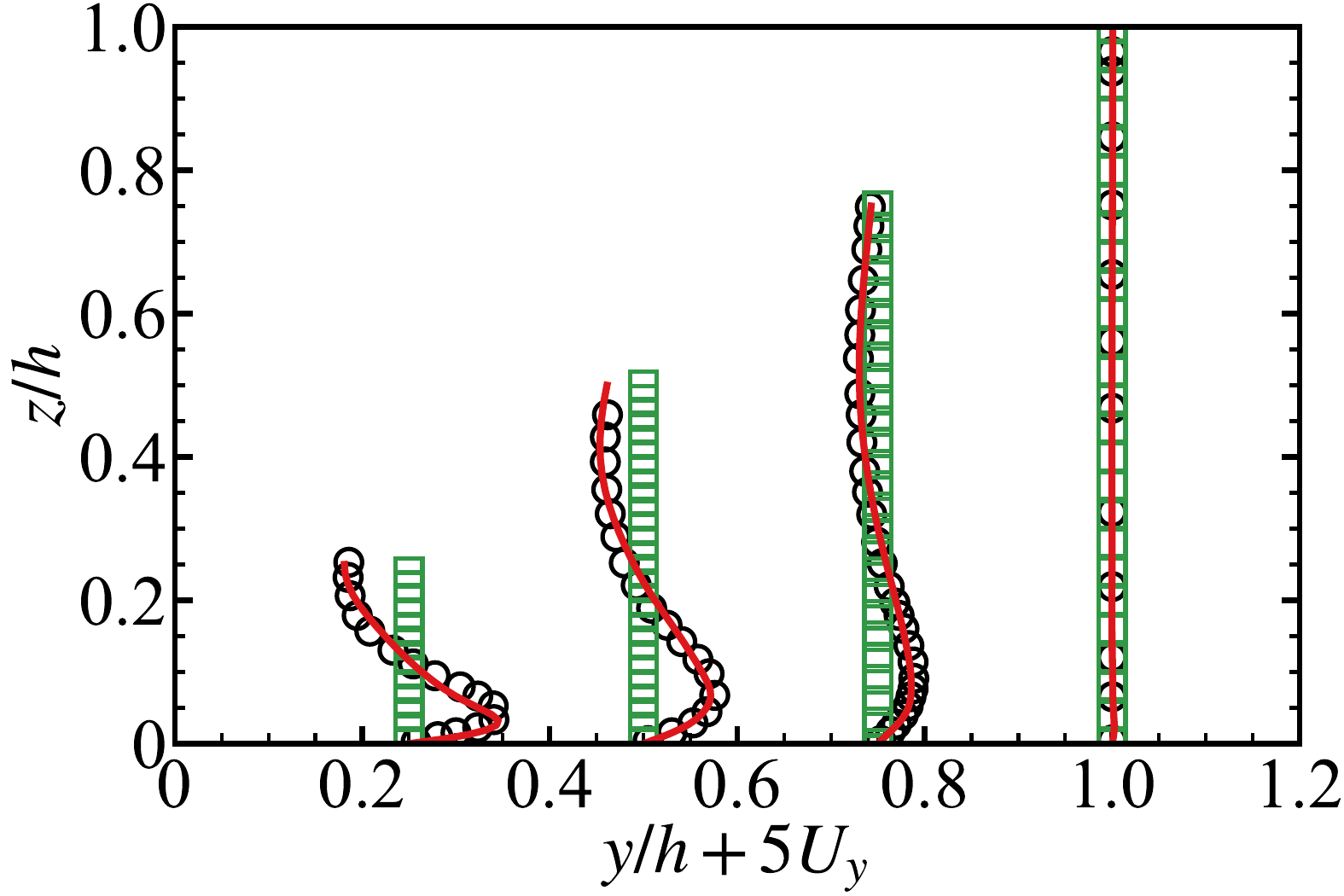} \put(13,66.4){$(g)$}
        \put(14.5,60){\small{$84.4\%$}}
    	\end{overpic}
        }  
    \hspace{0.08cm}
    \subfigure{
        \begin{overpic}[height=0.313\textwidth]{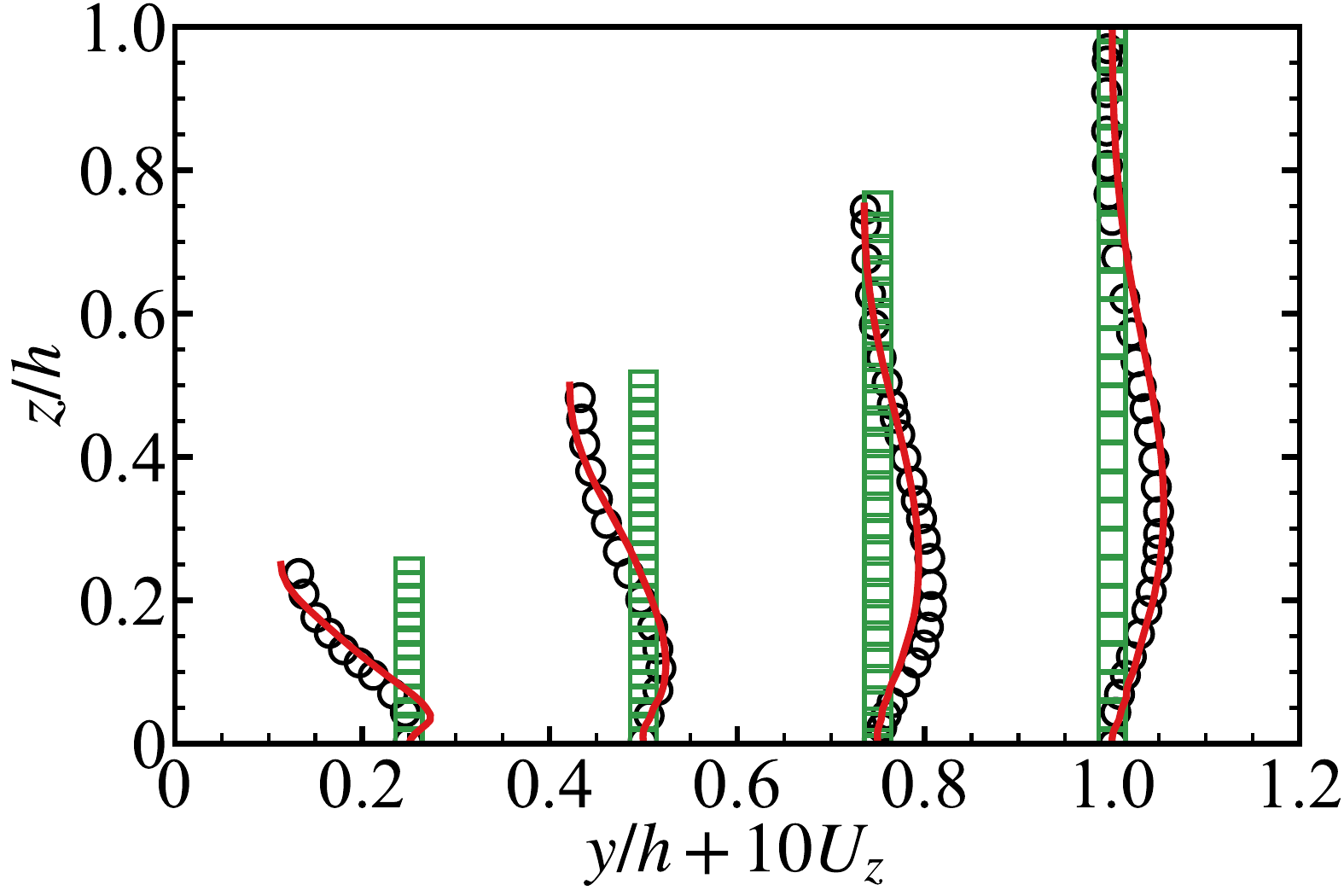} \put(13,66.4){$(h)$}
        \put(14.5,60){\small{$76.5\%$}}
    	\end{overpic}
        }

    \captionsetup{justification=justified, singlelinecheck=true}
    \caption{Mean velocity profiles for unseen cases with different operating conditions. (a) C8000. (b) PH0p8. (d) PH1p2. (f) PH1p5. (c), (e), (g), (h) SD3500.
    Symbols denote: 
    $\circ$ DNS/Experimental data; 
    \textcolor{green}{$\Box$} Baseline SA; 
    \textcolor{red}{\rule[0.5ex]{1.5em}{0.8pt}} PMoE-S3.
    The number shown in the top-left corner of each panel is the relative error defined by equation~\ref{RMSE_equ}.
    }

    \label{fig:validation_unseen}
\end{figure}

The robustness of the PMoE-S3 model is further tested against cases with different operating conditions never seen during training. 
The mean velocity profiles of all the cases are shown in figure~\ref{fig:validation_unseen}. 
Compared with the cases in \textsection~\ref{subsection:validation_trained}, these new tests evaluate the model's generalization ability to different Reynolds numbers and periodic hill geometries.

The model is applied to a channel flow at $Re_\tau=8000$  and a square duct at $Re=3500$. In both cases, the Reynolds numbers are different from the specific training conditions. The router successfully maps these flows to Experts $E_1$ and $E_3$, respectively, as shown in table~\ref{tab:router_results}. The velocity profiles show excellent agreement with DNS data, indicating that the learned corrections, including the damping function modification in $E_1$ and the QCR coefficient field in $E_3$, capture the underlying physics rather than merely overfitting the training data.

We further test the periodic hill flows with varying slopes $\alpha = \{0.8, 1.2, 1.5\}$. These geometric changes significantly alter the adverse pressure gradient and the separation bubble size. 
As shown in figures \ref{fig:validation_unseen}(b, d, f), the PMoE-S3 model consistently outperforms the baseline, despite that the model has only been trained for the $\alpha=1.0$ case. 
Similar to the conclusion drawn from \textsection~\ref{subsection:validation_trained}, the baseline SA model overestimates the size of the separation bubble in the lower hill area, while the PMoE-S3 model shows much better agreement with the result of DNS as shown in table~\ref{tab:Separation_location_PH}.
This suggests that the FIML-trained expert $E_2$ has learned a generalized correction for separation control that is robust to geometric deformation.

Up to now, the PMoE-S3 model has shown satisfactory predictive accuracy in a series of testing cases with different operating conditions, successfully overcoming catastrophic forgetting. 
These results indicate that, although each expert model is trained for one or two operating conditions, it can actually handle a relatively wide range of operating conditions within the same geometry. 
For extreme operating conditions that exceed the generalization boundaries of the expert models, however, the PMoE framework can be extended by fine-tuning only the relevant expert and its router, a far more efficient approach than retraining a monolithic model, highlighting its practical advantage.

\subsection{Computational efficiency}

A key practical consideration for the PMoE framework is its computational cost during inference. 
The complete inference pipeline consists of three stages: (i) a baseline SA simulation, from which input features are extracted; (ii) autoencoder-based routing; and (iii) a corrected RANS simulation using the activated expert. In many industrial workflows, a baseline RANS solution is already available as part of the standard design or analysis process. In such cases, only stages (ii) and (iii) constitute additional cost. When the router assigns the case to Expert $E_0$ (the standard SA model), the baseline solution is itself the final prediction, and no additional simulation is required.

To quantify the computational overhead, we performed wall-clock timing tests using both the baseline SA model and the PMoE-S3 model on identical hardware\footnote{All tests were conducted on a system equipped with an Intel\textsuperscript{\textregistered} Xeon\textsuperscript{\textregistered} Gold 6530 CPU and an NVIDIA\textsuperscript{\textregistered} AD102 (GeForce RTX 4090) GPU.}. The results are summarised in table~\ref{tab:cpu_time}.
Across all tested cases, the routing time remains below 10 seconds, confirming that the autoencoder-based router is sufficiently lightweight for its overhead to be negligible. The full-pipeline overhead, which accounts for the baseline SA run, ranges from approximately $100\%$ to $104\%$, reflecting the cost of effectively running RANS twice. This represents the worst-case scenario in which no prior baseline solution exists. 

Importantly, owing to the sparse-activation mechanism, model expansion through continual learning does not lead to a proportional increase in inference cost. Regardless of the total number of experts in the PMoE model, only one expert is activated per case, ensuring that the per-case simulation cost scales with the size of a single expert rather than the entire ensemble.

\begin{table}
  \begin{center}
    \begin{tabular}{
      >{\centering\arraybackslash}p{1.2cm}
      >{\centering\arraybackslash}p{2.7cm}
      >{\centering\arraybackslash}p{1.6cm}
      >{\centering\arraybackslash}p{2.5cm}
      >{\centering\arraybackslash}p{1.6cm} 
      >{\centering\arraybackslash}p{2.4cm}
    }
      \textbf{Case}  &\makecell{$T_{\mathrm{base}}$ \\ \textbf{Baseline Simulation}}   &\makecell{$T_{\mathrm{route}}$ \\ \textbf{Routing}}    &\makecell{$T_{\mathrm{exp}}$ \\ \textbf{Expert Simulation}}   &\makecell{$T_{\mathrm{tot}}$ \\ \textbf{Total}} &\makecell{{$(T_{\mathrm{tot}}-T_{\mathrm{base}})/T_{\mathrm{base}}$} \\ \textbf{Increase Ratio}}\\
      \hline
      ANW            & $1943s$                  & $9.82s$            & $-$               & $1952.82s$    & $0.51\%$ \\
      C5200          & $425s$                   & $8.01s$             & $434s$               & $867.01s$    & $104.00\%$ \\
      PH1p5          & $2810s$                  & $6.80s$             & $2884s$               & $5700.80s$   & $102.88\%$  \\
      SD5693         & $2077s$                  & $9.29s$             & $2124s$               & $4210.29s$    & $102.71\%$  \\
      
    \end{tabular}
    \captionsetup{justification=justified, singlelinecheck=true}
    \caption{Comparison of computational wall-clock time between the baseline SA and PMoE-S3 models. 
Here, $T_{\mathrm{tot}} = T_{\mathrm{base}} + T_{\mathrm{route}} + T_{\mathrm{exp}}$.}
    \label{tab:cpu_time}
  \end{center}
\end{table}

\section{Discussion} \label{sec:discussion}

The results in \textsection~\ref{sec:validation} have established that the PMoE framework achieves accurate predictions across canonical flows with different operating conditions while effectively avoiding catastrophic forgetting through its modular design. 
Two key questions arise when considering practical deployment: (i) how the model generalises to flow configurations that differ from the training set not only in operating conditions but also in geometry and boundary-condition type; and (ii) how the current routing strategy performs when multiple flow regimes coexist within a single computational domain, and what alternative strategies may be considered. 
To address these questions, \textsection~\ref{sec:discussion_new_cases} presents additional test cases with geometrically distinct configurations, and \textsection~\ref{sec:discussion_top1} investigates routing strategies for multi-regime flows using a representative three-dimensional configuration.

\subsection{Generalisation to geometrically distinct configurations} \label{sec:discussion_new_cases}

\begin{figure}
    \hspace*{-1.2cm}
  \begin{overpic}[height=0.44\textwidth]{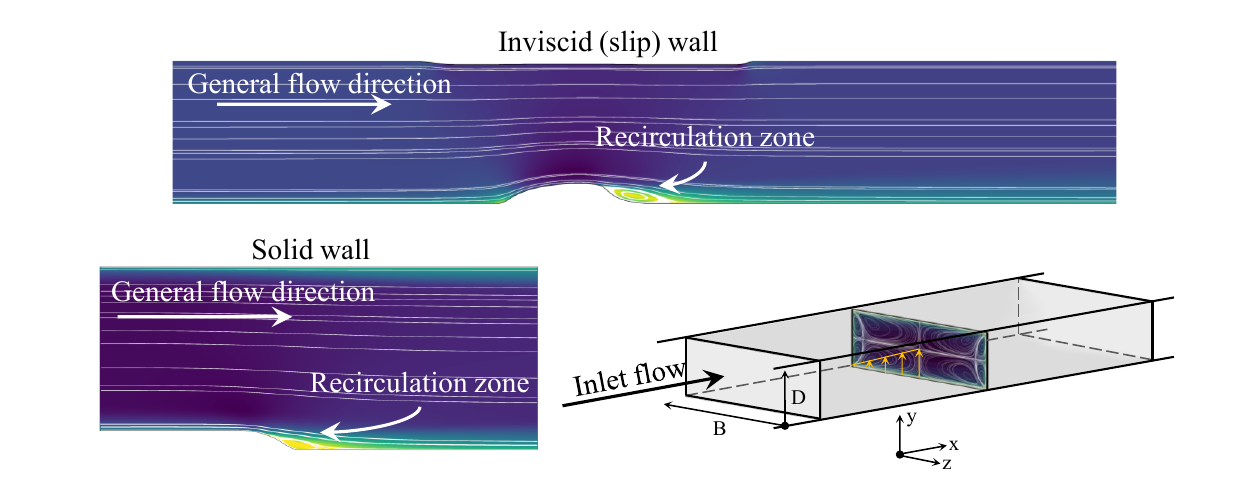} 
        \put(14.8,35){$(a)$}
        \put(55.5,18.7){$(c)$}
        \put(7.8,18.7){$(b)$}
    \end{overpic}
  \captionsetup{justification=justified, singlelinecheck=true}
  \caption{Schematic diagrams of additional test cases with different configurations. (a) 2DWMH. (b) CBFS. (c) RD2500.}
\label{fig:flow_regimes_new}
\end{figure}

The PMoE framework's generalisation capability arises from two complementary mechanisms: the router, which identifies the most relevant expert for a given flow based on learned feature-space representations, and the intrinsic generalisation of the expert models themselves, which encode physically meaningful corrections rather than case-specific data fitting. 
To probe the limits of both mechanisms, we introduce three additional test cases spanning distinct geometrical configurations as shown in figure~\ref{fig:flow_regimes_new}, including the 2D NASA wall-mounted hump separated flow (2DWMH), the curved backward-facing step at $Re = 13,700$ (CBFS), and the rectangular duct flow at $Re = 2500$ with an aspect ratio of three (RD2500). 
These cases are deliberately chosen to test the framework on geometries and boundary conditions that were not encountered during training, while retaining physical mechanisms, i.e., separation under adverse pressure gradients and corner-induced secondary flow, that overlap to varying degrees with the trained regimes. 
Notably, the 2DWMH case also incorporates an upstream zero-pressure-gradient boundary layer developing beneath a free stream, thereby simultaneously testing the coupling between free-stream, wall-attached, and separated flow regions within a single configuration.

\subsubsection{Router discrimination on unseen geometries} 

Table~\ref{tab:router_results_new} summarises the confidence distribution of the PMoE-S3 router for the three additional cases. The confidence levels span a considerable range, from a maximum of $90.8\%$ for RD2500 to $54.7\%$ for the most challenging case (2DWMH). 
These differences can be understood by examining how the geometric and physical characteristics of each new case align with the feature signatures learned during training, as characterised by the permutation feature importance analysis in appendix~\ref{appA}.

The RD2500 rectangular duct preserves the most similar features of the training square duct flow. 
It is a fully developed, pressure-driven wall-bounded flow with no-slip conditions on all walls and corner-induced secondary motions generated by Reynolds-stress anisotropy. 
The only systematic deviation is the aspect ratio, which changes from unity to three, modifying the relative strength and spatial extent of the corner vortices. 
As a result, the flow retains the same wall-driven shear representation, and the router assigns the highest confidence ($90.8\%$) to the square duct class $C_3$.

The CBFS case shares the dominant physics of the periodic hill training case, an adverse pressure gradient driving shear-layer separation followed by reattachment, but introduces a sharp-to-curved geometric expansion that alters the pressure-gradient distribution and shear-layer inception. 
Although the geometric expansion differs, the resulting flow structure remains governed by pressure–shear interaction. 
Consequently, the feature representation remains close to that of the periodic hill, and the router identifies class $C_2$ with a confidence of $76.7\%$, reflecting a substantial overlap in the underlying flow dynamics.

The 2DWMH case likewise involves separation driven by an adverse pressure gradient, yet the separation occurs over a convex hump surface in an external-flow setting rather than within a confined channel, introducing greater geometric and boundary-condition dissimilarity relative to the periodic hill. 
In particular, the presence of an upstream attached boundary layer and the absence of confinement lead to a more mixed flow character. This reduces the similarity in the encoded representation, resulting in a lower router confidence of $54.7\%$ for $C_2$, with a secondary attribution of $35.2\%$ to the channel class $C_1$.

The pattern emerging from table~\ref{tab:router_results_new} is physically intuitive. 
Flows that retain stronger similarity to the training regimes yield higher router confidence, whereas configurations with more pronounced geometric and boundary-condition deviations exhibit lower and more distributed confidence.

\begin{table}
  \begin{center}
    \begin{tabular}{
      >{\centering\arraybackslash}p{1.6cm}
      >{\centering\arraybackslash}p{1.6cm}
      >{\centering\arraybackslash}p{1.7cm}
      >{\centering\arraybackslash}p{2.6cm}
      >{\centering\arraybackslash}p{2.2cm} 
      >{\centering\arraybackslash}p{2.4cm} 
    }
      \textbf{Case} & \textbf{Unknown} &\makecell{$C_0$ \\ \textbf{Wake}} & \makecell{$C_1$ \\ \textbf{Channel}} & \makecell{$C_2$ \\ \textbf{Periodic hill}} & \makecell{$C_3$ \\ \textbf{Square duct}} \\
      \hline
      \textbf{2DWMH}      & $10.0\%$         & $0.0\%$                       & $35.2\%$                  & $\boldsymbol{54.7\%}$                & $0.0\%$  \\
      \textbf{CBFS}       & $0.9\%$         & $0.0\%$                       & $22.4\%$                  & $\boldsymbol{76.7\%}$                & $0.0\%$  \\
      \textbf{RD2500}     & $0.0\%$         & $0.0\%$                       & $1.8\%$                 & $7.4\%$                & $\boldsymbol{90.8\%}$  \\
    \end{tabular}
    \captionsetup{justification=justified, singlelinecheck=true}
    \caption{Confidence distribution of the PMoE-S3 router over different components for additional flow cases.
    }
    \label{tab:router_results_new}
  \end{center}
\end{table}

\subsubsection{Expert prediction} 
\begin{figure} 
  \hspace{-0.1cm}
    \vspace{0.1cm}
    \centering
    \subfigure{
        \begin{overpic}[height=0.323\textwidth]{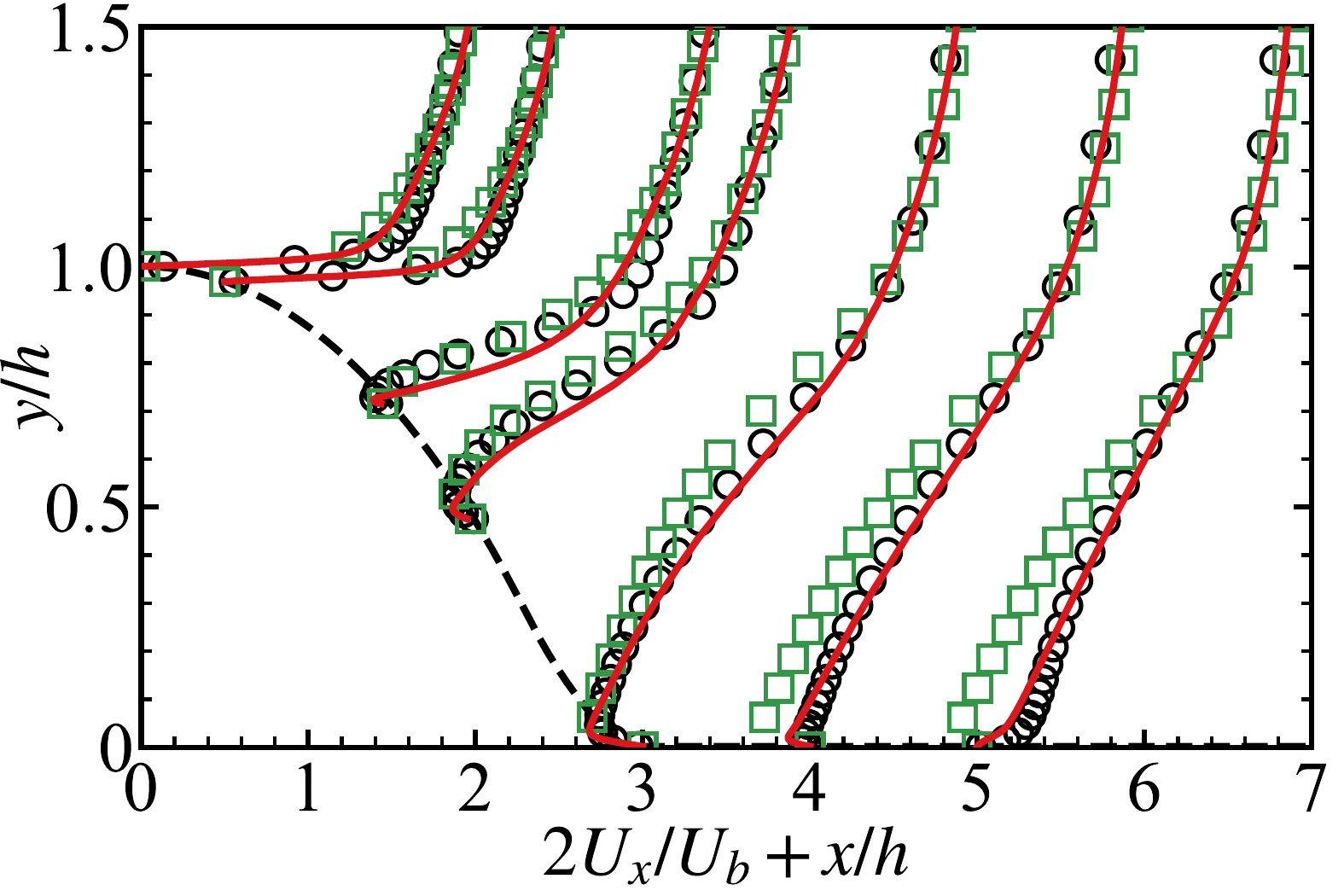} \put(10,66.4){$(a)$}
        \put(12,60){\small{$60.1\%$}}
    	\end{overpic}
        }  
    \subfigure{
        \begin{overpic}[height=0.323\textwidth]{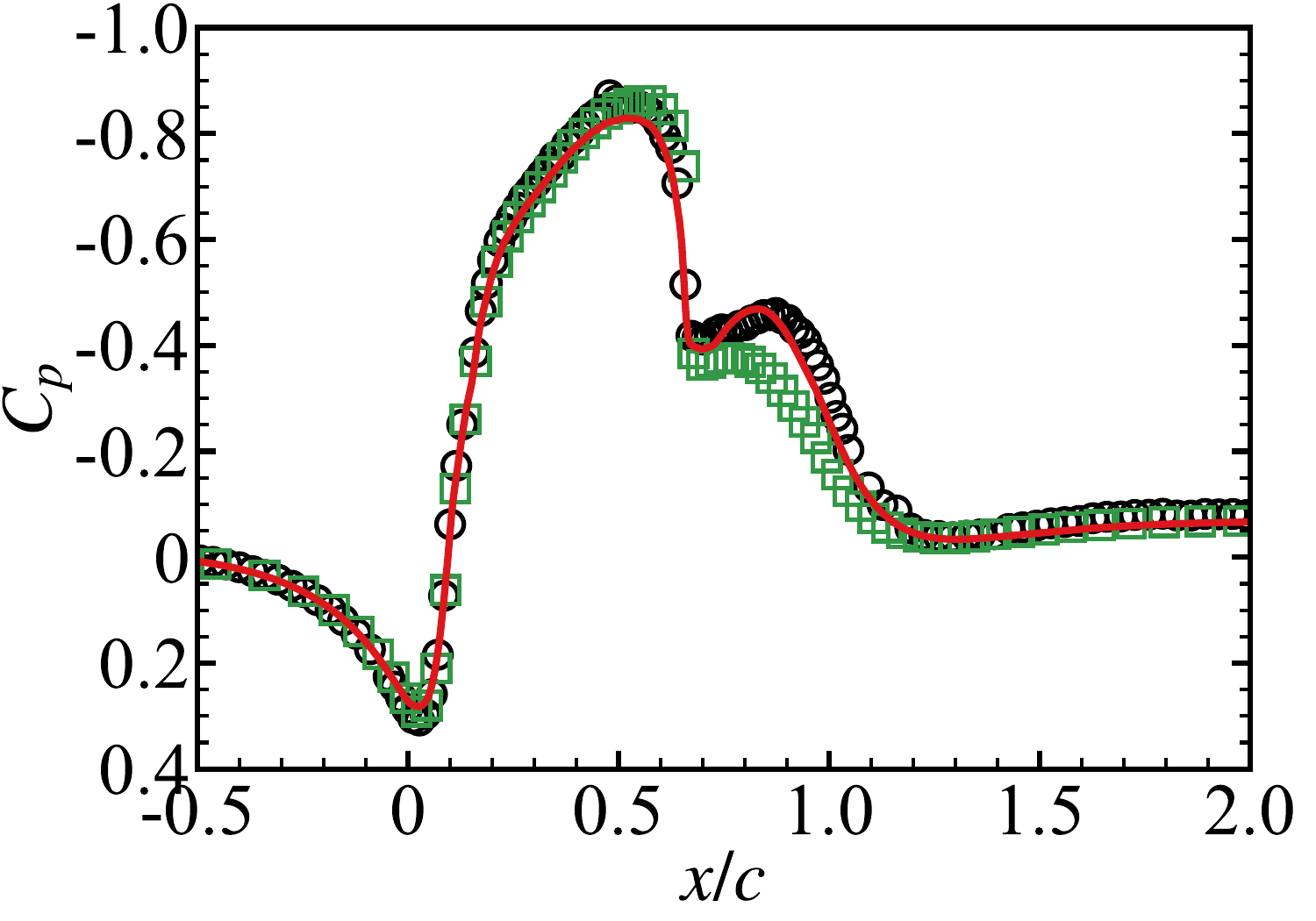} \put(14,69.4){$(b)$}
        \put(17,63){\small{$58.8\%$}}
        \end{overpic}
        }  

    \subfigure{
        \begin{overpic}[height=0.313\textwidth]{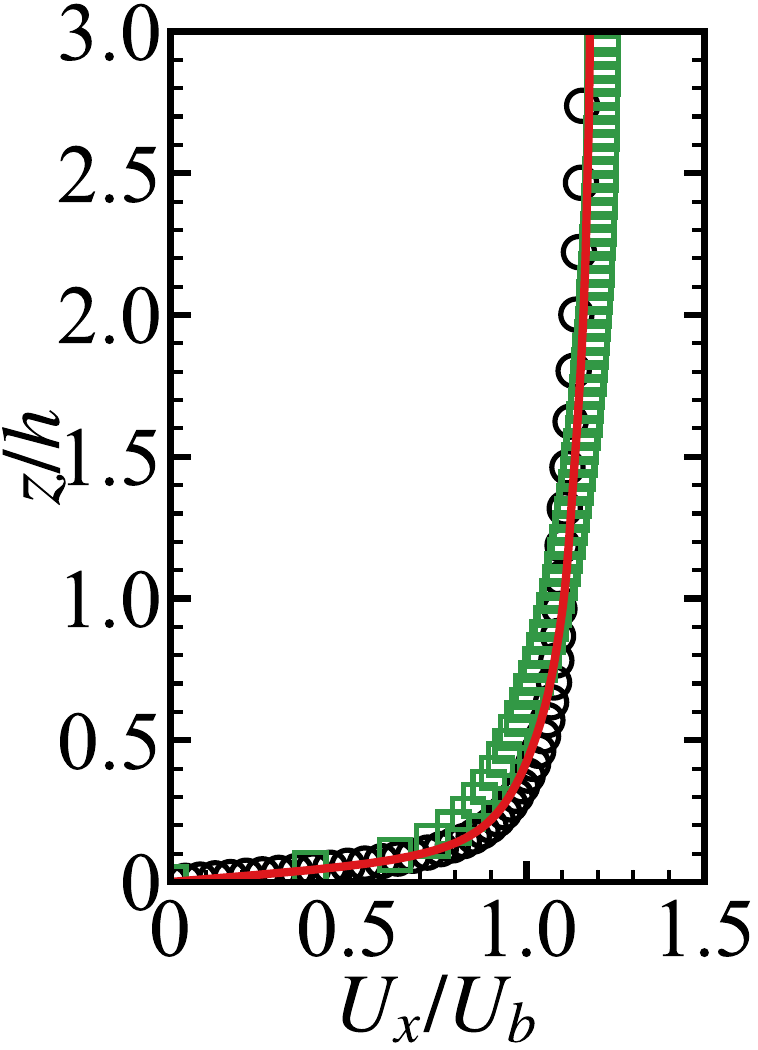} \put(15.5,100){$(c)$}
        \put(18,90){\small{$77.9\%$}}
    	\end{overpic}
        }  
    \subfigure{
        \begin{overpic}[height=0.313\textwidth]{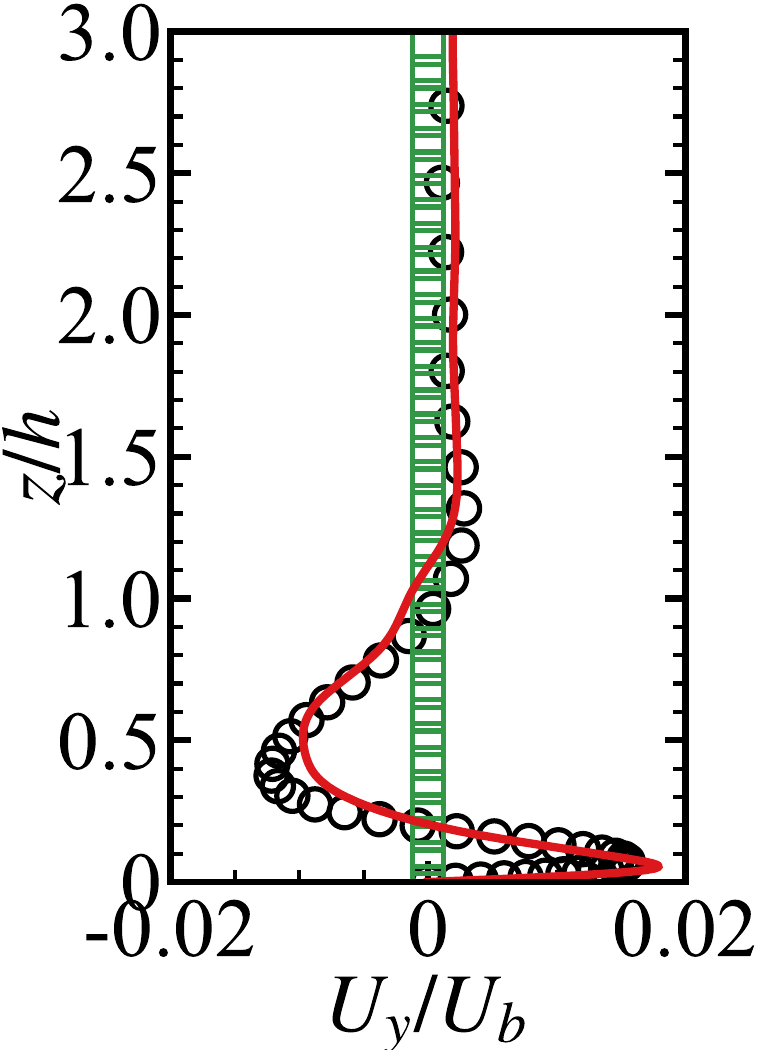} \put(15.5,100){$(d)$}
        \put(18,90){\small{$78.6\%$}}
    	\end{overpic}
        }  
    \subfigure{
        \begin{overpic}[height=0.313\textwidth]{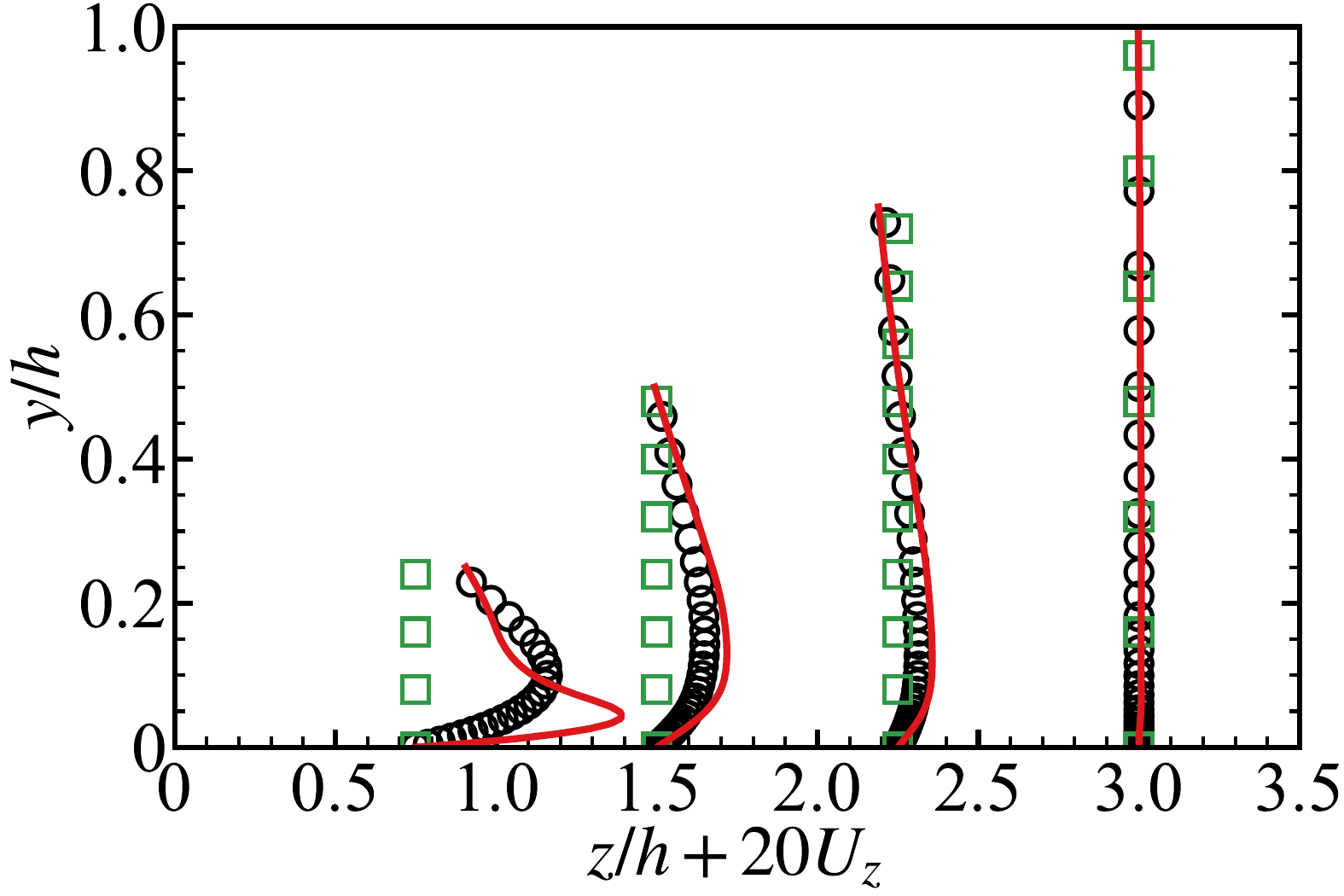} \put(12,66.4){$(e)$}
        \put(15,60){\small{$37.7\%$}}
    	\end{overpic}
        }

    \captionsetup{justification=justified, singlelinecheck=true}
    \caption{
    Predictions obtained with the highest-confidence experts for additional test cases. (a) CBFS. (b) 2DWMH. (c, d, e) RD2500.  
    Symbols denote: 
    $\circ$ DNS/LES/Experimental data; 
    \textcolor{green}{$\Box$} Baseline SA; 
    \textcolor{red}{\rule[0.5ex]{1.5em}{0.8pt}} Expert model.
    The number shown in the top-left corner of each panel is the relative error defined by equation~\ref{RMSE_equ}.
    }
    \label{fig:validation_new}
\end{figure}

Beyond the router's discriminative capability, the expert models themselves exhibit strong generalization when activated on the basis of the highest confidence assignment.  
For the CBFS case, the periodic hill expert $E_2$ substantially reduces the overprediction of the separation bubble extent produced by the baseline SA model, yielding corrected velocity profiles that faithfully capture the interaction between the adverse pressure gradient and shear-layer growth and accurately locate the reattachment point as shown in figure~\ref{fig:validation_new}(a). 
The reference data are obtained from the LES of \cite{bentaleb2012largeCBFS}.
The underlying correction, a recalibration of the non-equilibrium balance between turbulence production and dissipation under adverse pressure gradients, transfers effectively to this geometrically related but distinct separated-flow configuration. 

The 2DWMH case provides an even more stringent test: although the router confidence is appreciably lower, activating $E_2$ still yields improved pressure coefficient (figure~\ref{fig:validation_new}(b)) distributions, with particular gains in the separated region and the downstream recovery zone where the baseline model fails to predict the correct separation onset and pressure recovery. 
The reference data are obtained from the experiment of \cite{naughton2006skinHUMP,greenblatt2006experimentalHUMP2,greenblatt2006experimentalHUMP1}.
This result demonstrates that the FIML-trained expert encodes transferable corrections for the fundamental mechanisms governing turbulent separation.
Specifically, the suppression of excessive turbulence production in adverse-pressure-gradient regions that remain effective even when the external geometry and boundary-layer state differ substantially from those of the training case. 

Finally, for the RD2500 case, the square duct expert $E_3$, which predicts a spatially varying nonlinear stress coefficient $C_\mathrm{cr1}(\boldsymbol{x})$, accurately reproduces the corner-induced secondary vortices driven by anisotropic Reynolds stresses and cross-stream pressure gradients, and improves the streamwise velocity profile relative to the baseline SA prediction, despite the change in aspect ratio. 
This confirms that the expert has encoded the essential physics of corner-induced anisotropy rather than a geometry-specific correction. 
The reference data are obtained from the DNS of \cite{vinuesa2018SquareDuct}.
The mean velocity profiles shown in figures~\ref{fig:validation_new}(c,d) are taken at $y/h = 0.5, h=D/2$, where $D$ denotes the duct height, while those in figure~\ref{fig:validation_new}(e) are extracted along the lines indicated in figure~\ref{fig:flow_regimes_new}(c).

\subsubsection{Synthesis} 
Collectively, these results demonstrate that the generalization capability of the PMoE framework arises from the combined action of its two core components. 
The router provides a physically interpretable confidence measure that quantifies the relevance of each trained expert to an unseen flow and serves as an indicator of similarity in the learned feature space. 
The expert models encode physically meaningful corrections rather than case-specific data fitting, and these corrections remain effective across variations in geometry and boundary conditions as long as the dominant physical mechanism is preserved. 
Even when the router confidence does not reach the acceptance threshold of $90\%$, the expert with the highest confidence can still provide noticeable improvement over the baseline model, as demonstrated by the 2DWMH case. 
Conversely, when no expert is sufficiently relevant, the acceptance threshold $T_\mathrm{accept}$ ensures that the framework defaults to the unmodified baseline model, avoiding the introduction of spurious corrections. 
This combination of a discriminative router and robust physics-based experts provides a degree of generalization and fault tolerance that is not typically achieved by monolithic data-driven models. 
It also offers a practical strategy for extension. 
When the limit of an expert is reached for a substantially different flow configuration, only the relevant expert and its associated router component need to be refined, rather than retraining the entire model.

\subsection{Routing strategies for more complex flows}
\label{sec:discussion_top1}
\subsubsection{Assumptions and limitations of the current routing strategy}

The present PMoE router classifies flow types based on the baseline SA model’s predictions. 
The effectiveness generally rests on several key assumptions.
Specifically, the flow under consideration can be adequately characterized by the features listed in table~\ref{tab:features}; and the baseline SA model provides a qualitatively correct flow field, from which meaningful features can be extracted. As demonstrated in the preceding cases, even when the SA model exhibits deviations, such as overestimating the separation bubble in the periodic hill case or failing to capture the secondary flows in the square duct, these inaccuracies have only a limited effect on the routing outcomes. 
This robustness stems from the design of the router, which considers a comprehensive set of normalized invariant features that characterize the local flow. 

Conversely, the framework may become unreliable when these assumptions are violated. 
This can arise if the baseline model produces a qualitatively incorrect flow topology, such as failing to capture separation or transition, thereby corrupting the extracted features.  

Furthermore, the results presented in \textsection~\ref{sec:validation} and \ref{sec:discussion_new_cases} employ a top-$1$ (winner-takes-all) routing strategy, in which the expert with the highest router confidence is exclusively activated for the entire computational domain. 
This approach is effective when the flow field is dominated by a single physical regime, as demonstrated for the canonical test cases.
For flows closely resembling the trained cases (e.g., RD2500), the corresponding expert can be directly activated for optimal corrections. 
For moderately similar flows (e.g., CBFS and 2DWMH), the router’s top recommendation provides effective corrections even at reduced confidence. 
For unknown cases identified by the PMoE router, the baseline SA model will be invoked, ensuring physically consistent results.
Note that the top-1 routing is similar to industrial strategies such as GEKO \citep{menter2025GEKO}. Applying a single expert globally ensures numerical stability.
While a single expert is applied globally, the corrections it produces are inherently locally determined by the pointwise input features, so the expert remains quiescent in regions where the local flow state does not trigger its correction (e.g., $\beta(\boldsymbol{x}) \approx 1$ in equilibrium regions for Expert $E_2$).

For complex industrial configurations, when multiple flow regimes coexist with comparable importance within a single domain, a single top-1 routing decision may no longer be sufficient to represent the flow physics. 
In such situations, additional reference data or more advanced routing strategies may be required to ensure reliable performance.

\subsubsection{Multi-regime test: the 3D diffuser}

\begin{figure}
  \hspace{-0.2cm}
    \vspace{0.1cm}
  \centering
  \subfigure{
        \begin{overpic}[height=0.335\textwidth]{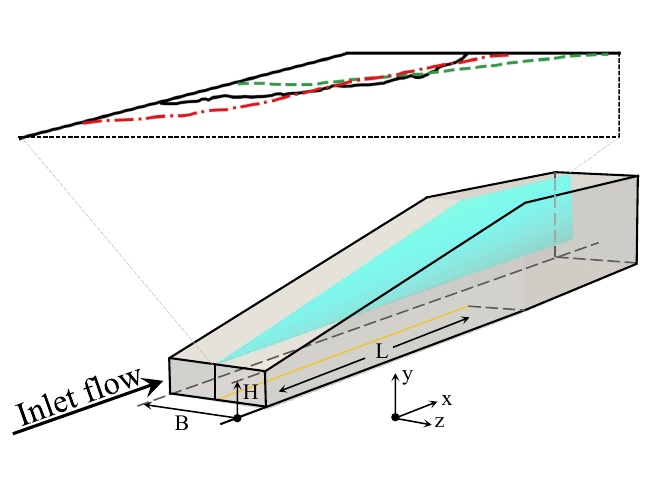}
        \put(7,69){$(a)$}
    	\end{overpic}
        }  
  \subfigure{
        \begin{overpic}[height=0.31\textwidth]{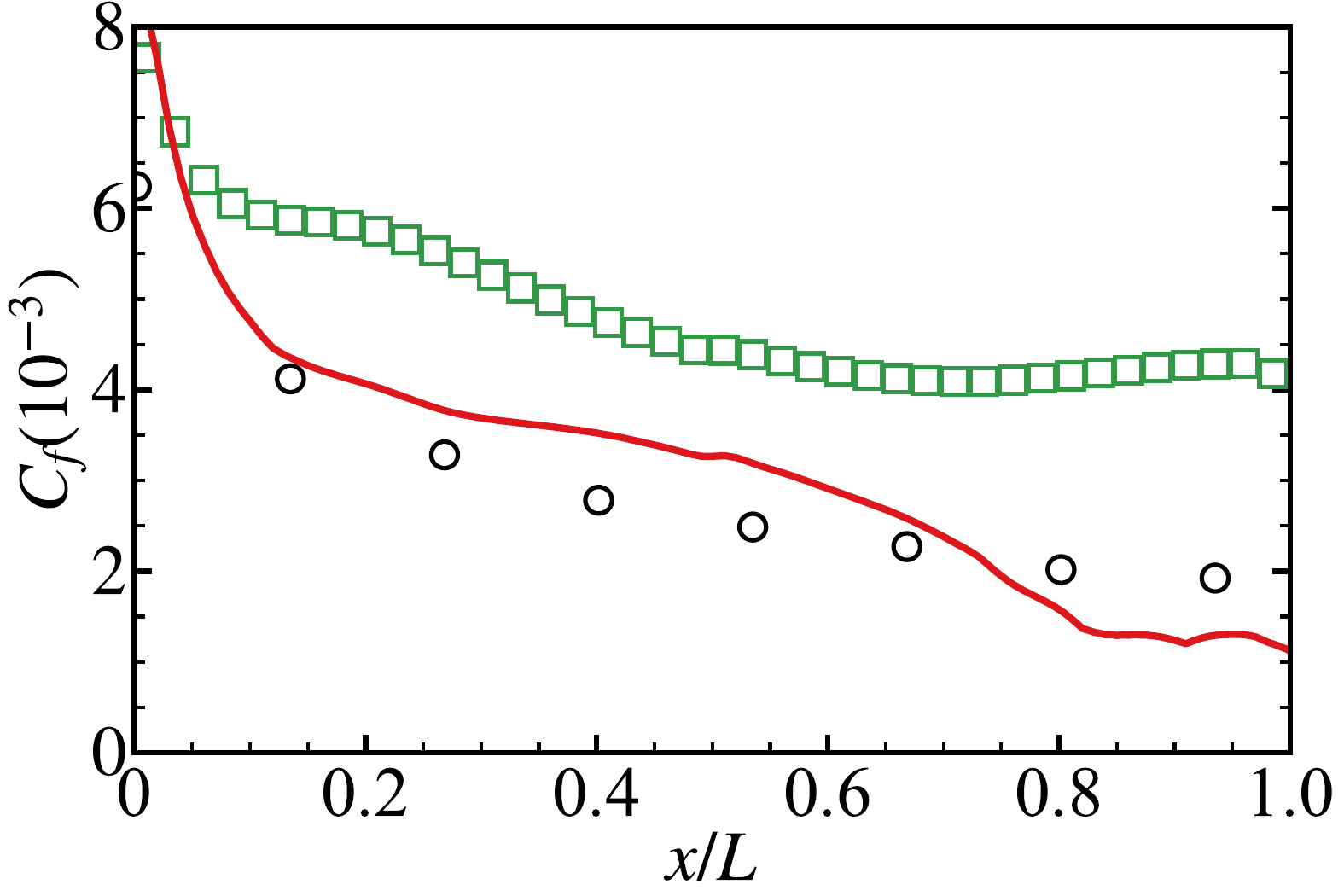}
        \put(9.8,66){$(b)$}
        \end{overpic}
        } 
        
  \subfigure{
        \begin{overpic}[height=0.235\textwidth]{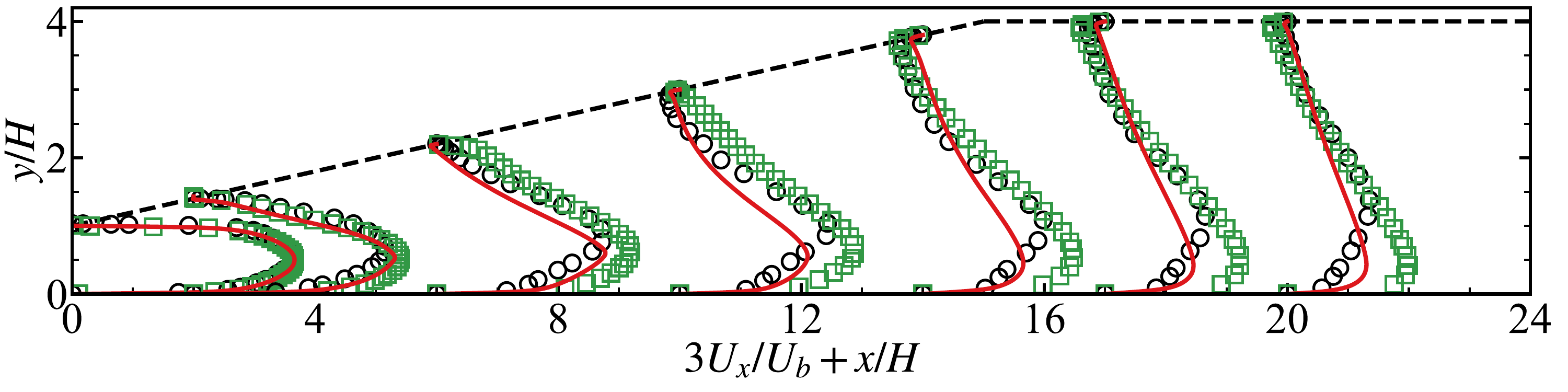}
        \put(4,26){$(c)$}
    	\end{overpic}
        }  
    
  \captionsetup{justification=justified, singlelinecheck=true}
  \caption{Performance evaluation of the mixed model for the 3D diffuser case.
    (a) Zero streamwise-velocity contours on the $x$–$y$ plane at $z=B/2$ (midplane), where black, green, and red denote DNS, baseline SA, and the PMoE experts, respectively.
    (b) Skin-friction coefficient ($C_f$) distribution along the bottom-wall centreline ($z=B/2$), extracted along the yellow line (length is $L$) indicated in (a).
    (c) Mean velocity profiles on the midplane.
    Symbols denote: 
    $\circ$ DNS data; 
    \textcolor{green}{$\Box$} Baseline SA; 
    \textcolor{red}{\rule[0.5ex]{1.5em}{0.8pt}} Expert model.
  }
  \label{fig:3DDiffuser}
\end{figure}

To examine the behaviour of the PMoE framework under conditions of genuinely coexisting flow regimes, we consider the 3D diffuser case \citep{cherry2008Diffuser}. 
As shown in figure~\ref{fig:3DDiffuser}(a), the geometry consists of a straight square duct inlet section that transitions into a rectangular expansion section, where one wall diverges at a prescribed angle while the opposite wall remains flat. 
The boundary conditions include a fully developed duct inflow, no-slip conditions on all walls, and a zero-gradient outflow at the diffuser exit.
The upstream section features fully developed square duct flow with corner-induced secondary motions, while the expansion region introduces an adverse pressure gradient, streamline curvature, and ultimately three-dimensional flow separation from the diverging wall. 
The interplay between the corner vortices carried from the inlet section and the incipient separation renders this flow particularly challenging for RANS turbulence models \citep{abe2010investigationRANSDiffuser}.

This configuration is obviously beyond the scope of the current top-$1$ strategy. 
The router identifies three competing regimes with comparable confidence: the periodic hill class $C_2$ at $40.7\%$, the channel class $C_1$ at $36.5\%$, and the square duct class $C_3$ at $21.5\%$. 
No single component achieves a dominant confidence, reflecting the genuinely multi-physics nature of the case.

As a preliminary exploration of strategies beyond the top-$1$ paradigm, we conducted a test in which all three major expert models ($E_1$, $E_2$, and $E_3$) were simultaneously activated across the entire 3D diffuser domain. 
This combined application exploits the inherently local nature of the expert corrections.
Specifically, $E_1$ primarily acts in near-wall regions through the damping function modification, $E_2$ adjusts the production term via $\beta(\boldsymbol{x})$ in zones of strong adverse pressure gradients and recirculation, and $E_3$ introduces nonlinear stress corrections through $C_\mathrm{cr1}(\boldsymbol{x})$ in corner regions. 
In flow regions that do not exhibit strong similarity to the training cases, i.e., where the local features do not trigger significant corrections, the predicted corrections remain small and the model behavior effectively reverts to that of the baseline SA model. 
As shown in figure~\ref{fig:3DDiffuser}(b), this simple combined strategy already yields encouraging results. 
In particular, the skin-friction coefficient $C_f$ along the bottom wall shows clear improvement compared to the baseline SA prediction and is notably closer to the DNS data of \citet{ohlsson2010directDNSDiffuser}. 
This demonstrates that the modular expert design, while developed for single-regime flows, has the potential to be meaningfully extended to multi-regime configurations.

It should be noted, however, that the 3D diffuser involves a substantially higher degree of geometric and physical complexity than any individual training case, including three-dimensional corner-driven separation interacting with pressure-induced detachment. 
Consistently, the results reveal mixed performance: while the adverse pressure gradient effects near the bottom wall are better captured, as reflected by both the improved $C_f$ distribution and the near-wall velocity profiles (figure~\ref{fig:3DDiffuser}(c)).
In contrast, the velocity profiles in the core region show larger deviations compared to the baseline SA prediction, indicating that the concurrent activation of multiple experts does not uniformly improve the solution throughout the flow field. 
Furthermore, the separation extent inferred from the $u=0$ contour indicates that directly activating multiple experts shifts the separation location from the baseline’s overly downstream prediction to an overcorrected upstream position.
The current expert library, designed and trained on canonical two-dimensional and simple three-dimensional configurations, is not expected to fully capture such interactions. 
The improvement observed here should therefore be viewed as a demonstration of the framework's extensibility and of the transferability of the expert corrections, rather than as a claim that the current model is sufficient for this class of flows.

\subsubsection{Toward region-aware routing}

The 3D diffuser results motivate several directions for extending the current routing strategy. 
A natural next step is region-wise application, in which the computational domain is partitioned into subregions and each subregion is assigned to the expert whose local confidence is highest.
Conceptually, this approach exploits the spatial distribution of the router's reconstruction error to create a patchwork of expert-corrected zones. 
This strategy has the potential to improve predictions in complex flows by locally matching the correction to the prevailing physics. 
However, it introduces practical challenges, including the treatment of interfaces between expert regions, where discontinuities in the correction field may arise and compromise solver convergence, and the sensitivity of the partitioning to the threshold and sampling density.

An alternative and potentially more robust approach is a top-$k$ ($k > 1$) gating strategy with soft blending, in which the corrections from the $k$ most confident experts are combined using continuous weights derived from the reconstruction-error-based confidence $p_k$. 
Because $p_k$ varies spatially through its dependence on local features, such a strategy would enable smooth transitions between regimes without the hard boundaries inherent in region-wise partitioning. 
The modular autoencoder architecture of the PMoE router is inherently compatible with such extensions, as each component can independently evaluate reconstruction error at the point level.

It is important to note, however, that the current heterogeneous expert formulations, which range from symbolic expressions modifying $f_{\nu 1}$ to neural-network-based corrections of different target quantities ($\beta$ for production, $C_\mathrm{cr1}$ for the constitutive relation), preclude straightforward soft blending, since the expert outputs are not directly commensurable. 
A future implementation targeting smooth region-wise routing would benefit from a unified expert format, for example all experts predicting the same correction type such as a Reynolds stress anisotropy tensor or a common multiplicative field. 
This represents a fundamental design trade-off. 
The heterogeneous formulations adopted in the present work maximize physical fidelity by allowing each expert to employ the correction form best suited to its regime, whereas homogeneous formulations would sacrifice some of this flexibility in exchange for the numerical smoothness required for blending. 
Resolving this trade-off, and developing the associated interface-treatment and blending algorithms, will be a key focus of future work. 
We anticipate that such developments will enable improved performance not only for the 3D diffuser case but also for other complex three-dimensional industrial flows in which multiple physical mechanisms interact.

\section{Conclusions}\label{sec:conclusion}

In this work, we propose a progressive modular mixture-of-experts framework, termed PMoE, to address generalization and continual learning in data driven RANS turbulence modelling. 
The framework is based on a successively trainable autoencoder based similarity router that enables the progressive introduction of specialised turbulence model experts as new flow regimes emerge. 
Using training data covering four representative regimes, namely airfoil wakes, channel flows, periodic hill flows, and square duct flows, we construct the PMoE-S3 model and demonstrate its ability to deliver accurate predictions for both seen and unseen test cases.

A key advantage of the proposed framework lies in its support for continual learning without catastrophic forgetting. 
New flow regimes can be incorporated by adding dedicated experts and router component, while previously trained components remain unchanged, preserving established predictive capabilities. 
Moreover, the modular design of PMoE allows experts of fundamentally different forms to coexist within a single framework, offering the flexibility to integrate turbulence models based on different formulations and assumptions as needed. 
Importantly, the training procedure relies on unsupervised feature extraction through the autoencoder-based router, making the framework particularly well suited to industrial workflows where labeled data and repeated full retraining are often impractical.
Beyond its demonstrated performance, PMoE provides a scalable and extensible platform for RANS modelling. 
Owing to the sparse activation mechanism, model expansion does not lead to a proportional increase in computational cost during inference, which is essential for industrial CFD applications.
Its plug-and-play structure enables modular updates and reduces both development and deployment costs, while its extensibility allows the turbulence modelling community to contribute new experts targeting previously unexplored flow regimes, establishing a living framework that can evolve alongside advances in modelling strategies.

Building on the modular and extensible nature of the framework, future developments may incorporate experts based on alternative turbulence closures, such as the $k$–$\omega$ shear stress transport (SST) model \citep{menter1994SST} or Reynolds stress models \citep{launder1975RSTM}, to broaden its applicability. For more complex flows, region aware routing is expected to become increasingly important for applying different experts within different spatial zones. Although the current set of input features is sufficient for distinguishing the flow regimes examined here, further enrichment of the feature representation may be required as additional flow types are considered.
\backsection[Acknowledgements]{This work has been supported by the National Natural Science Foundation of China (Grant Nos.~12588201, 12572247 and 12432010).}

\backsection[Declaration of interests]{The authors report no conflict of interest.}

\appendix
\begin{appen}

\section{Feature importance analysis}\label{appA}

Feature importance analysis is commonly employed to improve the interpretability of neural networks by quantifying the contribution of each input feature to the model prediction \citep{breiman2001random_MDI_ori}.
In this work, to explore whether the importance distribution of different features varies in various flow regimes and thereby demonstrate the feasibility of identifying flow types based on feature importance, it is necessary to conduct a feature importance analysis.

Among such approaches, PFI provides a model‐agnostic assessment of the sensitivity of a fitted model to individual features using a prescribed tabular dataset \citep{fisher2019PFI_all,mandler2023PFI}.
It is achieved by shuffling values of a single feature and subsequently assessing the consequent deterioration in the model performance. 
This approach can determine the extent to which the model depends on a specific feature by disrupting the intrinsic connection between the input features and the model prediction. 
The procedure of the PFI method is illustrated in figure~\ref{fig:PFI}. 
The inputs include a neural network used for reconstruction, an input feature matrix $\mathsfbi{X}\in\mathbb{R}^{N\times M}$ comprising $M$ features and $N$ sampling points, and the corresponding targets, which coincide with $\mathsfbi{X}$ for this reconstruction task.
For each feature $q_i$, a permuted matrix $\mathsfbi{X}_i^S$ is generated by shuffling the $i^\mathrm{th}$ column of $\mathsfbi{X}$, and the reconstruction error $e_i^S$ is recomputed as $\left\|\mathsfbi{X}_{Ri}^S-\mathsfbi{X}\right\|$.
The importance associated with each feature $q_i$ is defined as $\Psi_i$, the change in error induced by permutation, providing a quantitative measure of the model’s sensitivity to that feature.
All features are analyzed in this manner, and the ranking is obtained from repeated trials with averaged results to minimize statistical variability.

\begin{figure}
\hspace{-0.2cm}
  \centerline{
  \begin{overpic}[height=0.41\textwidth]{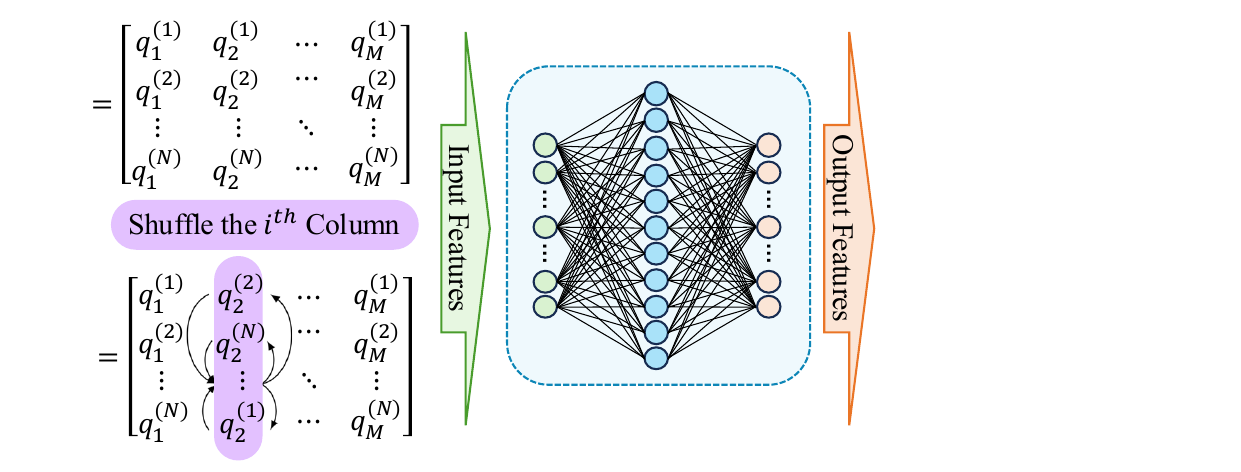} 
        \put(4.1,28.5){$\mathsfbi{X}$}
        \put(4.2,7.9){$\mathsfbi{X}_i^S$}
        \put(71,22){$\mathsfbi{X}_R$}
        \put(71,16){$\mathsfbi{X}_{Ri}^S$}
        \put(79.4,22){$e=\left\|\mathsfbi{X}_R-\mathsfbi{X}\right\|$}
        \put(78.5,16){$e_i^S=\left\|\mathsfbi{X}_{Ri}^S-\mathsfbi{X}\right\|$}
        \put(77,10){$\Psi_i=\left | e_i^S-e \right | $}
        \put(38.5,3.5){Network for Reconstruction}
        \put(69.2,3.5){Feature Importance Score}
    \end{overpic}
  }
  \captionsetup{justification=justified, singlelinecheck=true}
  \caption{Schematic illustration of the formulation and computational procedure of the PFI.}
\label{fig:PFI}
\end{figure}

To eliminate the interference caused by network capacity, 
we choose networks with sufficient parameters to test whether the four types of flowing data we adopt have different order of feature importance in the reconstruction task.
To facilitate the comparison of relative feature importance both within and across heterogeneous datasets, the raw PFI scores were normalized by the sum of seven feature importance scores. 
As shown in figure~\ref{fig:PFI_results_norm}, the bar plots display the normalized mean PFI values, with error bars representing the standard deviation computed from $20$ repeated permutation trials, indicating the stability of the importance estimates.
Features with higher importance scores are considered more informative.
The four flow regimes exhibit distinct importance profiles, indicating that different subsets of features govern the reconstruction.
A more detailed inspection of the dominant feature combinations reveals that each canonical flow type admits a distinct low-dimensional physical representation in the SA-generated solution space. 
In the following, the four baseline flow categories are discussed separately.
For the wake flows, the reconstruction is mainly governed by $q_2$, $q_1$ and $q_3$, indicating that the flow is encoded by shear–rotation dynamics and their interaction with the pressure gradient, consistent with a free shear-layer structure.
For the channel flow, the dominant features are $q_5$, $q_4$ and $q_7$, corresponding to eddy viscosity, wall distance, and shear anisotropy. This indicates that the flow is fully characterized by wall-bounded turbulence scaling, with negligible influence from rotational or pressure-driven effects.
For the periodic hill flow, the importance is distributed across $q_5$, $q_2$, $q_3$ and $q_7$, reflecting the coupled roles of turbulent transport, shear-layer instability, pressure gradient, and recirculation. This confirms that separated flows require a multi-physics representation.
For the square duct flow, the dominant features remain $q_5$, $q_4$ and $q_7$, indicating a wall-driven anisotropic shear system. The weak contribution of $q_2$ and $q_3$ suggests that secondary-flow-related rotational and pressure effects are not explicitly represented in the SA solution.
These class‐dependent importance patterns lead to substantially different parameter distributions when training the modular experts of the router network.
This explains why the approach of using reconstruction errors as the identification basis can effectively distinguish different flow types of data, and still maintain a high accuracy rate after modular expansion.

\begin{figure}
\vspace{0.59cm}
\hspace{0.09cm}
  \centerline{
  \begin{overpic}[width=0.9\textwidth]{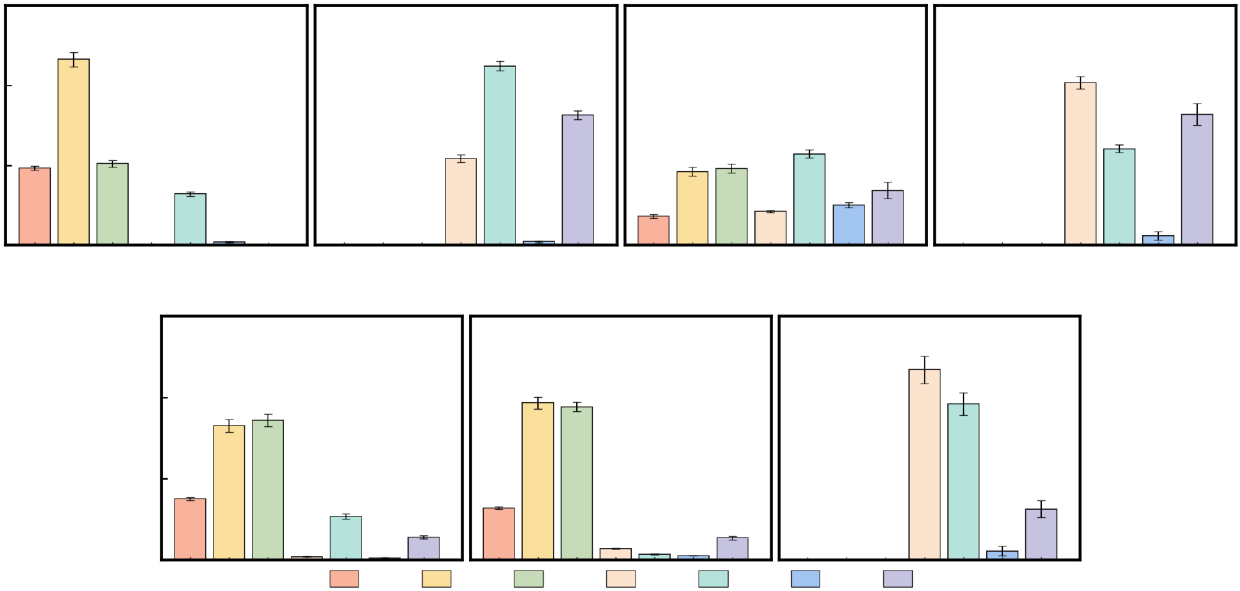} 
        \put(0.3,48){$(a)$}
        \put(9.1,48.4){Wake}
        \put(25,48){$(b)$}
        \put(32.3,48.4){Channel}
        \put(50,48){$(c)$}
        \put(54.8,48.4){Periodic hill}
        \put(75,48){$(d)$}
        \put(80.3,48.4){Square duct}  
        \put(13,23){$(e)$}
        \put(20.2,23){2DWMH}
        \put(37.6,23){$(f)$}
        \put(46,23){CBFS}
        \put(62.5,23){$(g)$}
        \put(70,23){RD2500}

        \put(29.5,0.5){$q_1$}
        \put(37,0.5){$q_2$}
        \put(44.3,0.5){$q_3$}
        \put(51.8,0.5){$q_4$}
        \put(59.3,0.5){$q_5$}
        \put(66.7,0.5){$q_6$}
        \put(74.1,0.5){$q_7$}
        \put(-1.2,27.1){$0$}
        \put(-3.6,33.3){$0.2$}
        \put(-3.6,40){$0.4$}
        \put(-3.6,46.3){$0.6$}
        \put(-4.3,36.5){$\Psi$}

        \put(11,1.8){$0$}
        \put(9,8.3){$0.2$}
        \put(9,14.8){$0.4$}
        \put(9,21.3){$0.6$}
        \put(8.3,11.5){$\Psi$}

    \end{overpic}
  }
  \captionsetup{justification=justified, singlelinecheck=true}
  \caption{Normalized feature importance scores obtained from the PFI for the the training data and additional test cases. (a) Wake. (b) Channel. (c) periodic hill. (d) square duct. (e) 2DWMH. (f) CBFS. (g) RD2500.}
\label{fig:PFI_results_norm}
\end{figure}

For the CBFS case, the dominant contributions from $q_2$ and $q_3$, together with secondary roles of $q_1$ and $q_7$, place it within the same feature subspace as the periodic hill flow. 
Accordingly, the router identifies the periodic hill class $C_2$ with a confidence of $76.7\%$, reflecting a strong overlap in the pressure–shear dominated representation of separated flows.
For the 2DWMH case, the PFI distribution remains dominated by $q_2$ and $q_3$, accompanied by non-negligible contributions from $q_1$ and $q_5$. This indicates that, while the flow retains a separation mechanism governed by pressure–shear coupling similar to the periodic hill, the role of turbulent transport becomes more pronounced. At the same time, the presence of an upstream attached boundary layer introduces additional dependence on wall-related structures, leading to a more distributed feature representation. As a result, the router confidence for the periodic hill class $C_2$ decreases to $54.7\%$, with a secondary attribution of $35.2\%$ to the channel class $C_1$, reflecting the mixed character of external separation and partial wall-bounded behavior.
For the rectangular duct (RD2500), the dominant features remain $q_5$, $q_4$ and $q_7$, closely matching the square duct flow. This consistency in feature space explains why the router assigns the highest confidence ($90.8\%$) to the square duct class $C_3$, indicating that the SA solution preserves the same wall-driven anisotropic shear representation.



\section{Case settings}\label{appB}
Numerical simulations in this paper are performed using the open-source computational framework OpenFOAM (Open Source Field Operation and Manipulation, version \href{https://www.openfoam.com/documentation/guides/v2012/doc/index.html#main-about-openfoam}{2012}). 
The SA baseline model mentioned in this article refers to the model and default parameter settings presented by \cite{spalart1992SA}.
Grids and computation domain are shown in table~\ref{tab:case_setting}.
The numerical setups exhibit minor variations across cases due to differences in geometry and flow conditions, but follow consistent practices based on standard OpenFOAM configurations for canonical flows. A representative setup employs second-order spatial discretization, with Gauss linear schemes for gradients and diffusion terms, and a linear upwind scheme for the convective term of velocity. Turbulence transport equations are discretized using bounded upwind schemes. Pressure–velocity coupling is handled using the SIMPLE algorithm with one non-orthogonal corrector. Standard under-relaxation factors are applied. Convergence is assessed based on residual reduction and stabilization of key flow quantities.

\begin{table}
  \begin{center}
    \begin{tabular}{
      >{\centering\arraybackslash}p{2.8cm}
      >{\centering\arraybackslash}p{3.0cm}
      >{\centering\arraybackslash}p{6.0cm}
    }
       \textbf{Case}   &\textbf{Box}                               &\textbf{Grid}  \\
      \hline
      \textbf{ANW}     &$1000c\times1000c$  & $337\times449, 513\times225$                  \\
      \hline
      \textbf{C2000}   &$4\pi h\times 2h$             & $250\times 100$                   \\
      \textbf{C5200}   &$4\pi h\times 2h$           & $500\times 200$               \\
      \textbf{C8000}   &$4\pi h\times 2h$           & $800\times 320$                   \\
      \hline
      \textbf{PH0p8}   &$8.23h\times3.036h$                 & $99\times149$                   \\
      \textbf{PH1p0}   &$9.0h\times3.036h$                 & $99\times149$                   \\
      \textbf{PH1p2}   &$9.77h\times3.036h$                 & $99\times149$                   \\
      \textbf{PH1p5}   &$10.93h\times3.036h$                 & $99\times149$                                     \\
      \hline
      \textbf{SD2500}  &$2.5D\times D\times D$                 & $75\times48\times48$                  \\
      \textbf{SD3500}  &$2.5D\times D\times D$                 & $75\times48\times48$                  \\
      \textbf{SD5693}  &$2.5D\times D\times D$                 & $75\times48\times48$                  \\
      \hline
      \textbf{2DWMH}   &$6c\times c$                    & $137\times424$                  \\
      \textbf{CBFS}    & $22.7h\times9.5h$             & $712\times104$                   \\
      \textbf{RD2500}  &$2.5D\times D\times 3D$                 & $75\times48\times144$                  \\

    \end{tabular}
    \captionsetup{justification=justified, singlelinecheck=true}
    \caption{Computational domains and grid resolutions for all cases considered in this study.}
    \label{tab:case_setting}
  \end{center}
\end{table}

\section{Training details}\label{appC}

In this appendix, we provide the detailed training procedures for the three experts, $E_1$, $E_2$, and $E_3$, which are used in the PMoE framework. 
These procedures include the generation of training data, the optimisation objectives, and the implementation details of the symbolic regression and neural network models. 
The description is organized sequentially from $E_1$ to $E_3$, highlighting the specific methodologies adopted for each expert. 

\textbf{Expert $E_1$: Symbolic regression of the damping function}

The damping function correction $f_{\nu1}$ for $E_1$ is obtained via offline symbolic regression following the methodology of \cite{Weatheritt2016GEP}. 
Candidate expressions are generated and evolved using gene expression programming (GEP) under predefined function and terminal sets. 
The function set includes the basic operators $\{+, -, \times, \div, \text{exponentiation}, \text{square}, \tanh\}$, while the terminal set comprises the variable $\chi = \tilde{\nu}/\nu$, the constant $1.0$, and randomly generated constants sampled from the interval $[-5, 5)$. 
The GEP algorithm is initialized with a population of $1000$ expressions and evolved over $10000$ generations through mutation and recombination.  

Each candidate expression is evaluated on the training data using the objective
\begin{equation}
\mathcal{J} = \frac{1}{N} \sum_{i=1}^{N} \left( f_{\nu1}^{\text{pred}}(\chi_i) - f_{\nu1}^{\text{target}}(\chi_i) \right)^2,
\end{equation}
where $f_{\nu1}^{\text{pred}}$ denotes the symbolic prediction and $f_{\nu1}^{\text{target}}$ is the corresponding reference value. 
The training data are extracted from the corrected $f_{\nu1}(\chi)$ profiles reported by \cite{bin2023AIAA_SA}, which were originally generated using a neural-network-based approach. 
These profiles capture the near-wall behaviour of the turbulence eddy viscosity, which is critical for accurately predicting the mean velocity distribution in wall-bounded flows. 
At the end of this offline regression stage, the optimal expression $f_{\nu1}(\chi)$ is selected according to the defined objective and is subsequently hard-coded into the SA solver as $E_1$ for RANS simulations, improving near-wall predictions without additional tuning.  

\textbf{Expert $E_2$: Learning the correction of production term for flow separation}

Expert $E_2$ is trained to capture the non-equilibrium physics in the recirculation bubble using a two-step FIML procedure. 
First, an adjoint-based field inversion is performed on the periodic hill case PH1p0 using the DAFoam framework \citep{he2018dafoam,he2020dafoam}. 
The optimisation objective is to minimise the discrepancy between the RANS-predicted velocity field and DNS reference data \citep{xiao2020PH} by adjusting a spatially varying production-term multiplier $\beta(\boldsymbol{x})$, yielding an optimal $\beta$ field on the computational grid.  

Second, a neural network is trained to learn the mapping from the input features $\boldsymbol{x} = [q_1, \ldots, q_7]$ to the inverted $\beta$ field. 
The dataset comprises grid-point-level pairs $\{(\boldsymbol{x}_j, \beta_j)\}_{j=1}^{N_\text{grid}}$ extracted from the PH1p0 case and is randomly divided into $80\%$ training and $20\%$ testing subsets.
Note that no extra sampling is applied at this stage.
The network is implemented as a three-layer fully connected feed-forward network in PyTorch \citep{paszke2019pytorch}. 
All input features are further normalized to the range $\left[0, 1\right]$ using min–max scaling, which is a standard preprocessing step to improve numerical conditioning during neural network optimization.
The network has seven input nodes, two hidden layers with $64$ and $32$ neurons, and a single scalar output. Rectified Linear Unit (ReLU) \citep{glorot2011RELU} activation functions are applied between layers to introduce nonlinearity. 
The network is trained using the Adam optimizer with a cosine-annealed learning-rate schedule and an adaptive reduction strategy triggered by validation-loss plateaus to enhance convergence robustness. The mean-squared-error (MSE) loss function is used during training. 
This compact architecture provides sufficient representational power for nonlinear feature interactions while limiting the risk of overfitting. 
The trained network subsequently injects the predicted $\beta(\boldsymbol{x})$ into the production term of the SA model, as defined in equation \ref{SA_equation_beta}, for subsequent RANS simulations. 

\textbf{Expert $E_3$: Learning the constitutive relation modification for secondary flows.}

Expert $E_3$ is trained following the same two-step FIML procedure as $E_2$, but applied to the square duct cases SD2500 and SD5693. 
First, an adjoint-based field inversion optimises the spatially varying non-linear stress coefficient $C_\text{cr1}(\boldsymbol{x})$ in the QCR formulation (equation \ref{QCR_equation}), with the objective of minimising the discrepancy between the RANS-predicted velocity field—including both streamwise and secondary-flow components—and DNS reference data \citep{pinelli2010SquareDuct3500,vinuesa2018SquareDuct}.  

Second, a neural network with the same architecture and training procedure as $E_2$ is employed to learn the mapping $\boldsymbol{x} \to C_\text{cr1}(\boldsymbol{x})$. 
For brevity, we do not repeat the details of the network architecture and training procedure, which are identical to those described for $E_2$ above. 
The trained network subsequently provides the $C_\text{cr1}(\boldsymbol{x})$ field for injection into the QCR-based RANS simulations.  

Overall, the training of these three experts combines data-driven model extraction and physics-informed optimisation, enabling the progressive mixture-of-experts framework to accurately capture both near-wall and non-equilibrium turbulent flow features across a variety of canonical cases.

\end{appen}\clearpage

%
%
%
%

\bibliographystyle{jfm}
\bibliography{jfm}



\end{document}